# The nontrivial role of interfacial or film-thickness in a magnetic field at a one-electron and a one-Composite Fermion level

G. Konstantinou[1] and K. Moulopoulos[2*]

*Department of Physics, University of Cyprus, PO Box 20537, 1678 Nicosia, Cyprus*

[1]ph06kg1@ucy.ac.cy, [2]cos@ucy.ac.cy

By developing a canonical approach that is exact, physically transparent and subtly different in its line of reasoning from standard approaches (that are usually blended with semiclassical methods), we present a systematic study with exact analytical calculations based on a Landau Level (LL) picture of the energetics of a many-electron system in an interface (or film) and in the presence of a uniform and perpendicular magnetic field, by seriously taking into account the finite thickness of the Quantum Well (QW) in the direction parallel to the field. We find ñinternalò phase transitions (i.e. at *partial* (fractional or irrational) LL filling) for the global magnetization and magnetic susceptibility that are not captured by other approaches, and that give rise to nontrivial corrections to the standard de Haas-van Alphen periods (but in a manner that reproduces the exact quantal deviations from the standard semiclassical periodicity in the limit of the full three-dimensional (3D) space, a problem mostly discussed in astrophysical applications and which we independently solve analytically as well for comparison). Additional features upon inclusion of Zeeman splitting are also found (such as certain energy minima that originate from the interplay of QW, Zeeman and LL Physics in the full 3D problem), while a corresponding calculation in a Composite Fermion picture (with ş -Levels) leads to new universal predictions on magnetic response properties of a fully-interacting electron liquid in a finite-thickness interface**;** these exhibit a richer and more delicate structure than the mere monotonic reduction of gaps with thickness reported long ago, a structure possibly detectable with present day technology. Finally, by pursuing the same line of reasoning for a topologically nontrivial system (with a relativistic spectrum, spin-orbit interactions and strong coupling between thickness and planar motion) we find evidence that similar effects may be operative in the dimensionality crossover of 3D strong topological insulators ($Bi_2Se_3$) to 2D topological insulator (HgTe/CdTe) quantum wells.

## 1. Introduction

There has recently been a surge of interest in the new area of topological insulators [1,2], namely electronic systems characterized by a bulk insulating gap, but also possessing topologically-protected gapless edge (or surface) states ï i.e. dissipationless conducting surface modes, immune to nonmagnetic impurity scattering and geometrical defects. The simplest example of such a phase (with broken time-reversal symmetry) can be found in a two-dimensional (2D) electron gas under a strong perpendicular magnetic field in the Quantum Hall regime. Through a very general bulk-edge correspondence [3] it has been well-established that the number of dissipationless edge states is equal to the integer that comes out from the so called TKNN invariant [4] (or the 1$^{st}$ Chern number in a fiber bundle language [5]) of the occupied energy bands, that is a bulk property (related to the ñvorticityò of the wavefunctions in the magnetic Brillouin zone)**;** in a jellium model picture, the 1$^{st}$ Chern number (or the number of edge states) turns out to be equal to the number of completely filled Landau Levels (LLs) in the Integral Quantum Hall Effect (IQHE) regime. If one now wanted to include the 3$^{rd}$ dimension (i.e. take into account the thickness of the macroscopic quasi-2D sample (interface or film) with i.e. open (rigid) boundary conditions), then a treatment of the above mathematical (topological) properties would be a formidable task (in fact it would spoil the beauty of the standard topological arguments normally applied to the 2D Brillouin zone). Here we point to an alternative general procedure, that is rigorous and based on *physical* (as opposed to purely mathematical) arguments and that seems to have not been discussed in the past; it is based on energy interplays in a one-electron (or one-Composite Fermion) picture, leading to the possibility (in fact, showing the existence) of abrupt changes in the occupancy of transverse (i.e. thickness-related) modes in the ground state, which occur *at partial LL filling* and are accompanied by associated changes in thermodynamic (and possibly transport) properties ï changes that, as it turns out, happen to occur in an interesting (nonintegrable in a certain sense) fashion as the thickness is varied.

The method we are presenting is a canonical ensemble approach (fixed number of particles) and is subtly different from standard (canonical or grandcanonical) approaches (that at some point invoke semiclassical approximations and that usually have mathematical difficulty in dealing exactly with the zero-temperature limit)**;**

**Corresponding Author*

i.e it does not anticipate (or assume) a Fermi sphere in the 3D zero-field limit *as part of the quasi-2D calculation*, but naturally *derives it* in a direct and rigorous manner. The method is exact (no approximations involved whatsoever), it describes the zero-temperature case (although this is immediately generalizable if Fermi factors are included), and ï what is most important ï it is physically transparent at every step of the procedure (hence rather easy to use for other, more involved or exotic, systems as well). The method works directly in *k*-space (by taking careful advantage of anisotropies in different directions) by *not using at all the density of states* (DOS) (the key quantity in all other approaches that, however ï by reducing everything to the energy variable ï basically masks the Physics (i.e. the intermediate physical steps) that take place in *k*-space and that depend on the geometry of each system)**;** it is also not necessary to go through the rather difficult step of first finding the DOS by determining the exact energy spectrum, and this is advantageous, especially if we want to have as much of an analytical control on our solution as possible. Moreover, a physical criterion (of ñequilibriumò) applied to the occupation procedure of a strongly anisotropic system is shown from the results to be superior to the usual semiclassical treatments that lead to the standard ñmagnetic oscillationsò [6]**;** unlike those methods, the present approach leads to exact quantal violations of the de Haas-van Alphen (dHvA) periodicities (in the quasi-2D interface or film), that become smooth quantal deviations (from the dHvA periodicity) in the 3D limit.
The method can actually be useful in a wide range of applications, as the precise role of thickness in various quasi-2D systems seems to be currently attracting considerable attention. By way of an example, mention should be made of bulk Quantum Hall Effect (QHE) measurements in a 3D topological insulator [7] where Shubnikovï de Haas oscillations in highly doped $Bi_2Se_3$ give evidence for layered transport of bulk carriers, with the sample thickness playing an essential role on the quantization of magnetotransport ï but also of the more exciting thickness-related issue of 2D to 3D dimensionality crossover in topological insulators (an issue that is actually briefly touched upon in this paper as will be seen shortly). In the largest part of this work, however, we take a step back, and we present the method in the simplest possible (but still nontrivial) setting**:** we solve exactly thickness-related problems involving an electron gas system in the jellium model (without and with a magnetic field in various dimensionalities) demonstrating that, even in these simplest possible cases, the role of thickness is nontrivial and noteworthy. [The jellium model gives the luxury of dealing with simple LLs, with their number being automatically identified with the topological (Chern) number or the number of edge states (whenever the LLs are completely filled), this giving one the opportunity to possibly identify abrupt changes in the Chern number (when LLs are abruptly depopulated ï as will actually occur many times in this work), with possible interesting consequences on transport properties**;** these, however, deserve a separate article, as this one is focused on *thermodynamic* consequences (i.e. violations of dHvA periods).] Furthermore, since the largest part of our analysis utilizes a jellium model of electrons in extended states, mention should also be made of a 2D semimetal that has recently been observed in wide HgTe quantum wells (QWs) with a broad range of interesting properties [8], and with their thickness still being an important factor not yet seriously studied. Moreover, very recent works on the 5/2-Fractional Quantum Hall Effect (FQHE) [9,10,11] examine the stability of the effect in wide QWs against variation of their thickness and find anomalous features (and it is with this in mind that we have applied the same method by carrying out a thickness-adapted Composite Fermion calculation as will be seen shortly). Mention could also be made of recently studied highly quantum-confined nanoscale membranes, with their thickness being crucial for their (mostly optical) properties [12], as well as of the newly discovered almost free electron gases in oxide heterointerfaces [13]. Finally, and back again to the one-body Physics of the recently-discovered topological insulators, our approach and results may actually cast doubts on the *completeness* of recent findings on a simple oscillatory crossover from a 2D to a 3D topological insulator [14], where transitions between different *z*-modes (with *z* being the direction of the external magnetic field) may not have been treated entirely properly, as will be apparent from the present article ï the point being that, in that work, energy-comparisons are made under the assumption of a given (fixed) transverse mode, not taking into account the energetically favorable possibility of abrupt changes of such modes that may occur in nontrivial ways as the thickness is varied. As we will see in a preliminary study towards the end of this paper, although such transitions may occur at points located a little further than the õ-point in the Brillouin zone, their distance from the õ-point in *k*-space is actually quite small, so that such effects may be operative. (We will actually see that they may occur inside the *k*-space region where the low-energy approximation widely used (namely, a modified Dirac equation) is valid, and at points that are well within an estimated Fermi wavevector $k_f$ resulting from surface carriers).

In order to present our analysis in the jellium model, it is useful to first remind the reader of a little more traditional (in the sense of well-studied) systems than the above**:** e.g. the standard sawtooth behavior of the low-temperature magnetization of an electron gas in 2D interfaces and in the presence of an external perpendicular magnetic field is well-known both from experimental measurements [15] as well as from analytical calculations of the total energy of a noninteracting electron system with the use of a picture of LLs in a canonical ensemble approach (reviewed in Section 2). This sawtooth behavior occurs as a function of the magnetic field (while as a function of the inverse field, the ñsawò has periodic steps, signifying the appearance of (or actually defining) the standard dHvA effect). In this article we go further than these calculations by taking up the issue of nonzero thickness of the interface seriously and by making a systematic study of its role on the ground state energetics of the interface (by also commenting on transport properties). We present extensions of the above type of analytical calculations to a quasi-2D interface (with a finite-thickness QW in the *z*-direction, parallel to the magnetic field) by using rigid boundary conditions at the 2 edges of the QW (i.e. with an infinite potential barrier to represent the vacuum ï similar to the ñopen boundary conditionsò used in the area of 3D topological insulators). We also present independent analytical calculations (extensions of ones that have already been carried out earlier in systems of astrophysical interest) for a fully 3D quantum system of noninteracting electrons in infinite space and in an external magnetic field (now with periodic boundary conditions parallel to the field), all at zero temperature (T=0). Both systems, the quasi-2D interface and the full 3D space, seem to lead to previously unnoticed features in each system's magnetic response properties. For the interface the crucial point is the *single-particle energy competition* (between LLs and the QW-levels) for *the different types of occupation-scenarios* that are possible (and allowed by Pauli exclusion principle) when one attempts to determine the lowest *total* energy of the many-electron system**:** the basic physical reason is that each one-particle state is now characterized by 3 quantum numbers; there are then cases when the system energetically prefers to change (increase) a *z*-mode, and then it can (in fact it must) go back to lower quantum numbers of the 2D motion (in our case LLs) without violating Pauli principle, and in so doing it can acquire a lower (in fact the lowest possible) total energy. It is shown in this paper that the manner that occupancies (and transitions) occur according to the above criteria is an interesting and nontrivial exercise (with the total energy probably not reducible to closed analytical forms immediately when an arbitrary field and an arbitrary thickness are given**;** one has to actually run the occupation scenarios starting from special values of parameters (for which the problem is easy) and then vary these parameters in some well-defined manner until they assume their values under consideration). And when this exercise is carefully and properly solved it defines a sequence of critical fields (or correspondingly of QW thicknesses) where ñinternal transitionsò occur (in the sense that the highest LLs are only partially filled), which in turn lead to a number of new singular features in global magnetization and in magnetic susceptibility. As a result, nontrivial quantal corrections to (or, better, violations of) the standard dHvA periodicities are found. In the independent calculation in the full 3D infinite space, we determine the exact quantal behavior of magnetization that, in strong magnetic fields, is found to also deviate considerably from the standard semiclassical dHvA period (but is also found to rapidly converge to this semiclassical periodicity as the magnetic field is reduced). The complete solution of this latter problem, derived here in closed form, also demonstrates some interesting analytical patterns in terms of Hurwitz zeta functions that seem to have not been properly identified in earlier works. The mathematical problem of how to go analytically from the quasi-2D results to the results of the full 3D system (in the limit of infinite thickness) is also tackled, providing therefore a test and a proof of correctness and consistency of all analytical expressions found here to describe the quasi-2D interface problem. Additional features upon inclusion of Zeeman splitting are also highlighted (such as certain minima in total energy, that originate from the interplay of QW, Zeeman and LL Physics in the full 3D problem), that might possibly be useful for the design of stable 3D quantum devices (in cases i.e. that the magnetic field can be self-consistently considered as self-generated). Furthermore, a corresponding calculation (now with the so-called ş -Levels in place of LLs) in a Composite Fermion picture (in the approximation of noninteracting Composite Fermions) demonstrates the utility of our method, as it leads to new predictions on magnetic response properties of a fully-interacting electron liquid, possessing a certain form of universality, with the finite thickness of the interface playing a major role (although different from earlier works such as [11]). These predictions should be compared with the (much earlier reported) mere monotonic reduction of FQHE gaps with thickness (see [16] for conventional FQHE systems ï while for recent topologically nontrivial systems see [17])**;** in our results they exhibit a richer and more delicate structure (that could possibly be detectable with present day technology).

In the largest part of this article particles are assumed nonrelativistic (with a parabolic spectrum), although a similar procedure for a model system with the Relativistic energy spectrum of Graphene in the plane could be easily followed (something however that is not pursued here). Moreover, the method of energy-interplays presented in this work is immediately extendable to include Rashba or other types of spin-orbit coupling [18,19], although we will not consider this either in full generality in the present article. However, we *do* provide hints of relevance (or of applicability of the present method) to analogous systems, namely systems with topologically nontrivial *k*-space behaviors, such as the dimensionality crossover from a 3D to a 2D topological insulator (systems with strong spin-orbit coupling and with low-energy properties described by a Dirac-type of equation) towards the end of the article.

The paper is organized as follows: Section 2 briefly reviews the energetics (and QHE transport properties) of a 2D system of noninteracting electrons in a perpendicular magnetic field, by placing emphasis on the thermodynamic functions of the system and on how the dHvA periodicities directly come out (although a relevant discussion of transverse conductivity is also briefly made). Section 3 deals with the same system being confined in an interface of nonzero thickness *d*, with no magnetic field applied, and presents a systematic study of the energy behavior for several thicknesses. Even this seemingly trivial problem leads to interesting behaviors (such as a sequence of Fermi circles (or disks) associated with each QW-level, that are generally different from the circular cross-sections (of a 3D Fermi surface) that result from earlier semiclassical treatments (through intersection of $k_z$ with a predetermined Fermi sphere), reproducing those only when there is a large number of QW-levels involved**;** for small QW numbers it is shown that, when the thickness *d* is below a critical value, the system can be considered as two dimensional, while for very large *d* we recover the energy of 3D noninteracting electron gas. Moving forward, in Section 4 we apply on the interface a uniform perpendicular magnetic field *B*, and we study in detail all thermodynamic properties such as energy, magnetization and susceptibility for several values of *d* and *B*, or under combined variation of both, demonstrating that they exhibit a rich pattern of behaviors in a rather unpredictable manner. [Transport properties are also discussed, and they have a great resemblance with the corresponding 2D results, which is rather expected for such a conventional system (being essentially a multilayered QHE system).] An inclusion of Zeeman coupling modifies the results (they now depend strongly on gyromagnetic ratio) and an inclusion of interactions in a Composite Fermion picture gives further, not easily predictable corrections (and a type of universality). Section 5 presents the original electronic problem in full 3D space**:** the electrons are now confined in a large macroscopic cube with periodic boundary conditions along the field direction and we present there exact analytical expressions of all thermodynamic properties (with a method not usually applied to solid state systems but more often to astrophysical treatments). We find in this problem a sequence of Fermi lines (segments) associated with each LL, that again are generally different from results of semiclassical treatments (determined by semiclassical Landau tubes inside (and intersecting) a predetermined Fermi sphere)**,** reproducing those only when there is a large number of LLs involved. But what is more gratifying is that the results are analytically shown to be consistent with the limiting behavior of the corresponding results of the quasi-2D interface, when its thickness goes to infinity (with the fine details of the quasi-2D calculation being essential for reproducing this limit). We also recover (for the full 3D problem) the dHvA periodicities in the limit of weak *B*, while for large *B*ôs we provide the exact quantal violations of (or deviations from) these semiclassical periodicities, and we also give estimates of particle densities for which such violations might be detectable in 3D solid state systems. Finally, in Section 6 we turn our attention to the applicability of our method to the more interesting problem of the dimensionality crossover from a 3D topological insulator (possessing a single Dirac cone on its surface) to a 2D topological insulator quantum well**;** it is briefly demonstrated how this line of reasoning can be pursued even in this case (where the thickness-related modes are strongly coupled to the planar degrees of freedom), and argued that effects of the above type may also be present in these more exotic systems. Section 7 summarizes our conclusions, while an Appendix presents more esoteric mathematical details relevant to the rather complicated patterns appearing in Section 4.

## 2. Nonrelativistic electron gas in 2D in a perpendicular magnetic field

As a precursor to the main results of this work, we begin with the well-known problem of a system of many (*N*) noninteracting electrons (each with charge -*e*, effective mass *m* and spin *s*) that are free to move in a 2D plane in the presence of an external homogeneous magnetic field *B* perpendicular to the plane, at temperature T=0. Let us first for simplicity ignore the Zeeman splitting (i.e. take the gyromagnetic ratio $g^*$=0 ï note however that we consider particles that *do* have spin (i.e. *s* =1/2), providing therefore a slightly more complete treatment than the standard (academic) one with spinless fermions). As is well-known, this simple jellium model accounts for both the thermodynamic and transport properties of electrons as these are observed in experiments on QHE systems (in properties such as magnetization or Hall magnetoresistivities). We should state at the outset that, although these types of systems (interfaces or films) are not purely 2D, we can always reduce their thickness to achieve an effectively two-dimensional system (see Section 3 for the corresponding ñcritical thicknessò (that depends on the areal density of electrons), as this is rigorously determined (at T=0) by our analytical calculations).

It is well-known that the orbital motion of the noninteracting electrons (that satisfy the nonrelativistic Schrodinger equation) in this 2D problem is described by a Landau Level (LL) picture for the single-particle energy spectrum, namely

$$\varepsilon_n = (n + \frac{1}{2})\hbar\omega_c \,, \qquad (2.1)$$

where $\omega_c = {eB}/{mc}$ is the cyclotron frequency, *e* is the absolute value of charge of each electron, and *n* (the LL index) is a non-negative integer (that characterizes all LLs). It is also well-known that each LL has degeneracy 2ū /ū ɞ (accounting for the spin *s* = (1/2) of each electron ï more generally the prefactor being 2*s* + 1), where ū is the total magnetic flux passing through the system and ū ɞ the flux quantum (ū ɞ=*hc/e*). Each LL can then contain 2ū /ū ɞ electrons (due to Pauli principle at T=0) so that in the most general case, when there are Ɉ (a positive integer) LLs occupied by electrons (namely $\rho = n+1$, with $n$ the LL index of the highest occupied level) the following inequality is satisfied

$$2(\rho-1)\frac{\Phi}{\Phi_o} \le N \le 2\rho\frac{\Phi}{\Phi_o}, \tag{2.2}$$

or, equivalently (given that $\Phi = BS$, with $S$ being the total surface area of the sample)

$$\frac{1}{2(\rho-1)} n_A \Phi_o \ge B \ge \frac{1}{2\rho} n_A \Phi_o \tag{2.3}$$

where $N$ is the total number of particles and $n_A = N/S$ is their areal density. (Note that we follow a picture of constant number of electrons (canonical ensemble), although this does not hurt generality as we will see later). When the magnetic field varies in the above window, the electrons occupy $\rho$ LLs (with the last occupied level (of LL index $\rho$-1) not necessarily being completely filled up – a complete filling merely corresponding to an equality in the right side of (2.3)). First, if $\rho$ = 1, valid for $B \ge \frac{1}{2} n_A \Phi_o$, all electrons are accomodated in the lowest LL and the total energy is simply

$$E = N\frac{\hbar\omega_c}{2}$$

it is therefore linear in $B$. For many LLs ($\rho$ > 1) it is easy to sum over all occupied LLs to find the total energy of the system, namely

$$E = 2\frac{\Phi}{\Phi_o}\sum_{n=0}^{\rho-2}\left[\hbar\omega_c(n+\tfrac{1}{2})\right] + \left(N - 2(\rho-1)\frac{\Phi}{\Phi_o}\right)\hbar\omega_c(\rho-1+\tfrac{1}{2}) \tag{2.4}$$

By then using the sums

$$\sum_{n=0}^{\rho-2} n = \frac{(\rho-1)^2 - (\rho-1)}{2}, \tag{2.5}$$

and

$$\sum_{n=0}^{\rho-2}\frac{1}{2} = \frac{(\rho-1)}{2}, \tag{2.6}$$

we can determine the total energy in units of 2D Fermi energy ( $E_f = \frac{\hbar^2 k_f^{\,2}}{2m}$, $k_f = \sqrt{2\pi n_A}$ ), that has the following final form:

$$E = NE_f\left[2(-\rho^2+\rho)\left(\frac{B}{n_A\Phi_o}\right)^2 + 2\left(\frac{B}{n_A\Phi_o}\right)(\rho-\frac{1}{2})\right] \tag{2.7}$$

One immediately notes that the energy varies quadratically with respect to $B$ (for $\rho$>1), so one notes a linear behavior of the magnetization or a constant value of the magnetic susceptibility, quantities that are determined by derivatives of $E$ with respect to $B$, as discussed further below. (As already mentioned, for very strong $B$, i.e. for $B \ge \frac{1}{2} n_A \Phi_o$ (so that $\rho$=1) $E$ is given only by the last term in (2.7) and is linear in $B$, the magnetization being therefore constant and having the value $-N\mu_B$ with $\mu_B$ the Bohr magneton, an "atomic value" of magnetic moment that is expected for almost nonoverlapping particles in the strong field limit (see more general discussion below)).

From application of the first law of thermodynamics at T=0 one can then determine the global magnetization $M$ [it is actually the total magnetic moment of the system, i.e. an extensive quantity, but we will here follow the usual terminology] and magnetic susceptibility $\chi$ through simple derivatives of (2.7), namely

$$M = -\frac{\partial E}{\partial B}\bigg|_V \quad \text{and} \quad \chi = \frac{\partial M}{\partial B}\bigg|_V ,$$

and these turn out to give

$$M = N\frac{E_f}{(n_A\Phi_o)}\left[4(\rho^2-\rho)\frac{B}{(n_A\Phi_o)} - 2(\rho-\frac{1}{2})\right] \tag{2.8}$$

$$\chi = N\frac{E_f}{(n_A\Phi_o)^2}\left[4(\rho^2-\rho)\right] \tag{2.9}$$

(It should be noted that $\chi$ is always non-negative for this 2D case; it is probably useful to state early on that, when we later include a thickness for our interface we will find cases (ranges of parameters) where $\chi$ will also assume negative values). If the magnetization is measured in units of Bohr magneton ( $\mu_B = e\hbar/2mc = E_f/n_A\Phi_o$ ) etc., the above results are represented by the figures shown below

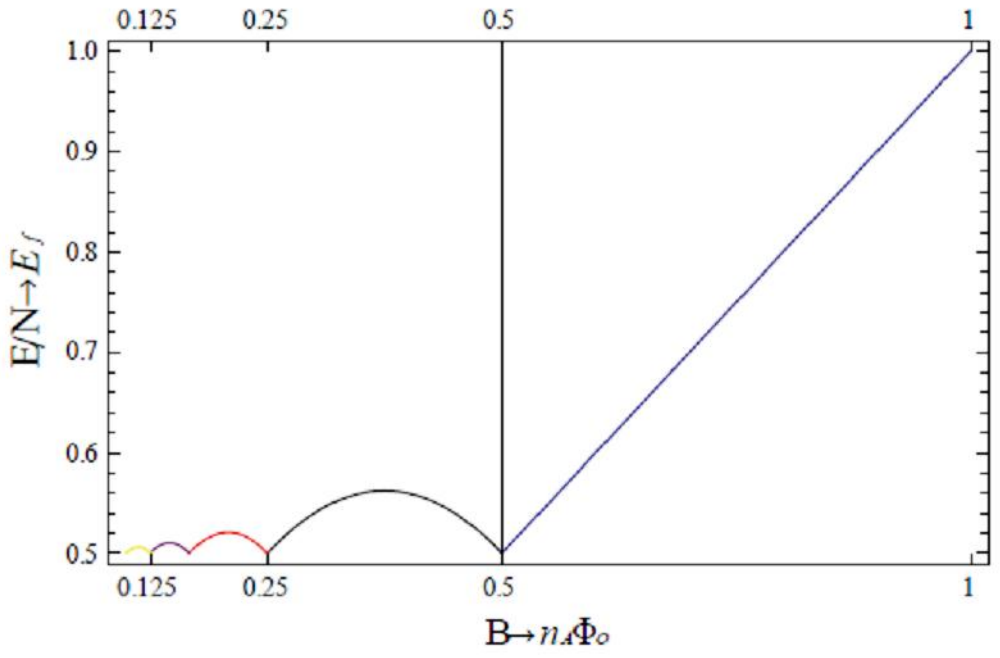

**FIG. 2.1:** Energy per electron (in units of 2D Fermi energy**)**

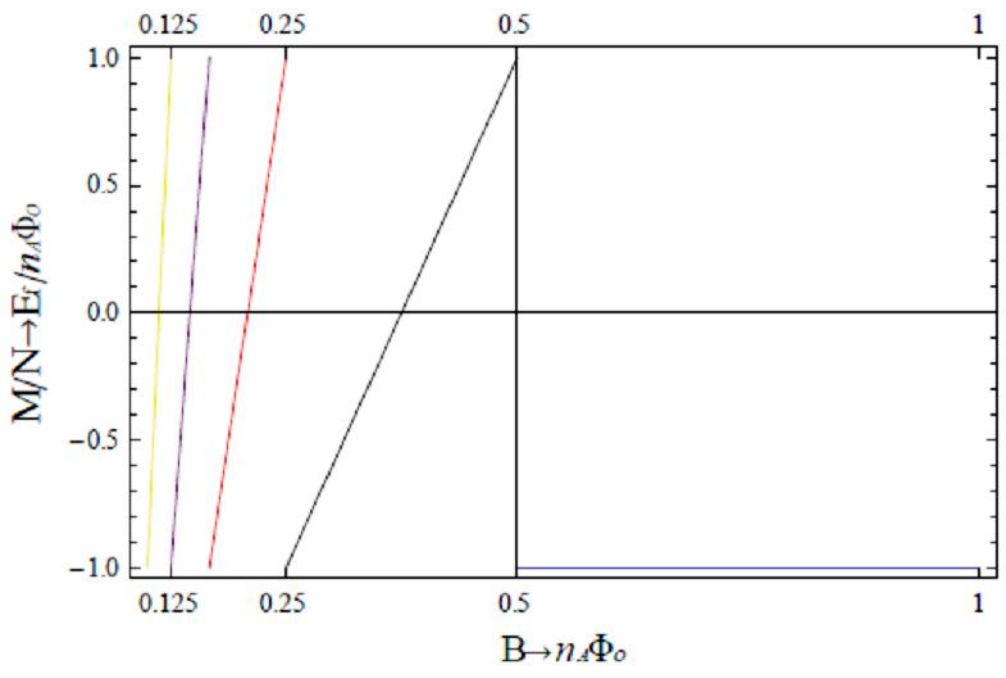

**FIG. 2.2:** Magnetization per electron (in units of $\varepsilon_{\Phi}$)

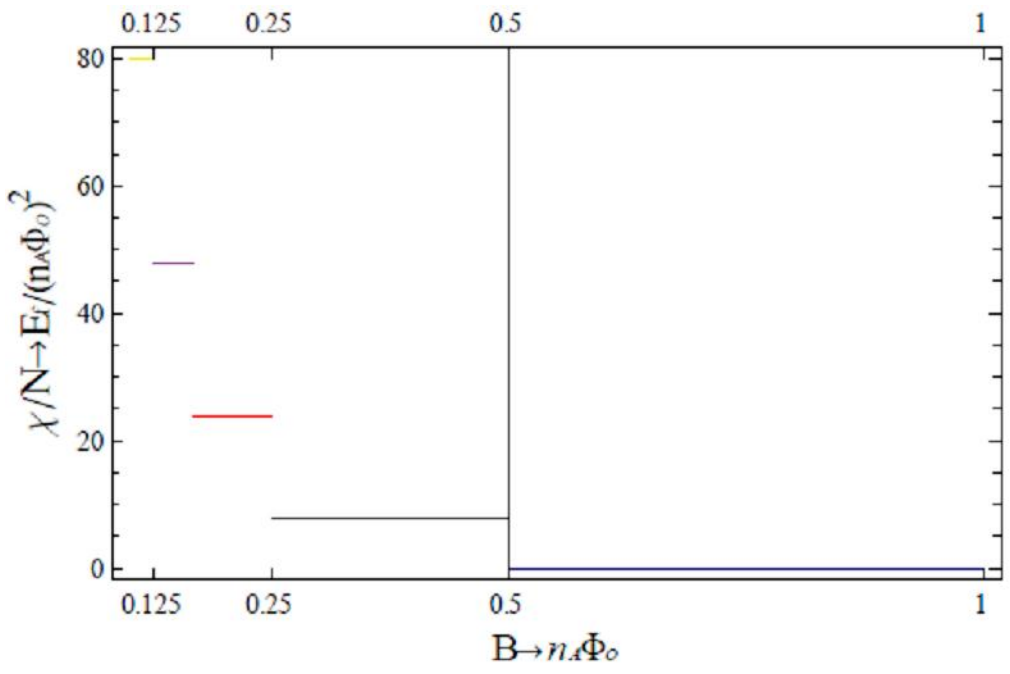

**FIG. 2.3:** Susceptibility per electron (in units of $\varepsilon_{\Phi}/n_A\Phi_0$)

One obtains therefore in this manner the well-known sharp sawtooth behavior of magnetization (in a system with constant number of electrons) measured in low-T experiments [15]. (If the above were plotted as a function of 1/*B*, then the above windows would be periodically repeated with a period $\Delta(1/\mathrm{B}) = 2/n_A\Phi_o$, which is compatible with the dHvA period $2\pi e/\hbar c A_f$ (with $A_f = \pi k_f^2$ and $k_f^2 = 2\pi n_A$) (see i.e. [6]). Also note that, for $B \to 0$, the above energy correctly reproduces the 2D noninteracting result $E/N = \frac{1}{2}E_f$ (i.e. in (2.7) take $B \to 0$ and $\rho \to \infty$ in such a way that the product $B\rho$ is fixed).

**Relation to transport properties – Hall conductivity**

It is useful to mention in passing a physical interpretation of the above thermodynamic results (at T=0) that has connection to transport properties, and that in particular relates the above magnetization discontinuities with diamagnetic currents**:** indeed, the discontinuities in $M$ can be associated with the abrupt change of chiral currents on the edges (in spite of the fact that edges did not directly enter anywhere in the above formulation). This connection is through the simple relation of the magnetization $M$ with the diamagnetic electric currents $I$ that flow around the edges (in opposite directions), namely $M=I\,S/c$ (as one can immediately see by comparing $M=-\partial E/\partial B$ with the Aharonov-Bohm formula $I=-c\partial E/\partial\Phi$ if $I$ is assumed flowing along the edges so that the flux $\Phi=BS$ can be viewed as an enclosed flux) – in combination with the quantized values of the Hall conductance ( $\sigma_{\mathrm{H}}=2\rho e^2/h$ for spinfull electrons) and the fact that, during the transitions to a different LL the current responds to a transverse potential that is equal to $\Delta\mu=\hbar\omega_c$ divided by $e$. We therefore expect to have (for the magnitudes of the various quantities involved)

$$I=\sigma_{\mathrm{H}}\frac{\Delta\mu}{e} \quad \text{with} \quad \Delta\mu=\hbar\omega_c \quad \rightarrow \quad \Delta\mathrm{M}=\sigma_{\mathrm{H}}\frac{\Delta\mu S}{ec}=2N\mu_B \tag{2.10}$$

where in the above, the values of $B=n_A\Phi_o/2\rho$ (where the transitions occur) have been used in the last step, giving therefore the correct magnitude of discontinuities $2N\boldsymbol{\varepsilon}_B$ for the magnetization that we see in fig.2.2 (that occur whenever we have complete filling of $\rho$ LLs). [For completeness we simply mention here that the above could have also been derived with the well-known Widom-Streda formula combined with a thermodynamic Maxwell relation, a more frequently followed procedure that gives $\Delta\mathrm{M}=N\Delta\mu/B$ (for the simultaneous discontinuities of $M$ and $\boldsymbol{\varepsilon}$) which turns out to be equivalent to (2.10), but the above given diamagnetic current interpretation is preferable if we want to later generalize in a similar line of reasoning to the finite-thickness case (see corresponding discussion of transport in Section 4).]
The above also shows immediately how the discontinuities of $M$ are directly related to the Hall conductance. One could i.e. determine $\sigma_{\mathrm{H}}$ from (2.10) by measuring the simultaneous discontinuities of $M$ and $\boldsymbol{\varepsilon}$ (a line that is actually going to be followed in the finite-thickness case of Section 4).

The electron gas in full 3D space inside a homogeneous magnetic field would normally be the next example to consider, and it will indeed be discussed in Section 5. This is a problem that has mostly been treated in astrophysical applications but here we want to place it in a framework interesting to fully 3D solid state systems. Although it might be useful to present it at this point (in order to see the rather large differences from the above 2D case – for example the smooth deviations from the above dHvA periods), we choose to present it *after* discussion of the quasi-2D cases that follow below. In this manner, we can study in detail the dimensionality crossover from 2D to 3D, addressing therefore issues regarding possible dHvA violations both in quasi-2D and in bulk 3D solids in a unifying manner.

**3. Finite-thickness interface (without magnetic field)**

Let us now consider an interface with a finite (nonzero and non-infinite) thickness $d$, but let us first begin with the simpler problem of vanishing magnetic field**:** even in this case we will see that the standard Fermi circle or disk (of 2D noninteracting electrons in the jellium model) will now be replaced by a sequence of many Fermi circles of appropriate radii, each one connected to a particular QW-level associated with the $z$-motion – the procedure of determining the appropriate radii being not so trivial and rather tedious as we shall see. (Once again we will work in the canonical ensemble with fixed number N of electrons (so that the surface areal density $n_A$ is the control parameter, although at the end this can be relaxed – the results can recover those that would have been obtained if the control parameter were the volume density $n_V=n_A/d$, see later below, and especially so in the limit $d\rightarrow\infty$ )).
Indeed, consider an interface (or film) that again extends in a macroscopically large area $S$ in $x$ and $y$ directions, while in the $z$ direction it is characterized by a width $d$, which we can initially consider as very small (of the order of nanometers, i.e. a few atomic layers thick). In the jellium model that we consider here the Hamiltonian is effectively just a nonrelativistic kinetic energy term in 3D space, namely

$$\frac{P^2}{2m}\Psi = E\Psi \tag{3.1}$$

with $\vec{P}$ being the canonical momentum (we have obviously taken the simplest gauge $\vec{A}=0$). For a large system on a plane it is natural to impose periodic boundary conditions in $x$ and $y$ directions, but the $z$ axis can be treated like a 1D double quantum well with impenetrable walls at $z$=0 and $z$=$d$ (the simplest way to impose the spacial confinement). The eigenfunctions of (3.1) can then be written as simple product functions of the form

$$\Psi(x,y,z) \propto \sin(k_z z) e^{ik_y y} e^{ik_x x} \tag{3.2}$$

A quantum state is then characterized by the eigenvalues of the 3 Cartesian components of canonical momentum $\vec{P}/\hbar$ $\{k_x, k_y, k_z\}$ (or $\{n_x, n_y, n_z\}$ after quantization, see below). Pauli principle requires that each such orbital state (namely a triplet $\{n_x, n_y, n_z\}$) can be occupied at T=0 by only two electrons (of opposite spins), and this is a very important criterion that, for strongly anisotropic systems such as this one, must be imposed in a careful manner as we will see below and also in later Sections. The single-particle energy spectrum is

$$\varepsilon_{n_x,n_y,n_z} = \frac{\hbar^2 k^2}{2m} + \varepsilon_{n_z} \text{ where } \varepsilon_{n_z} = \frac{\hbar^2 k_z^2}{2m} \text{ and } k^2 = k_x^2 + k_y^2,\ k_x = 2\pi\frac{n_x}{L_x},\ k_y = 2\pi\frac{n_y}{L_y},\ k_z = \pi\frac{n_z}{d} \tag{3.3}$$
$$(n_x, n_y) = (0, \pm 1, \pm 2...),\ n_z = 1, 2, 3...$$

i.e. $k_x$ and $k_y$ are quasicontinuous variables (since $L_x$, $L_y \to \infty$) while $k_z$ is strongly quantized. For extremely small $d$ the variable $k_z$ is expected to take its lowest value (corresponding to $n_z$=1) for *all* electrons (the usual case, almost always discussed in the literature, where the particles are ñfrozenò at the lowest QW-level $n_z$=1 ï making therefore the system effectively 2D). And this is so because of the enormous energy-difference between $n_z$=2 and $n_z$=1 levels (that goes as $1/d^2$) and, therefore, because it is indeed energetically favorable to start filling states with increasing $|k_x|$ and $|k_y|$ (or equivalently $|n_x|$ and $|n_y|$, starting from 0 and gradually occupying higher numbers in a symmetric manner, with $n_z$ always being 1), forming therefore the standard Fermi circle of 2D noninteracting electrons. However the reader should note that, for any fixed nonzero $d$, even at T=0, the above mentioned 2D character *may be violated* for sufficiently large density (to be quantified below)**:** there may come a point (i.e. if the number of electrons to be accommodated in single-particle states is sufficiently large) when it is no longer favorable to continue increasing the Fermi circle and keep $n_z$=1**;** it may be favorable for the remaining electrons to start jumping to the $n_z$=2 QW-level, and then $k_x$ and $k_y$ can start taking values back at $|n_x|$=$|n_y|$=0 i.e. start forming a new Fermi circle, now associated with the level $n_z$=2. We emphasize that this occurs without violating Pauli principle, since in the triplet $\{n_x, n_y, n_z\}$ (that labels a single-particle state) $n_z$ has changed value, so that $n_x$ and $n_y$ can now acquire the same values as they had before this transition, starting again from 0. The transition to $n_z$=2 will of course occur whenever the ñinitialò Fermi circle (for $n_z$=1) will become so large (with such a long radius $k_{f1}$) that the single-particle energy $\hbar^2 k_{f1}^2/2m$ will become equal with (and from that point on it will exceed) the energy difference between the two QW-levels, or, equivalently, whenever the following equality holds

$$\frac{\hbar^2 k_{f1}^2}{2m} + \varepsilon_{nz=1} = \varepsilon_{nz=2} \tag{3.4}$$

The left-hand-side of (3.4) is the single-particle energy of an ñextraò electron that we wish to place on the perimeter of the Fermi circle (previously formed by all other electrons that were in the QW-level $n_z$=1), while the right-hand-side of (3.4) is the analogous single-particle energy if we were to put the ñextraò electron at the QW-level $n_z$=2 (and start a new Fermi circle from the beginning, namely from zero radius).
It is now important to note that eq. (3.4) provides a sense of ñequilibriumò in the occupation procedure. As stated, from that point on, there starts a 2nd Fermi circle being formed (corresponding to $n_z$=2), and, what is more important, the above sense of ñequilibriumò must be preserved during the entire occupation procedure that follows**:** if we still have excess of electrons and we keep occupying available (empty) single-particle states, then the extra electrons must be placed back and forth in both QW-levels $n_z$=1 and $n_z$=2 in a way that the Fermi radius associated with $n_z$=1 and the one associated with $n_z$=2 will both keep increasing and will at every point (for every density) be related with each other through (the ñequilibriumò relation)

$$\frac{\hbar^2 k_{f1}^2}{2m} + \varepsilon_{nz=1} = \frac{\hbar^2 k_{f2}^2}{2m} + \varepsilon_{nz=2} \tag{3.5}$$

so that the occupation procedure is ñfairò (guaranteeing that it will give the lowest possible *total* energy for the many-particle system). (3.5) demands that the extra electron to be placed anywhere at any moment of the occupation procedure must have the same single-particle energy in any of the possible occupational scenarios. [For the same line of reasoning as this is applied to different problems, see also (4.3) and (5.7))]. If equality (3.5) were not satisfied, and one side were larger than the other, it would mean that the procedure followed up to that point was *not* the optimal (energetically lowest) one, since we could always move electrons around in state-space to gain energy. [It can actually be shown variationally [20] that the above procedure is the lowest energetically.] The reader should notice that this ñfairnessò strategy is actually a *generalization* of the standard symmetric manner of occupation scenarios that are followed in the usual construction of the 3D Fermi surface (where this ñequilibriumò in the single-particle states occupation procedure is the usual *isotropic* filling in *k*-space that leads to the standard Fermi sphere); the above is a generalization of this *to a highly anisotropic system*. This optimal partitioning for our anisotropic problem (in the above described cases of 1 or 2 Fermi circles) is pictorially represented in Fig.3.1 and Fig.3.2.

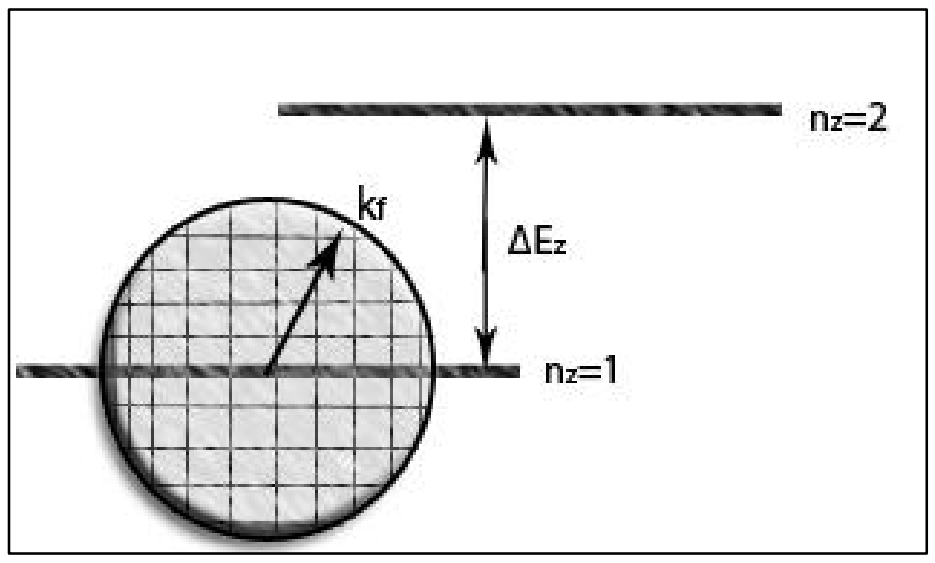

**FIG 3.1** Only one Fermi circle is created ($p$=1) when $d < d_{crit1}$

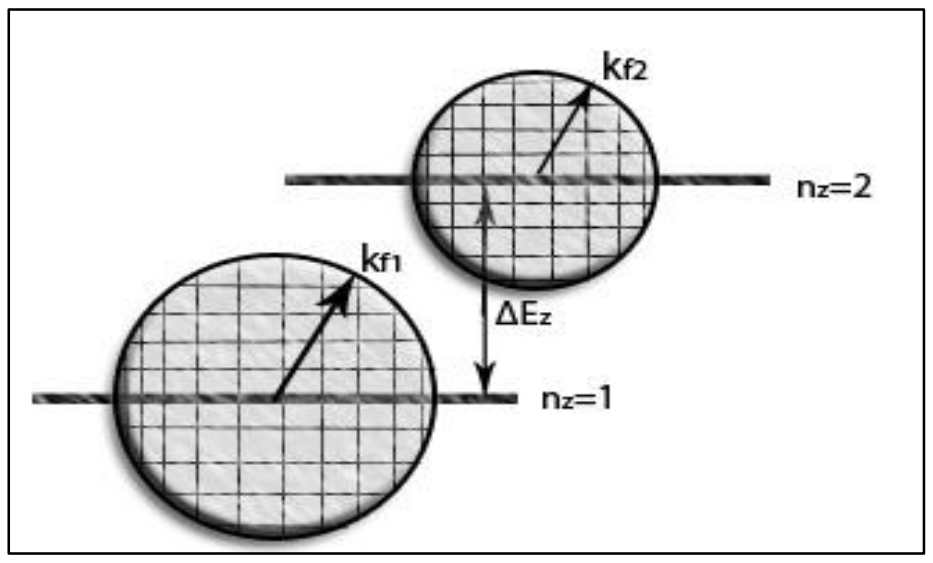

**FIG 3.2** Two Fermi circles are created ($p$=2) when $d_{crit1} < d < d_{crit2}$

In the (usual) case of only 1 QW-level being enough to accommodate all particles (the case depicted in Fig.3.1) it turns out from (3.4) that $d$ must be

$$d < d_{crit1} = \sqrt{\frac{3\pi}{2n_A}} \tag{3.6}$$

that gives a rigorous quantitative measure of what is meant by two-dimensionality, and in such case of sufficiently small $d$ the total energy per electron is simply

$$\frac{E}{N} = E_f \left( \frac{1}{2} + \frac{\pi}{2n_A d^2} \right) \tag{3.7}$$

the usual 2D result plus a constant term. As seen above $d_{crit1}$ depends on $n_A$: for $n_A = 10^{16} m^{-2}$ the critical thickness is $21.7nm$.

(Alternatively of course the result is that, for any given fixed fixed $d$, there is a critical areal density

$$n_{A_{crit}} = \frac{3\pi}{2d^2} \tag{3.8}$$

below which (i.e. for $n_A < n_{A_{crit}}$) the interface essentially behaves as 2D, having again the energy (3.7)).

[The reader should note that if the volume density $n_V$ were the good variable, dividing both sides of (3.4) by $d$ would instead give the result $d_{crit1}=(3´/2n_V)^{1/3}$ as the criterion for two-dimensionality; this might be more appropriate for systems with a constant volume density (as $d$ changes) [21] rather than constant particle number, or for systems that anticipate a 3D Fermi surface in some semiclassical approximation [22,23]; however we here follow a more appropriate procedure, and at the end we will recover the previous results in the appropriate limit.]

In the case of 2 (and only 2) QW-levels being necessary to accommodate all electrons (the case of Fig.3.2), it turns out (by solving (3.5) with respect to $k_{f1}$, $k_{f2}$ with the extra condition $n_A = n_{A_1} + n_{A_2}$ and use of (3.13) below) that the optimal partition in the two Fermi circles is described by the (partial) areal densities

$$n_{A_1} = \frac{n_A}{2} + \frac{3\pi}{4d^2} = \frac{n_A}{2} + \frac{n_{A_{crit}}}{2} \qquad (3.9)$$
$$n_{A_2} = \frac{n_A}{2} - \frac{3\pi}{4d^2} = \frac{n_A}{2} - \frac{n_{A_{crit}}}{2}$$

and that this occurs whenever $d_{crit1} = \sqrt{3\pi / 2n_A} < d < d_{crit2} = \sqrt{13\pi / 2n_A}$, with $d_{crit2}$ being determined by another equilibrium condition analogous to (3.4) above, namely

$$\frac{\hbar^2 k_{f1}^2}{2m} + \varepsilon_{nz=1} = \varepsilon_{nz=3} \ , \qquad \text{or} \qquad \frac{\hbar^2 k_{f1}^2}{2m} = \frac{8\hbar^2\pi^2}{2md^2} \qquad (3.10)$$

together with (3.13) below and in combination with

$$n_A = n_{A_1} + n_{A_2} \qquad (3.11)$$

And in the above case (of Fig.3.2) the total energy (that has contribution from 2 QW-levels and 2 Fermi circles) finally turns out to be

$$\frac{E}{N} = E_f \left( \frac{1}{4} + \frac{5\pi}{4n_A d^2} - \frac{9\pi^2}{16 n_A^2 d^4} \right) \qquad (3.12)$$

(valid when $d_{crit1} < d < d_{crit2}$). After having given the main physical idea with the above examples, let us in the following solve the problem in full generality (i.e. generalize this line of reasoning to *any* arbitrary number of QW-levels playing a role in the energy partition). In the sense discussed above, to each quantum number $n_z$ =1,2é there corresponds a different 2D Fermi circle (of radius $k_{f_{nz}}$ and an associated areal density of electrons $n_{A_{nz}}$ that satisfies

$$k_{f_{n_z}} = \sqrt{2\pi n_{A_{n_z}}} \qquad (3.13)$$

as is easy to show (with a standard 2D argument for spinfull electrons). The immediate question is how to determine in the most general case the proper (i.e. lowest-total-energy) *partition* of the total number (or density) of electrons to the correct values of $n_{A1}$, $n_{A2}$ etc, which are generally many, their actual number depending of course on the value of thickness *d*. (As *d* becomes exceedingly large we expect more and more QW levels to play a role, and in such case we expect the results of the above procedure to tend to previous semiclassical results with the relevant variable being the volume density $n_V$ [22,23] (indeed for a check and for exactly how we recover the correct limit see (3.30) below)).

The technique to determine the correct partition in the general case is rather simple (and it was already motivated for 2 QW-levels)**:** At every point we must have ñequilibriumò in the sense discussed above (but now for many (an arbitrary number of) *z*-levels). Let us suppose that the width *d* of the interface is such that all electrons occupy *p* *z*-axis levels (generalizing the earlier examples that would correspond to *p*=1 and *p*=2). This means that there is a total of *p* Fermi circles created in the system (each circle labeled by a particular value of the quantum number $n_z$). Now, for a given (constant) value *d* of thickness, the single particle energy of an extra electron that we wish to place at the perimeter of a Fermi circle (of a particular $n_z$) must be equal to the corresponding single-particle energy of the same electron if that were placed at the perimeter of any other Fermi circle (for different $n_z$ ôs), this being a reflection of the ñequilibriumò noted above (guaranteeing the lowest total energy), namely

$$\frac{\hbar^2 k_{f1}^2}{2m} + \frac{\hbar^2\pi^2}{2md^2} = \frac{\hbar^2 k_{f2}^2}{2m} + \frac{4\hbar^2\pi^2}{2md^2} = ... = \frac{\hbar^2 k_{fp}^2}{2m} + \frac{p^2\hbar^2\pi^2}{2md^2} \qquad (3.14)$$

Now, using the above relations (viewed as a system of $p$-1 equations for the $k_f$ôs), we can determine all the partial areal densities (corresponding to each $n_z$) as functions of the density of the electrons that belong to $n_z$=1. By solving the above system of equations, we find the optimal partition to be described compactly by

$$n_{Aj} = n_{A1} - \frac{(j^2-1)\pi}{2d^2}, \quad \text{where the index } j \text{ runs from 1 to } p. \tag{3.15}$$

However, the total sum of all partial densities must of course give the total areal density of the system

$$n_A = \sum_{j=1}^{p} n_{Aj} \tag{3.16}$$

Using (3.15) and (3.16), we determine the areal density corresponding to $n_z$=1 analytically, the result being

$$n_{A1} = \frac{n_A}{p} + \frac{\pi}{2d^2}\left[\frac{1}{p}\sum_{j=1}^{p} j^2 - 1\right] = \frac{n_A}{p} + \frac{\pi}{2d^2}\left[\frac{1}{6}(p+1)(2p+1) - 1\right] \tag{3.17}$$

Substituting then (3.17) into (3.15), we can find all partial densities (in the energetically optimal configuration, hence the ground state of the many-electron system) in closed form**:** all results can be finally expressed by

$$n_{Aj} = \frac{n_A}{p} + \frac{\pi}{12d^2}(p+1)(2p+1) - \frac{j^2\pi}{2d^2} \tag{3.18}$$

with $j$=1,é ,$p$. Notice that for $p$=1, $j$=1 we obtain $n_{A1} = n_A$, i.e. all electrons occupy only the lowest QW-level as assumed, while for $p$=2, $j$=1 and $j$=2, (3.18) reproduces both of (3.9) (observations that can be viewed as consistency tests). However, we have not yet retrieved the most useful information**:** it is also practically useful to calculate the range of values of thickness needed for the system to actually occupy exactly the above assumed $p$ levels of the QW. This however is not difficult to determine**:** the $p$ QW-states start all being necessary to be used whenever the $p^{th}$ Fermi circle is just about to form. Equilibrium condition then requires that (assume $p$>1):

$$\frac{\hbar^2 k_{f1}^2}{2m} + \frac{\hbar^2\pi^2}{2md^2} = \frac{p^2\hbar^2\pi^2}{2md^2} \tag{3.19}$$

By solving this equation with respect to the sample-thickness $d$ and by using (3.13) and (3.17), we find a series of critical values of thickness (for various values of $p$=1,2,3,é ), namely

$$d_{crit}(p) = \sqrt{\frac{p(p-1)(4p+1)\pi}{12n_A}} \tag{3.20}$$

That is, for values of thickness larger than (3.20) the system occupies $p$ QW-levels, until the ($p$+1) Fermi circle starts over. This of course happens (as can be seen by just replacing $p$ with $p$+1 in (3.20)) when $d$ is equal to

$$d_{crit}(p+1) = \sqrt{\frac{p(p+1)(4p+5)\pi}{12n_A}} \tag{3.21}$$

So, the conclusion is that when the thickness $d$ varies in the following window

$$d_{crit}(p) \le d \le d_{crit}(p+1) \tag{3.22}$$

then the system occupies $p$ (and no more than $p$) QW-levels.

The above results (3.20), (3.21) and (3.22) reproduce the previous examples for $p$=1 and 2: For

$$0 < d < d_{crit}(1) = \sqrt{3\pi / 2n_A} \qquad (p=1) \tag{3.23}$$

we have the case of Fig.3.1 (and the above (3.23) can be viewed as a criterion of 2-dimensionality). For

$$d_{crit}(1) = \sqrt{3\pi / 2n_A} < d < d_{crit}(2) = \sqrt{13\pi / 2n_A} \qquad (p=2) \tag{3.24}$$

we have the case of Fig.3.2, where 2 Fermi circles are present etc.

The last step is to determine in full generality the total internal energy of the electron gas (when $d$ lies in the range (3.22)). This is given by

$$E = \sum_{j=1}^{p} \left[ \frac{1}{2} N_j E_{fj} + N_j \frac{j^2 \hbar^2 \pi^2}{2md^2} \right] \tag{3.25}$$

with $N_j = n_{Aj} S$ the number of electrons at QW-level j (hence with $n_{Aj}$ given by (3.18)), and with $E_{fj} = \hbar^2 k_{fj}^2 / 2m = \hbar^2 2\pi n_{Aj} / 2m$ the corresponding 2D Fermi energy. We have therefore

$$E = \frac{\hbar^2}{2m} \sum_{j=1}^{p} \left[ \pi n_{Aj}^{\ 2} S + n_{Aj} \frac{j^2 \pi^2}{d^2} S \right] \tag{3.26}$$

which, after carrying out the sums, turns out to be

$$E/N = \frac{\hbar^2 (2\pi n_A)}{2m} \left[ \frac{1}{2p} + \frac{\pi(p+1)(2p+1)}{12 n_A d^2} - \frac{\pi^2 p(p^2-1)(2p+1)(8p+11)}{1440 n_A^{\ 2} d^4} \right] \tag{3.27}$$

that gives directly the total ground state energy per electron, when the thickness of the system lies between (3.20) and (3.21). This reproduces the earlier results (3.7) (for $p$=1) and (3.12) (for $p$=2). We note that, even in this rather trivial system, the role of thickness on the energetics is noteworthy.

Once again we should stress that, compared to earlier work [21,22,23], eq.(3.27) is *exact* and does not generally describe quantum oscillations with wavelengths that are governed by the extremal diameters of cross-sections with an anticipated 3D Fermi surface (those being expected only for a large number of QW states (i.e. with very large $p$ôs) involved)**;** in contradistinction to earlier work, (3.27) is also valid for any small value of $p$.

A final point that is of interest is to take the thickness of our system to infinity, but now keeping the volume density $n_V = n_A / d$ constant. One expects that in the limit of infinite space, the above expression will converge to the standard energy of noninteracting electrons in full 3D space (the standard result that comes out with the use of a macroscopically large cube). But this has rather to be checked, since the standard problem that leads to the symmetric spherical Fermi surface utilizes periodic boundary conditions in all Cartesian directions, while here we have infinite potentials (rigid boundary conditions) at 2 points of the $z$-axis. To examine if the above expectation is true, we choose to write the total energy in units of the 3D Fermi energy:

$$E_f(3D) = \frac{\hbar^2 (3\pi^2 n_V)^{2/3}}{2m}$$

We then have from (3.27)

$$E/N = Ef(3D) \left( \frac{8}{9\pi} \right)^{1/3} \left[ \frac{n_V^{1/3} d}{2p} + \frac{\pi(p+1)(2p+1)}{12 n_V^{2/3} d^2} - \frac{\pi^2 p(p^2-1)(2p+1)(8p+11)}{1440 n_V^{5/3} d^5} \right] \tag{3.28}$$

Substituting (3.20) into (3.28) we can plot the energy for large values of $p$ (keeping the volume density $n_V$ constant), see fig 3.3, from which it is readily seen that it indeed tends to the well-known energy of free electrons in full 3D space, namely

$$E/N = \frac{3}{5} Ef(3D)$$

To see this analytically, we need to now make explicit use of the volume density**:** from (3.20) and (3.21), after setting $n_A$=$n_V\, d$ and then solving for $d$), take the limits $d \to \infty$, $p \to \infty$ so that the $d$-window of values is then shrunk to only a single value:

$$d^3 = \frac{p^3 \pi}{3 n_V}, \qquad (3.29)$$

and from (3.28) we then have:

$$E/N = Ef(3D)\left(\frac{8}{9\pi}\right)^{1/3}\left[\frac{n_V^{1/3} d}{2p} + \frac{\pi p^2}{6 n_V^{2/3} d^2} - \frac{\pi^2 p^5}{90 n_V^{5/3} d^5}\right] \qquad (3.30)$$

Substituting then $(d/p)^3 = \pi / 3n_V$ (due to (3.29)) we finally have

$$E/N = Ef(3D)\left(\frac{8}{9\pi}\right)^{1/3}\left[\frac{n_V^{1/3}}{2}\left(\frac{\pi}{3n_V}\right)^{1/3} + \frac{\pi}{6 n_V^{2/3}}\left(\frac{3n_V}{\pi}\right)^{2/3} - \frac{\pi^2}{90 n_V^{5/3}}\left(\frac{3n_V}{\pi}\right)^{5/3}\right] \qquad (3.31)$$

which turns out to be

$$= \frac{3}{5} Ef(3D)$$

We have therefore given a full analytical treatment of the dimensionality crossover of nonrelativistic noninteracting electrons (in zero-field) from pure 2D to full 3D, passing through a sequence of quasi-2D well configurations. The above results can be viewed as an extension of (or, better, an exact quantal correction to) the extremal free-electron cross-sections picture, usually employed in this problem [21,22,23].

We can also note that with the above analytical solution one can extend calculations to the derivation of other (thermodynamic) properties of the interface, such as Pressure or Compressibility, by taking proper derivatives with respect to volume (for constant $N$), something, however, that we will not pursue here.

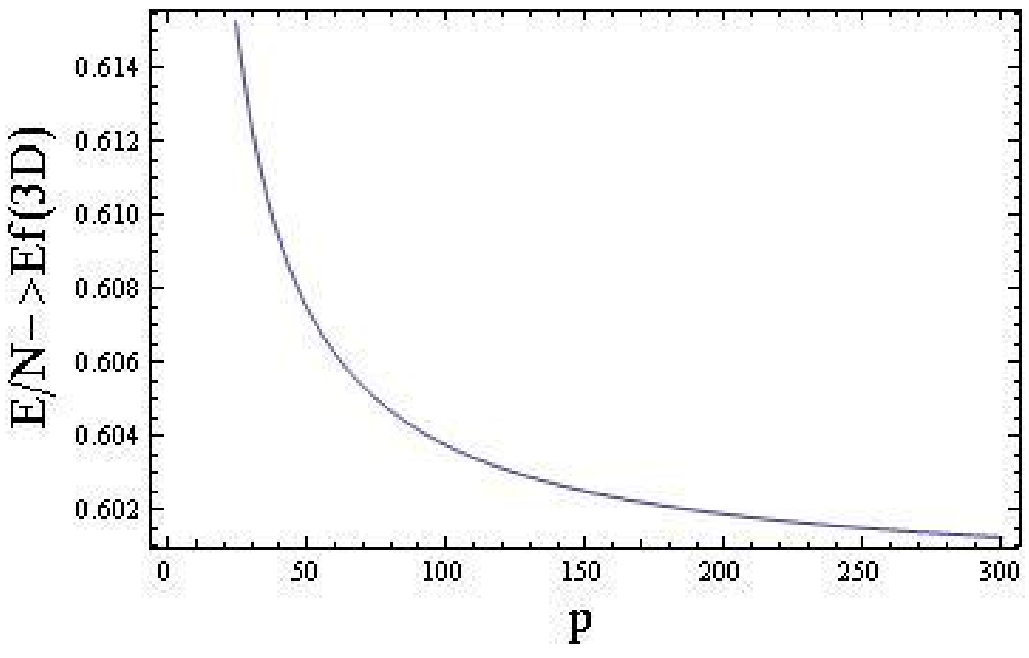

**Fig. 3.3:** Energy (in units of 3D Fermi energy) as a function of $z$-levels quantum number. Note that for approximately more than 300 $z$-axis occupied levels, the total energy tends rapidly to $E/N = 3/5\, Ef(3D)$.

## 4. Finite-thickness interface in a perpendicular magnetic field

In the previous example of the finite-*d* interface, the single-particle energies were quantized only in the *z* direction (and they were quasicontinuous in the *xy* plane). And later (in Section 5), where the full 3D problem in a magnetic field is considered, we will find single-particle energies that will only be quantized in the *xy* plane (Landau levels) and will be quasicontinuous in the *z* direction. In all these cases we have quasicontinuity in at least one direction, so that the above discussed ñequilibrium conditionsò can *continuously* be satisfied (giving at every point (i.e. for every density) the optimal partition or arrangement of our Fermionic particles in single-particle states). One may wonder then how the above method can be used if the single-particle energy is strongly quantized in *all* directions. This *is* actually the case of our main interest, namely when we consider a finite-thickness interface inside a perpendicular magnetic field. In such case, the previous equilibrium condition is not satisfied in a continuous way, as there are not any quasicontinuous Fermi circles (of Section 3) or Fermi line segments (of Section 5). We now do not quite have equilibrium equalities at every density as in the other 2 problems, but we rather have inequalities (that change directions in a discrete manner with variation of density) that actually determine the lowest-energy occupation scenario. However, we will still have distinct points of transition (into different occupation scenarios) whenever certain *equalities* are again satisfied, as we will see. Specifically, it will turn out that, to determine these equalities, requires a close and careful study, and that there is no simple analytic solution that can be written directly for a generic *B* and *d*, even though we are dealing with noninteracting electrons, at T=0 (i.e. the energy cannot directly be written in closed form for an arbitrary field and thickness ï one has to actually run the occupation procedure carefully for all ñpreviousò values of *B* and *d* (starting from easy limiting cases), unlike the other two problems). The interplay between the strong quantization in *xy* plane and the simultaneous strong quantization in *z* axis leads to rather unpredictable patterns (under combined variations of *B* and *d*) when one simply occupies one-electron states in a manner that maintains the lowest possible total energy.

The single-electron spectrum is now given by

$$\varepsilon(n,n_z)=\hbar\omega_c\left(n+\frac{1}{2}\right)+\varepsilon_{nz} \tag{4.1}$$

where *n* is again the Landau level index (*n*=0,1,2é ) , and the QW-levels are again represented by

$$\varepsilon_{nz}=\frac{\hbar^2k_z^{\,2}}{2m}\,,\quad \text{where}\quad k_z=\frac{\pi n_z}{d}\,,\quad n_z=1,2,3,\text{é} \tag{4.2}$$

Let us first see a simple example of the above mentioned competitions that are expressed with inequalities**:** if *d* is extremely small (to be further quantified below), then, for a given *B* (not very strong ï so that there are more than one LLs needed (see below)), it is energetically favorable for the electrons to be placed in several distinct LLs (consecutively, starting from the lowest and moving upwards in energy until all the electrons of the system are accommodated) and keep the system ñfrozenò in the $n_z=1$ QW-level**;** in such case the problem is essentially equivalent to the 2D problem of Section 2 (apart from an extra constant term in the energy (i.e. common to all electrons) due to the QW confinement). But if the thickness *d* starts increasing, then there might come a point (in density) when an extra electron would energetically prefer to be placed in $n_z=2$ (and start to occupy from the beginning a lower LL that is already occupied by other electrons (that correspond to $n_z=1$) without violating Pauli principle (note that, apart from *n* and $n_z$, the 3rd integer *l* is already implicitly used, counting the degenerate states for each combined pair ($n,\ n_z$), so it does not need to be mentioned in any special way). The simplest nontrivial case is when two lowest LLs (i.e. *n*=0 and *n*=1) are originally involved (for $n_z=1$), and then, upon increase of *d*, the above transition (to $n_z$=2 and back to *n*=0 only) takes place**;** this transition will happen when

$$\frac{\hbar\omega_c}{2}+\varepsilon_{nz\,=\,2}=\frac{3\hbar\omega_c}{2}+\varepsilon_{nz\,=\,1} \tag{4.3}$$

This is in the spirit of ñequilibriumò that was used earlier in (3.5), although here it occurs in steps (for discrete values of parameters)**:** once again, the extra particle that is about to be accommodated according to various possible occupation scenarios, has a single-particle energy that must be the same in all of them (otherwise the process would not be fair and it would lead to higher total energy [20]). Eq.(4.3) leads to a critical value of thickness *d* where the transition occurs (for a given *B*), namely

$$d_{crit} = \sqrt{\frac{3\pi\Phi_o}{4B}} \tag{4.4}$$

The above was only the simplest example (to stress the essential point and to motivate what follows), the general case (involving an arbitrary number of LLs and of QWs) needing to be worked out. One then wonders what one can say in full generality for the correct partition (in combined $n$ and $n_z$ states) for arbitrary values of $B$ and $d$ for this problem. It turns out in the general case that there is an ñasymmetryò in the manner that we need to treat the $d$-$B$ variations, if we want to have a good control on all possible cases (and a better understanding of the patterns that show up)**:** if, for example, one follows the route of having a fixed $d$ and varying $B$, analogous to what is done in the 2D case (section 2) but for a finite $d$ (the experimentally relevant route, of a given interface), then the problem is rather difficult to analyze systematically, with results that sometimes look ñsurprisingò (i.e. new transitions appear ***in the interior*** of certain windows of $B$-values (windows with ends that are consistent with dHvA variations), the origin and location of these ñinternal transitionsò not being easily identifiable). The point is that variation of $B$ (for fixed $d$) changes not only the energetic distance between LLs but also their degeneracies, and this interplay, together with the competition with the energetic distance between QW-levels, leads to a multitude of cases to be investigated (that do not seem to be easily subdued to a systematic control). It turns out, however, that the opposite route (of temporarily keeping $B$ fixed and varying $d$, and then change $B$ in a particular way and repeat the procedure of variation of $d$) offers a much better control in the theoretical treatment (basically because degeneracies of each LL are fixed and we only need to focus on competitions between LL-QW energetic distances)**;** although the results are of course equivalent with both methods, what we called ñsurprising resultsò of the 1st route will find a better understanding through the 2nd route, both in terms of origin and location). We will follow below the 2nd route, for theoretical convenience, but in the final figures we will also show results as these would appear from the 1st method, and we will also later provide 2D figures that show the full results under combined variation of $B$ and $d$ (the ordering then (of what is kept fixed) being not important).

Let us start being more quantitative and, in accordance to the mathematically 2D problem of Section 2, let us first assume that the number of electrons lies in the following window:

$$N \leq 2\frac{\Phi}{\Phi_o} \tag{4 .5}$$

(so that only a single LL is involved, although now combined with a QW-level (see below)). Treating always $N$ as fixed (so that $n_A = N/S$ is fixed as well), (4.5) is equivalent to

$$B \geq \frac{1}{2} n_A \Phi_o \, , \tag{4.6}$$

where it should be reminded that the effective areal density $n_A = N/S$ is related to the volume density $n_V$ through $n_A = n_V d$ ), and $\Phi_o = hc/e$ is the flux quantum. Now, each quantum state is again characterized by three quantum numbers, namely, $\{n, l, n_z\}$ with the positive integer $l$ counting the degenerate states inside a LL (or, better, inside a combined ($n$, $n_z$ )-pair) and taking 2ū /ū ɞ values so that each combined pair ($n$, $n_z$ ) can contain up to 2ū /ū ɞ electrons (according to Pauli exclusion principle). Then it is easy to see that, when (4.5) is satisfied, the electron system will occupy **only one** combined-pair, $\{n = 0, n_z = 1\}$ (while $l$ runs from 1 up to $N$, which is here less than the LL degeneracy 2ū /ū ɞ), and this will give a total energy

$$E = N\varepsilon\{n = 0, n_z = 1\} \, , \tag{4.7}$$

with $\varepsilon\{n = 0, n_z = 1\}$ the single particle energy (4.1) with $n$=0 and $n_z$=1. We can write this energy in terms of 2D Fermi energy (in the absence of magnetic field), as

$$\frac{E}{N} = E_f \left[ \frac{B}{n_A \Phi_o} + \left( \frac{\pi}{2 n_A d^2} \right) \right], \tag{4.8}$$

where $E_f = \frac{\hbar^2 k_f^2}{2m} = \frac{\hbar^2 2\pi n_A}{2m}$.

That is, if $B$ satisfies (4.6) then, for *every* value of thickness $d$, electrons occupy only the states with the lowest possible quantum numbers $n$ and $n_z$ (see Fig.1 – note that in this and all following figures we simply compare single-particle **energy-differences**, by always placing at the same level the beginning of each energy difference that needs to be compared (in such a way, the comparison is visually obvious); the placement of the levels does not therefore have an absolute meaning in energy, and it is only differences that matter).

**Fig 1:** Occupied States for every $d$

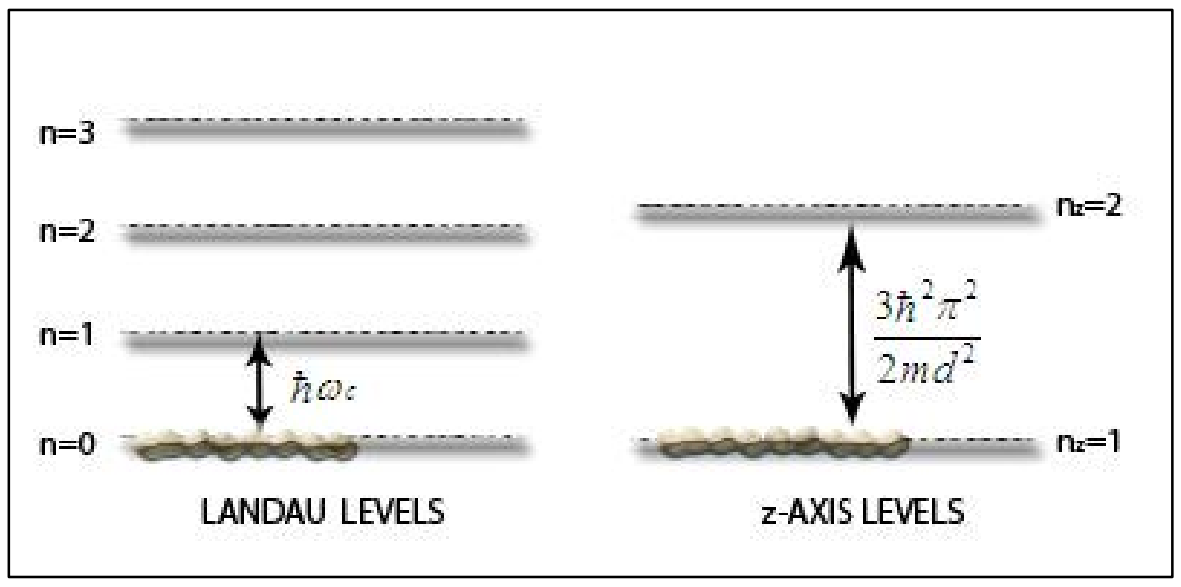

Let us now start lowering $B$**:** The next window of $B$-values (a natural choice if we follow the 2D paradigm of Section 2) is

$$4\frac{\Phi}{\Phi_o} \geq N \geq 2\frac{\Phi}{\Phi_o} \qquad \text{or} \qquad \frac{1}{4}n_A\Phi_o \leq B \leq \frac{1}{2}n_A\Phi_o \tag{4.9}$$

Now the usual occupation scenario would normally be the one in which the extra $N$-$2\Phi/\Phi_o$ electrons will need to be placed in the next LL (the one with $n$=1) as in Section 2 – but this is not necessarily true. It may be energetically favorable for some electrons to occupy another QW-level, with respect to Pauli principle (because of the extra degree of freedom provided by $n_z$). We can immediately see the possible options**:** the two appropriate possibilities are $\{n=1, n_z=1\}$ (increase $n$ by 1) or $\{n=0, n_z=2\}$ (increase $n_z$ by 1 and go back to the lowest LL). But which one is the correct one, and under what conditions? The answer is that this will be determined by *the thickness of the sample*. Let us try to find the critical thickness at which the two possibilities lead to the same single particle energy:

$$\varepsilon\{n=1, n_z=1\} = \varepsilon\{n=0, n_z=2\} \tag{4.10}$$

which is (4.3) that we saw earlier as a motivating example, or equivalently

$$\hbar\omega_c = \frac{3\hbar^2\pi^2}{2md^2} \tag{4.11}$$

which in turn leads to

$$\Rightarrow d_{crit} = \sqrt{\frac{3\pi\Phi_o}{4B}} \qquad \text{(the same as (4.4))} \tag{4.12}$$

Note that the critical thickness depends on the value of the magnetic field (provided that this field lies inside the window of (4.9)). It is easy to see that for values of thickness lower than eq. (4.12), it is the states $\{n=1, n_z=1\}$ (always meaning for all 0< $l$ <$N$-$2\Phi/\Phi_o$) that are occupied by the excess electrons (see fig.2) (the system behaving like a 2D system), and for values of $d$ greater than eq. (4.12) it is the states $\{n=0, n_z=2\}$ that are occupied by the excess electrons (because the energy-difference $\varepsilon_{nz=2} - \varepsilon_{nz=1}$ is smaller that $\hbar\omega_c$, see Fig. 3).

It is interesting to also note that for exactly $B = \frac{1}{2} n_A \Phi_o$ then $d_{crit} = \sqrt{3\pi / 2n_A}$ (which is exactly the critical thickness (3.6) of Section 3 that gives the criterion for the 2-dimensionality of the system).

In the figures below, where the relevant information is visually presented, the arrows denote the LLs that are combined with QW levels (and their *common* filling is represented by filling of corresponding boxes). [Note that the *number of arrows* that combine states is the same for every window of *B*-values that we study (as fixed) in what follows (and is not necessarily equal to the number of LLs and/or the number of QWs involved, as will be seen by later examples) ï see also Appendix that goes deeper in the mathematical structure of the results of this Section.]

**Table 1: Occupied States**

| $\frac{1}{4} n_A \Phi_o \le B \le \frac{1}{2} n_A \Phi_o$ | |
|---|---|
| **Thickness values** | **Occupied states** |
| $d \le \sqrt{\frac{3\pi\Phi_o}{4B}}$ | $\{n = 0, n_z = 1\}$, $\{n = 1, n_z = 1\}$ |
| $d \ge \sqrt{\frac{3\pi\Phi_o}{4B}}$ | $\{n = 0, n_z = 1\}$, $\{n = 0, n_z = 2\}$ |

**Schematic representation:**

**Fig. 2** $d \le \sqrt{\frac{3\pi\Phi_o}{4B}}$

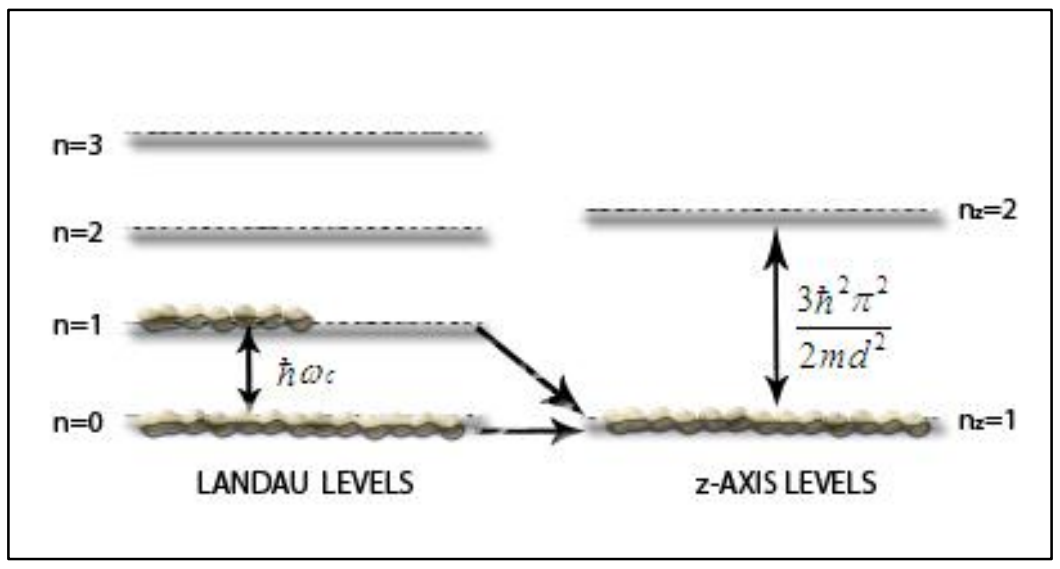

**Fig.3** $d \ge \sqrt{\frac{3\pi\Phi_o}{4B}}$

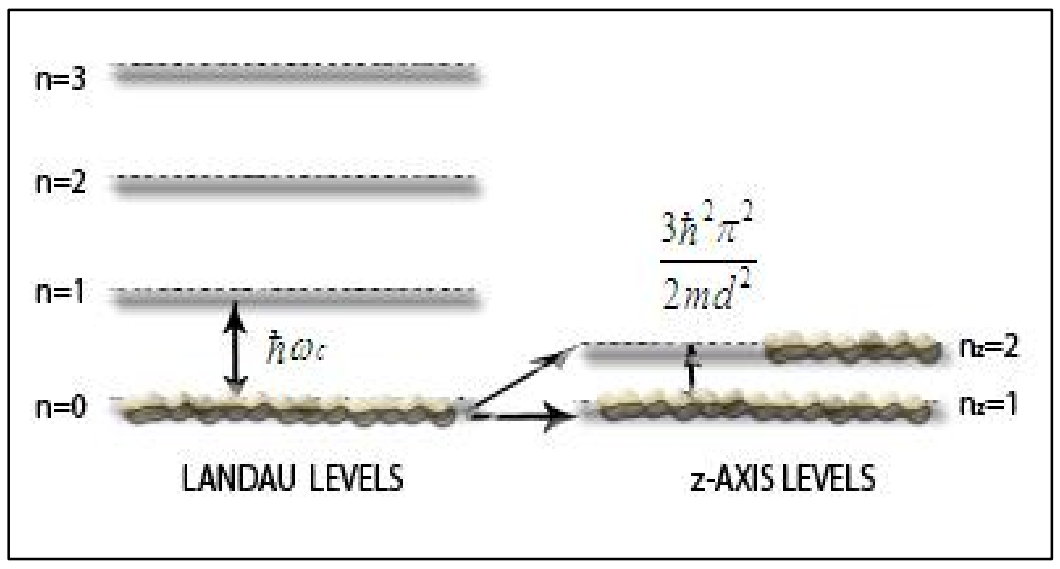

Following the above, it is now straightforward to write the total energy of the system for $\frac{1}{4} n_A \Phi_o \le B \le \frac{1}{2} n_A \Phi_o$. For every value of *B* in this window, the total energy depends on *d*, according to:

$$d \le \sqrt{\frac{3\pi\Phi_o}{4B}} \qquad E = \frac{2\Phi}{\Phi_o}\varepsilon\{n = 0, n_z = 1\} + \left(N - \frac{2\Phi}{\Phi_o}\right)\varepsilon\{n = 1, n_z = 1\} \tag{4.13}$$

$$d \ge \sqrt{\frac{3\pi\Phi_o}{4B}} \qquad E = \frac{2\Phi}{\Phi_o}\varepsilon\{n = 0, n_z = 1\} + \left(N - \frac{2\Phi}{\Phi_o}\right)\varepsilon\{n = 0, n_z = 2\} \tag{4.14}$$

or in units of 2D Fermi energy:

$$d \le \sqrt{\frac{3\pi\Phi_o}{4B}}: \qquad \frac{E}{N} = E_f\left[-4\left(\frac{B}{n_A\Phi_o}\right)^2 + 3\left(\frac{B}{n_A\Phi_o}\right) + \left(\frac{\pi}{2n_A d^2}\right)\right] \tag{4.15}$$

$$d \geq \sqrt{\frac{3\pi\Phi_o}{4B}}: \qquad \frac{E}{N} = E_f\left[\left(\frac{B}{n_A\Phi_o}\right) - 6\left(\frac{B}{n_A\Phi_o}\right)\left(\frac{\pi}{2n_A d^2}\right) + \left(\frac{4\pi}{2n_A d^2}\right)\right] \tag{4.16}$$

Let us now proceed further and present the third window of *B*-values, namely

$$\frac{1}{6}n_A\Phi_o \leq B \leq \frac{1}{4}n_A\Phi_o \tag{4.17}$$

A similar line of reasoning must then be followed**:** starting with small values of thickness *d* the system occupies only distinct LLs and is restricted to the lowest state in *z* axis. But how small must the thickness be for this to be the case? The answer is when the energy gap of the first two QW-levels is larger than $2\hbar\omega_c$ (since then the system is energetically favored to occupy only distinct LLs). [Note the difference from the previous cases where the QW difference should be compared with $\hbar\omega_c$ rather than $2\hbar\omega_c$, see (4.11).] The first critical value of thickness where this is violated is determined by

$$2\hbar\omega_c = \frac{3\hbar^2\pi^2}{2md^2} \tag{4.18}$$

and is equal to

$$d_{crit1} = \sqrt{\frac{3\pi\Phi_o}{4.2.\mathrm{B}}} \tag{4.19}$$

Note that it again depends on the value of the magnetic field (and it is also interesting to note that, if the field is exactly $B = \frac{1}{4}n_A\Phi_o$ then we find again:

$$d_{crit}(2D) = \sqrt{\frac{3\pi}{2n_A}}\,, \tag{4.20}$$

which is nothing but the 2D criterion (3.6) that we found earlier).

**Fig. 4** $d < d_{crit1}$

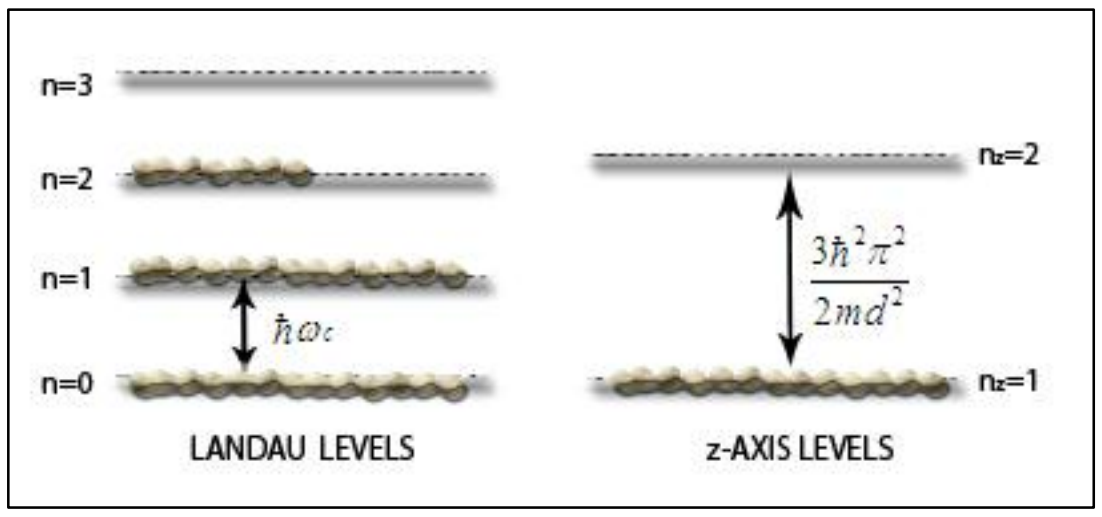

**Fig. 5** $d = d_{crit1}$

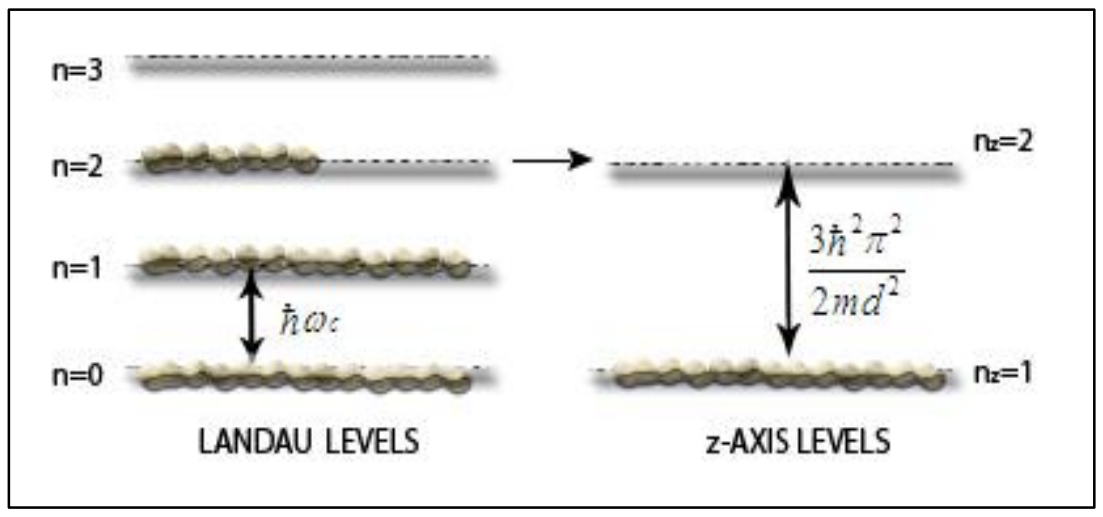

If we continue increasing the thickness beyond this first critical value, the electrons start the occupation of the second QW-level, and this happens when:

$$2\hbar\omega_c > \frac{3\hbar^2\pi^2}{2md^2} \tag{4.21}$$

In this case the electrons in the (incompletely filled) LL state *n*=2 are falling in the state *n*=0, losing energy $2\hbar\omega_c$. Simultaneously, the same electrons are excited from QW-level $n_z = 1$ to $n_z = 2$, gaining energy $3\hbar^2\pi^2/2md^2$. The energy gained by this procedure is of course lower than the one lost, due to (4.21), making therefore this transition energetically favored (see fig. 6).

The next transition (upon further increase of $d$) occurs when the gap between the two first QW levels drops to the value $\hbar\omega_c$ of the Landau gap, namely:

$$\hbar\omega_c = \frac{3\hbar^2\pi^2}{2md^2} \Rightarrow d_{crit2} = \sqrt{\frac{3\pi\Phi_o}{4\mathrm{B}}} \qquad (4.22)$$

**Fig. 6** $d > d_{crit1}$

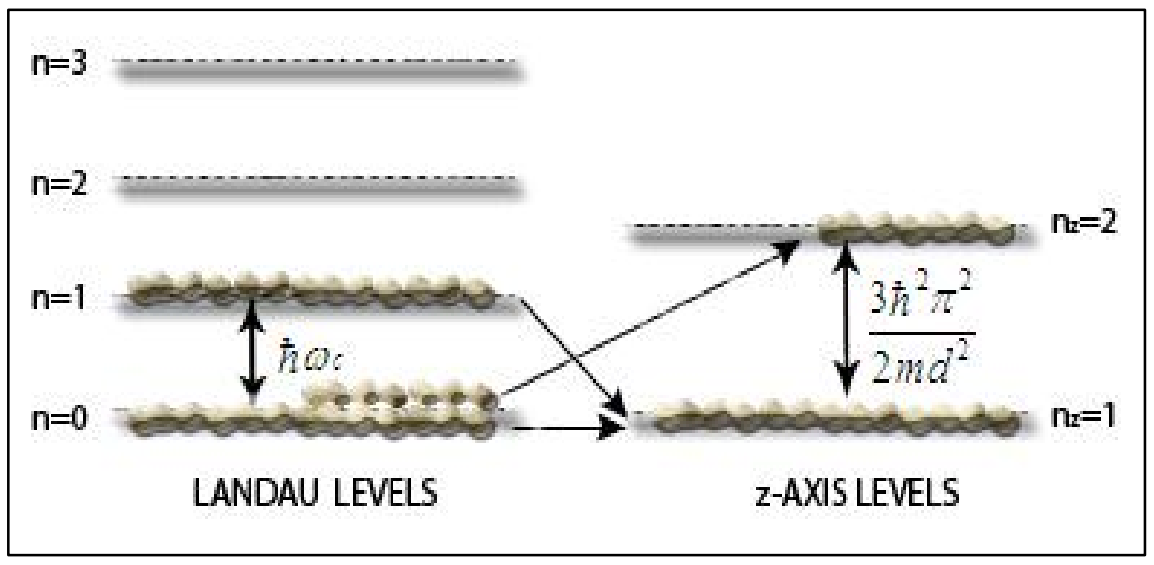

**Fig. 7** $d = d_{crit2}$

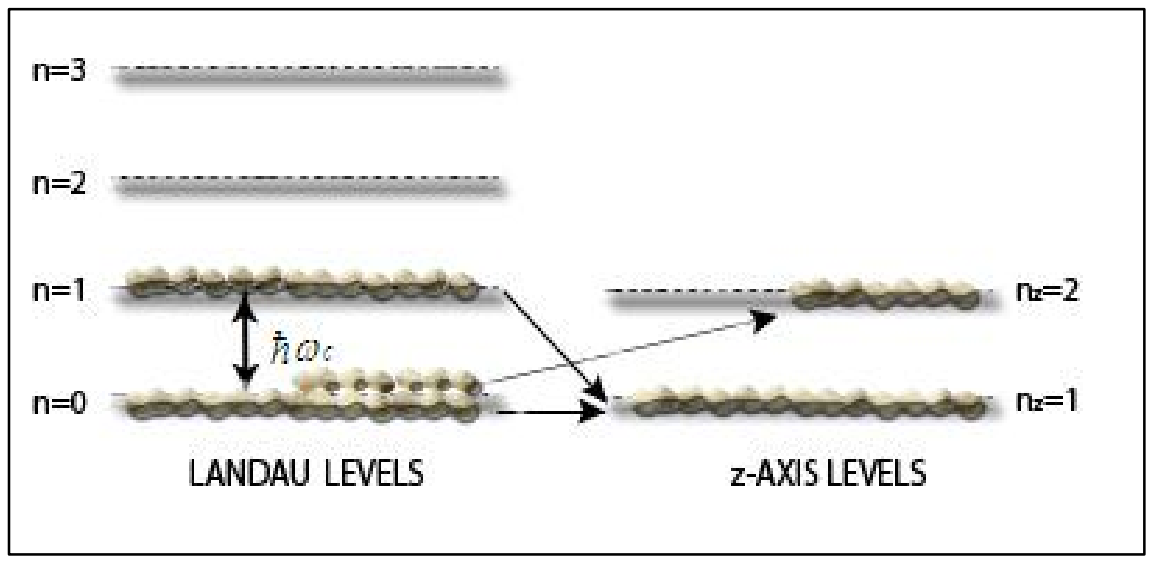

If the thickness is further increased, then the following relation holds:

$$\hbar\omega_c > \frac{3\hbar^2\pi^2}{2md^2} \qquad (4.23)$$

Some of the electrons of the $n$=1 LL will then fall on the $n$=0 LL (by now *fully* occupying it) and at the same time they are excited in the $n_z = 2$ QW-level (leaving the $n$=1 LL partially occupied, see fig. 8).

But there is still one more qualitatively distinct scenario before we get to the end of this procedure: with further increase of thickness, we find the following relation:

$$\hbar\omega_c = \frac{8\hbar^2\pi^2}{2md^2} \Rightarrow d_{crit3} = \sqrt{\frac{8\pi\Phi_o}{4\mathrm{B}}}\ , \qquad (4.24)$$

when it has happened that the $n_z = 3$ level has fallen so low that the difference $\varepsilon_{nz\,=\,3}$ ï $\varepsilon_{nz\,=\,1}$ is equal to $\hbar\omega_c$. Then, above this value of thickness, we only have the lowest LL combined with the 3 lowest QW-levels (see fig. 10).

**Fig. 8** $d > d_{crit2}$

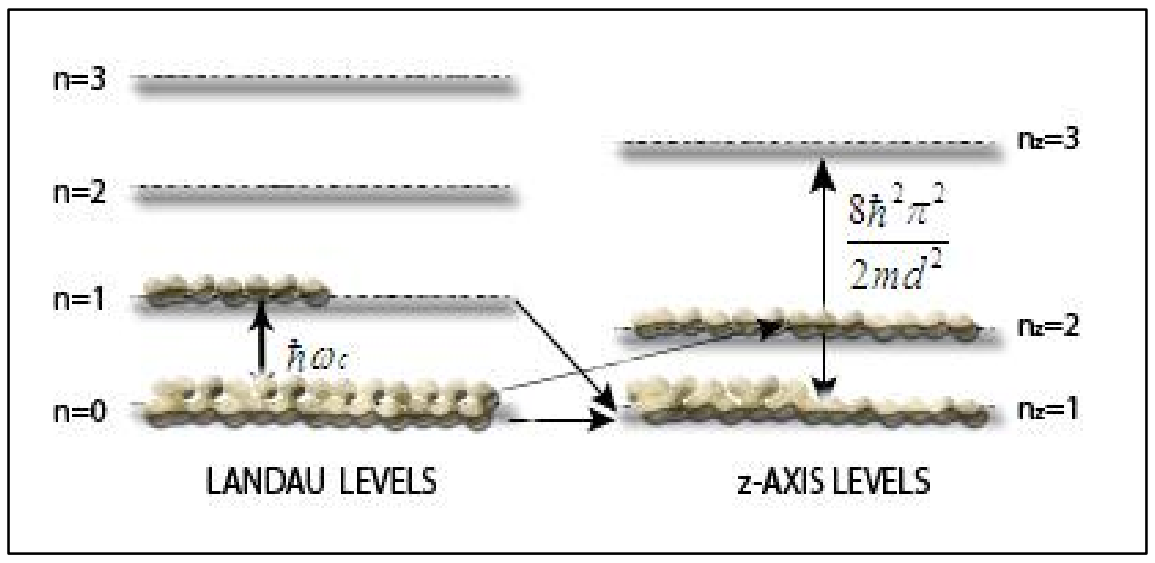

**Fig. 9** $d = d_{crit3}$

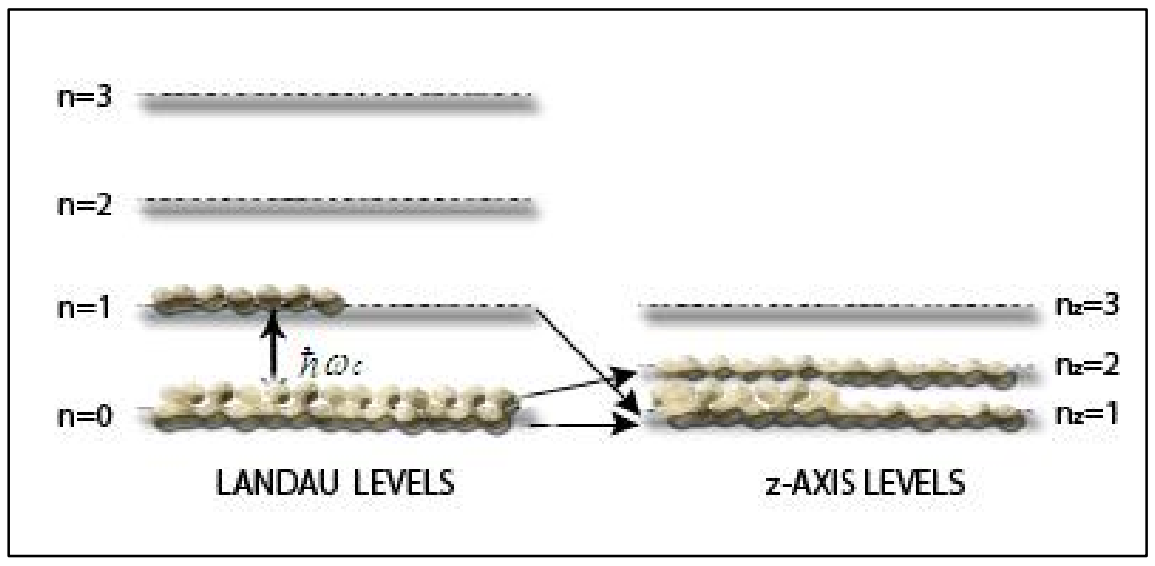

**Fig. 10** $d > d_{crit3}$

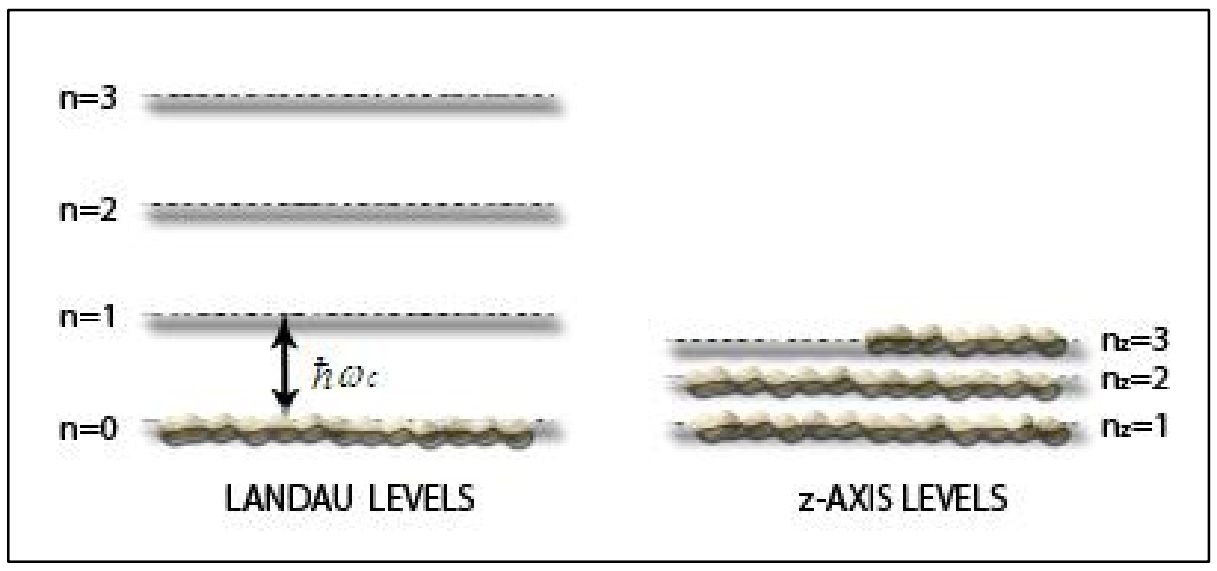

A summary of all the occupation scenarios (for $\frac{1}{6} n_A \Phi_o \le B \le \frac{1}{4} n_A \Phi_o$) is shown in Table 2, and the corresponding total energies for all the various windows of $d$-values are given in Table 3.

**Table 2:** Occupied States

| $\frac{1}{6} n_A \Phi_o \le B \le \frac{1}{4} n_A \Phi_o$ | |
|---|---|
| Window of $d$-values | Occupied States |
| $d \le \sqrt{\frac{3\pi\Phi_o}{4.2.B}}$ | $\{n=0, n_z=1\}$, $\{n=1, n_z=1\}$, $\{n=2, n_z=1\}$ |
| $\sqrt{\frac{3\pi\Phi_o}{4.1.B}} \ge d \ge \sqrt{\frac{3\pi\Phi_o}{4.2.B}}$ | $\{n=0, n_z=1\}$, $\{n=1, n_z=1\}$, $\{n=0, n_z=2\}$ |
| $\sqrt{\frac{8\pi\Phi_o}{4B}} \ge d \ge \sqrt{\frac{3\pi\Phi_o}{4.1.B}}$ | $\{n=0, n_z=1\}$, $\{n=0, n_z=2\}$, $\{n=1, n_z=1\}$ |
| $d \ge \sqrt{\frac{8\pi\Phi_o}{4B}}$ | $\{n=0, n_z=1\}$, $\{n=0, n_z=2\}$, $\{n=0, n_z=3\}$ |

**Table 3:** Total energies

| $\frac{1}{6} n_A \Phi_o \le B \le \frac{1}{4} n_A \Phi_o$ | |
|---|---|
| Window of $d$-values | Total energy (in units of 2D Fermi energy) |
| $d \le \sqrt{\frac{3\pi\Phi_o}{4.2.B}}$ | $\frac{E}{N} = E_f\left[-12\left(\frac{B}{n_A\Phi_o}\right)^2 + 5\left(\frac{B}{n_A\Phi_o}\right) + \left(\frac{\pi}{2n_A d^2}\right)\right]$ |
| $\sqrt{\frac{3\pi\Phi_o}{4.1.B}} \ge d \ge \sqrt{\frac{3\pi\Phi_o}{4.2.B}}$ | $\frac{E}{N} = E_f\left[4\left(\frac{B}{n_A\Phi_o}\right)^2 + \left(\frac{B}{n_A\Phi_o}\right) - 12\left(\frac{B}{n_A\Phi_o}\right)\left(\frac{\pi}{2n_A d^2}\right) + \left(\frac{4\pi}{2n_A d^2}\right)\right]$ |
| $\sqrt{\frac{8\pi\Phi_o}{4B}} \ge d \ge \sqrt{\frac{3\pi\Phi_o}{4.1.B}}$ | $\frac{E}{N} = E_f\left[-8\left(\frac{B}{n_A\Phi_o}\right)^2 + 3\left(\frac{B}{n_A\Phi_o}\right) + 6\left(\frac{B}{n_A\Phi_o}\right)\left(\frac{\pi}{2n_A d^2}\right) + \left(\frac{\pi}{2n_A d^2}\right)\right]$ |
| $d \ge \sqrt{\frac{8\pi\Phi_o}{4B}}$ | $\frac{E}{N} = E_f\left[\left(\frac{B}{n_A\Phi_o}\right) - 26\left(\frac{B}{n_A\Phi_o}\right)\left(\frac{\pi}{2n_A d^2}\right) + \left(\frac{9\pi}{2n_A d^2}\right)\right]$ |

Let us also study the next window of $B$-values, namely

$$\frac{1}{8}n_A\Phi_o \le B \le \frac{1}{6}n_A\Phi_o \tag{4.25}$$

since there are some special elements showing up, signifying the nonintegrability of this problem. We will now be more compact and will show in figures essentially an animation of what happens as the thickness $d$ is continuously increased (always for a fixed value of $B$, that lies inside the window (4.25)).

**Fig 11:** $d \le \sqrt{\frac{3\pi\Phi_o}{4.3.B}}$

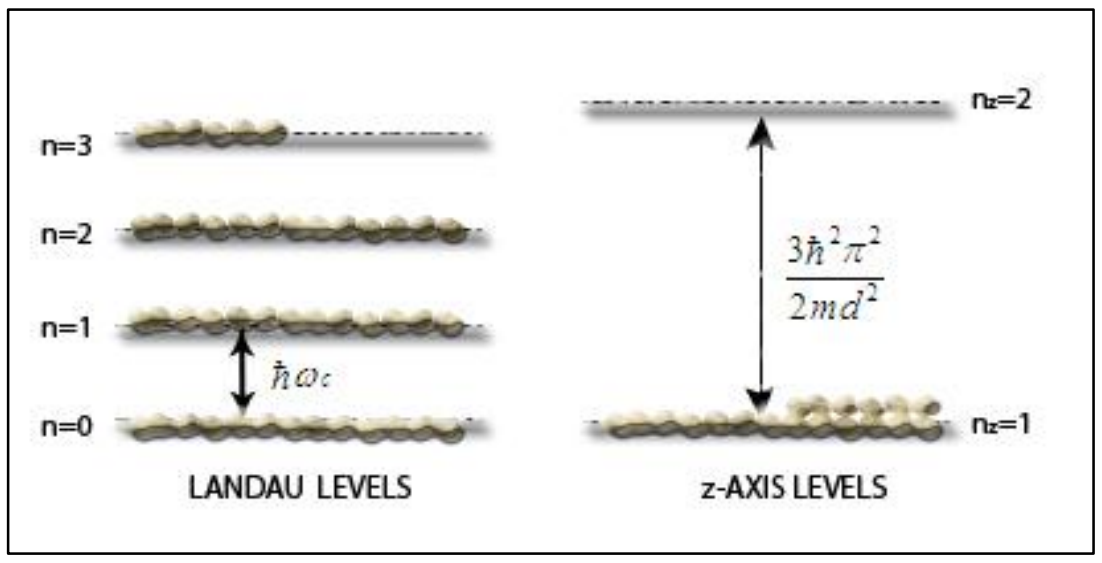

Only the lowest QW is occupied because $\Delta E_z\{1,2\} > 3\hbar\omega_c$

**Fig 12:** $d = \sqrt{\frac{3\pi\Phi_o}{4.3.B}}$

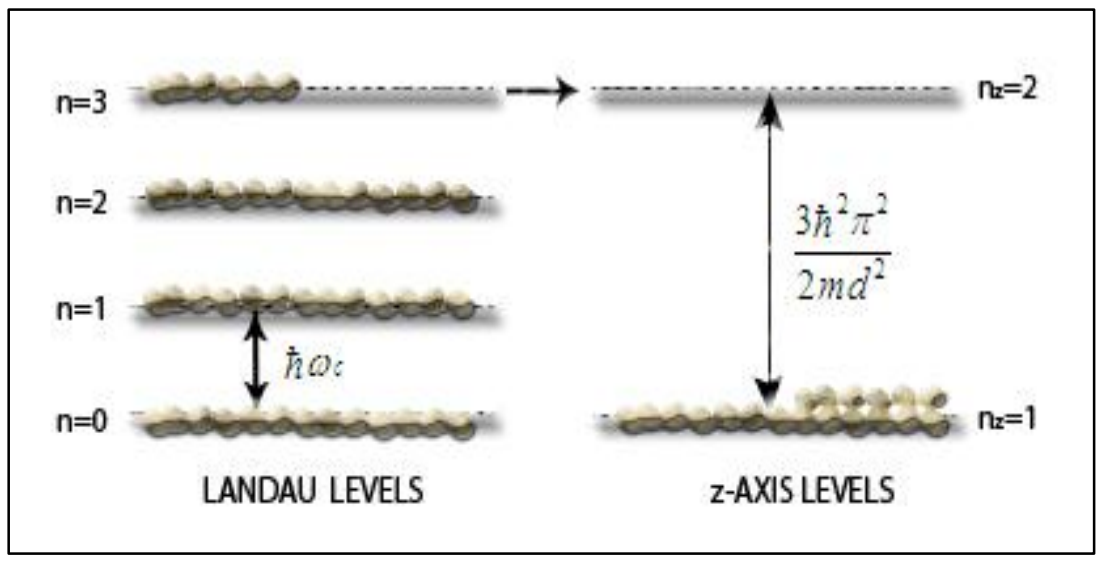

The equality $\Delta E_z\{1,2\} = 3\hbar\omega_c$ is satisfied

**Fig 13:** $d > \sqrt{\frac{3\pi\Phi_o}{4.3.B}}$

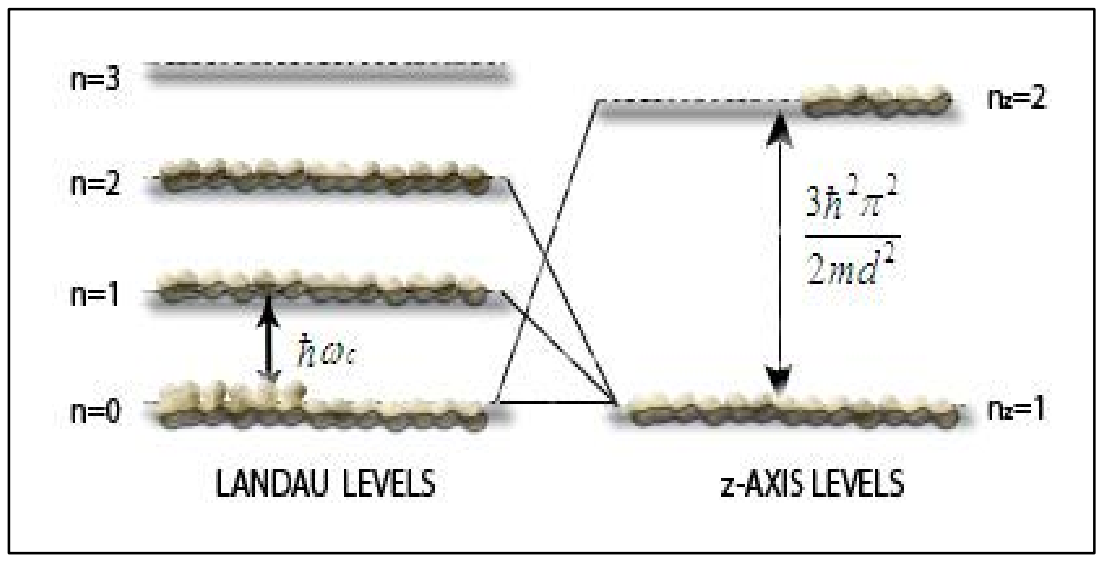

States $\{n=3, n_z=1\}$ are abandoned and $\{n=0, n_z=2\}$ are partially occupied, because: $\Delta E_z\{1,2\} < 3\hbar\omega_c$

**Fig 14:** $d = \sqrt{\frac{3\pi\Phi_o}{4.2.B}}$

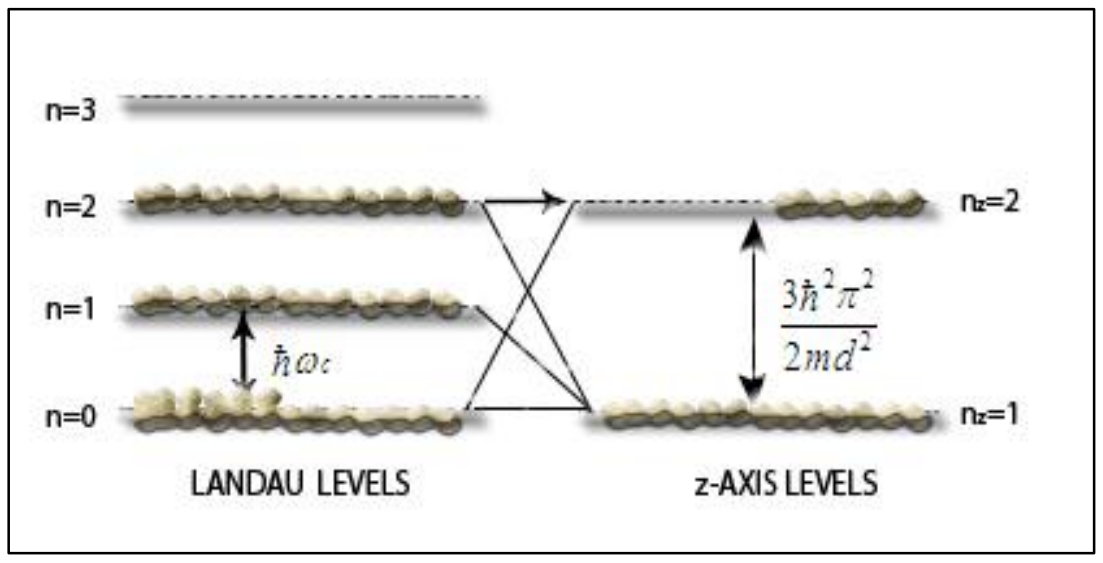

The equality $\Delta E_z\{1,2\} = 2\hbar\omega_c$ is satisfied

**Fig 15:** $d > \sqrt{\frac{3\pi\Phi_o}{4.2.B}}$

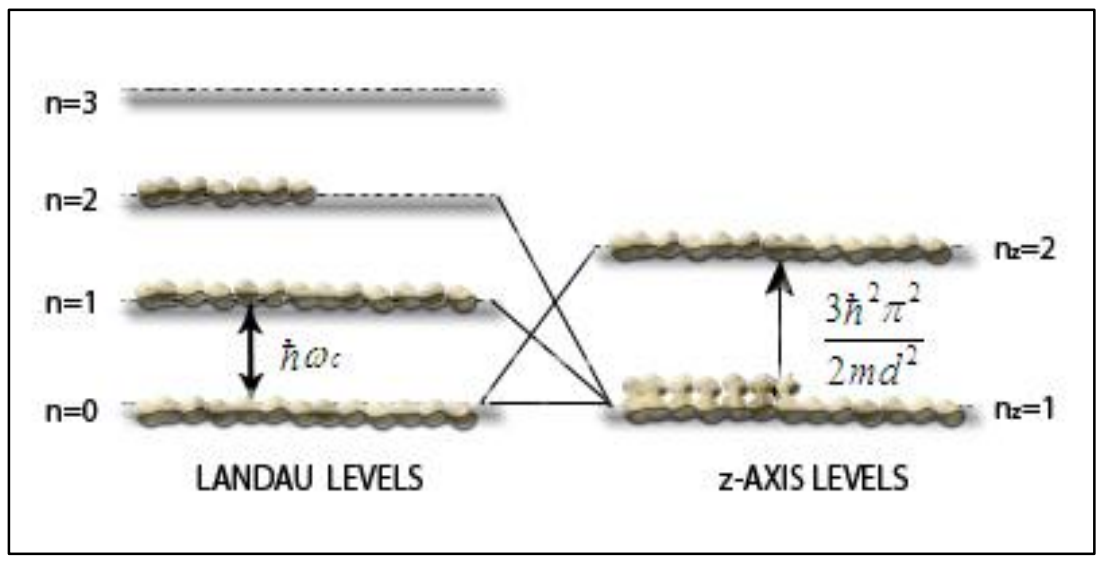

States {$n$=0,$n_z$=2} are fully occupied, and {$n$=2, $n_z$=1} are only partially filled, because: $\Delta E_z\{1,2\} < 2\hbar\omega_c$

**Fig 16:** $d = \sqrt{\frac{3\pi\Phi_o}{4.1.B}}$

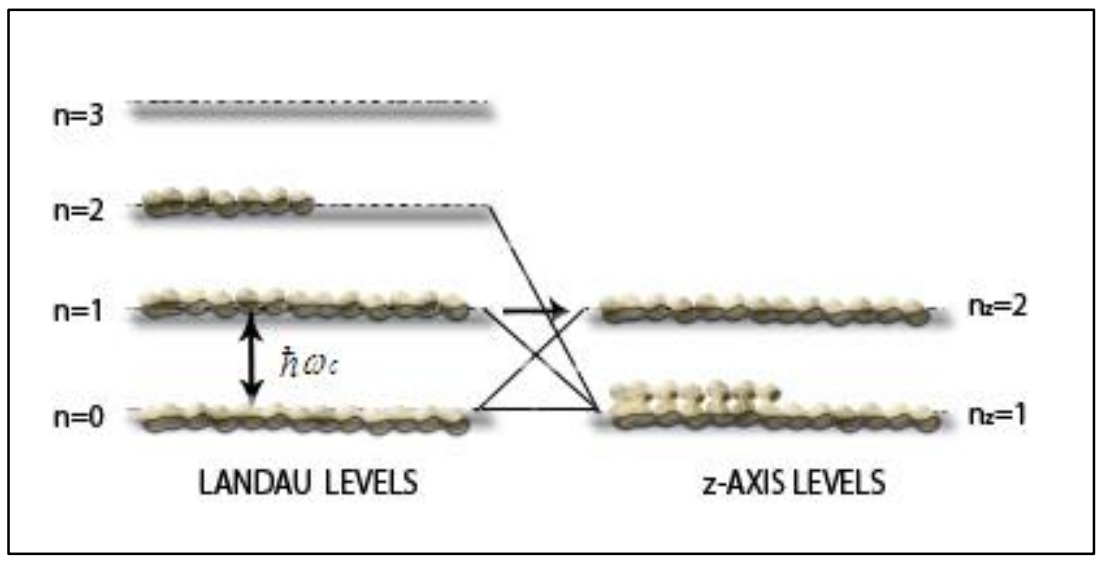

The equality $\Delta E_z\{1,2\} = \hbar\omega_c$ is satisfied

**Fig 17:** $d > \sqrt{\frac{3\pi\Phi_o}{4.1.B}}$

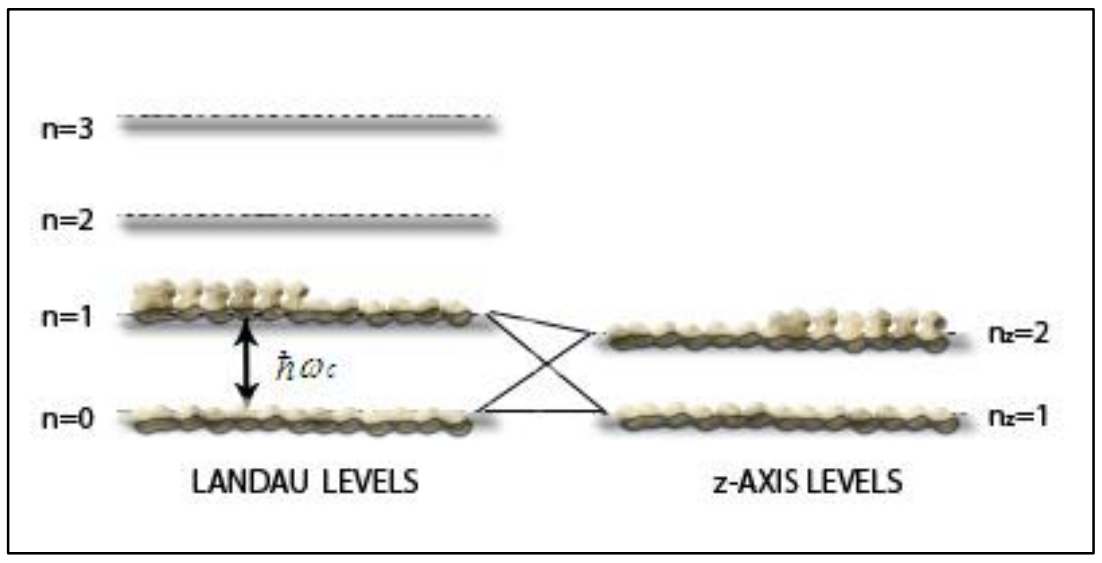

States {$n$=2, $n_z$=1} are abandoned and {$n$=1, $n_z$=2} are partially occupied, because: $\Delta E_z\{1,2\} < \hbar\omega_c$

**Fig 18:** $d = \sqrt{\frac{5\pi\Phi_o}{4B}}$

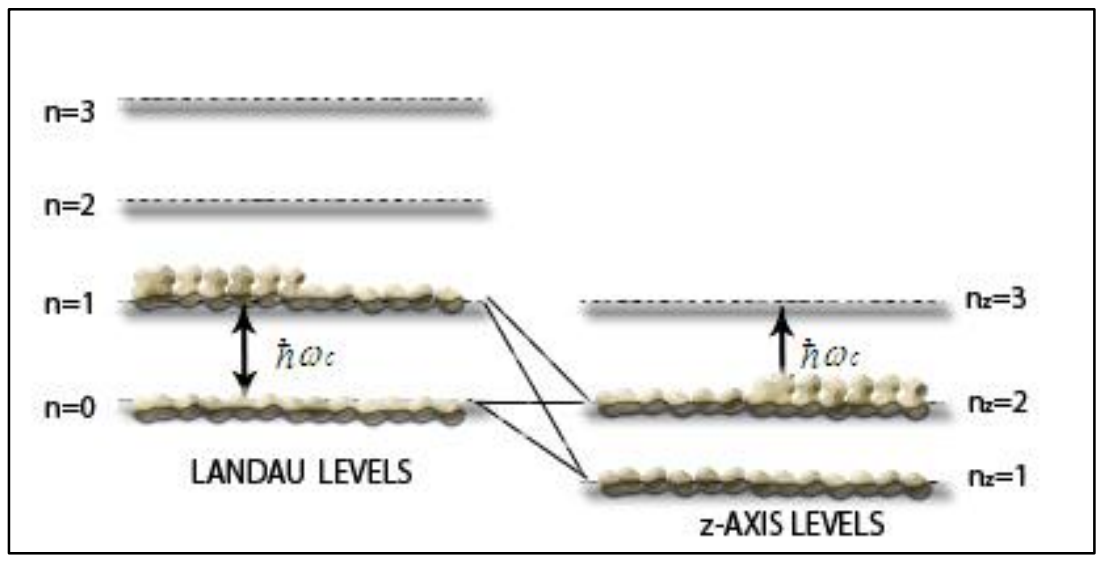

The equality $\Delta E_z\{3,2\} = \hbar\omega_c$ is satisfied. (Note that we now have to compare between states that have $n_z$ greater than 1)

**Fig 19:** $d > \sqrt{\frac{5\pi\Phi_o}{4B}}$

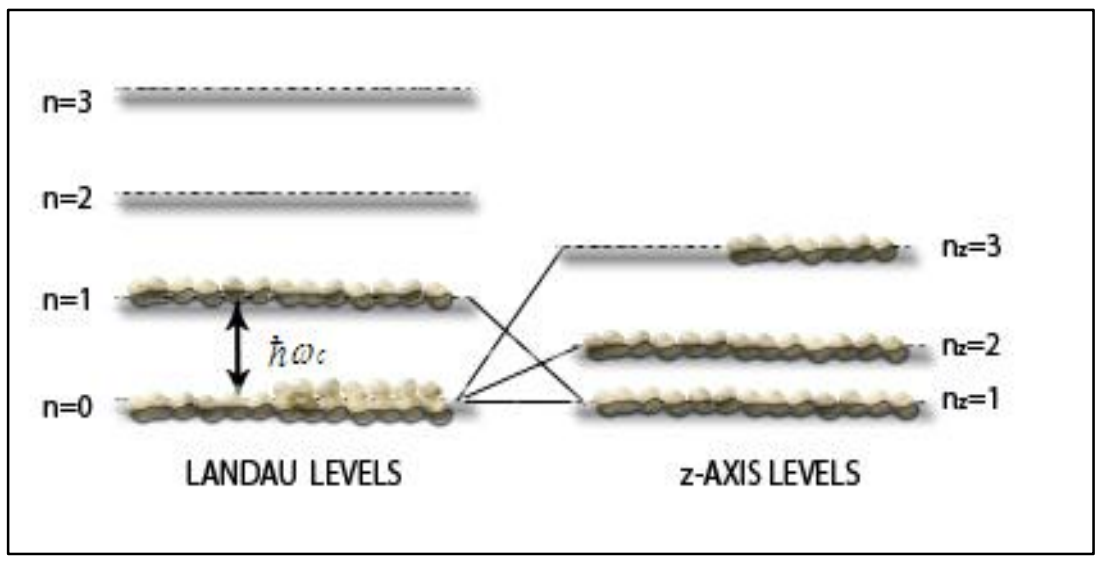

States {$n$=1, $n_z$=2} are abandoned and {$n$=0, $n_z$=3} are partially occupied, because: $\Delta E_z\{3,2\} < \hbar\omega_c$

**Fig 20:** $d = \sqrt{\frac{8\pi\Phi_o}{4B}}$

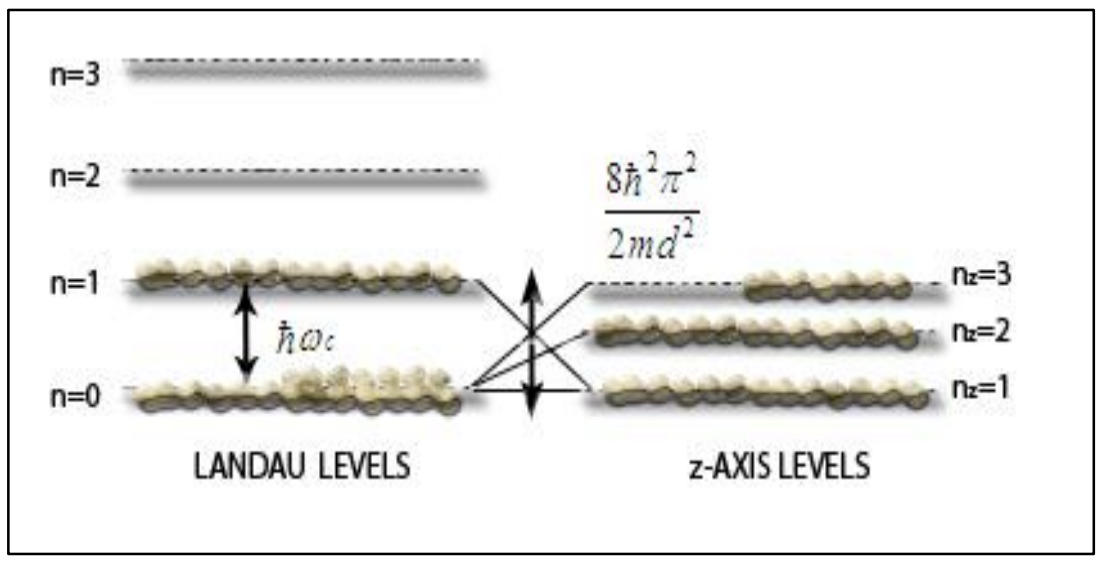

The equality $\Delta E_z\{3,1\} = \hbar\omega_c$ is satisfied

**Fig 21:** $d > \sqrt{\frac{8\pi\Phi_o}{4B}}$

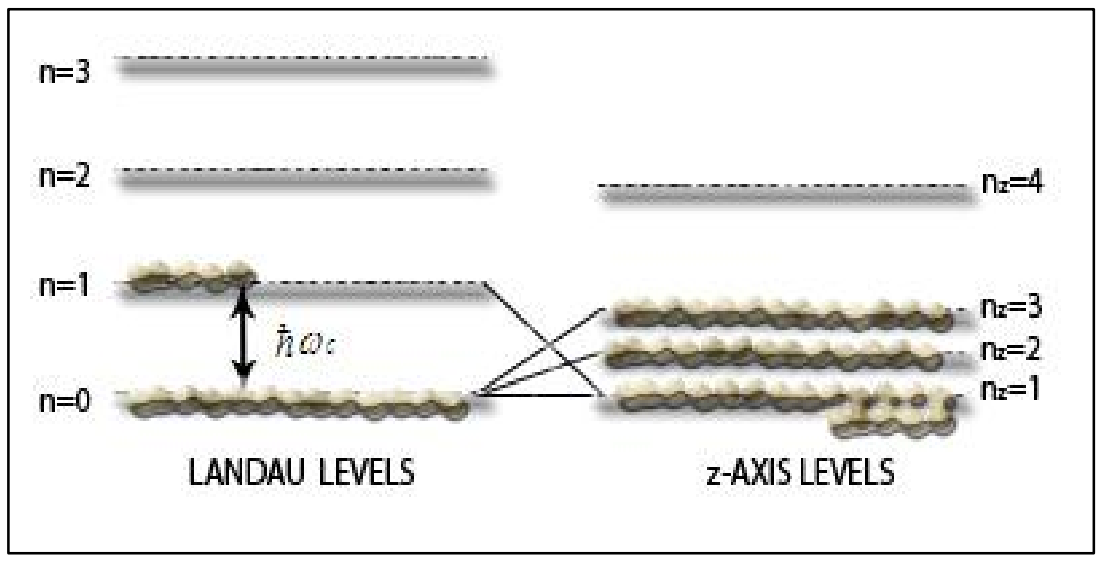

States {$n$=0, $n_z$=3} are fully occupied, and {$n$=1, $n_z$=1} are partially filled because: $\Delta E_z\{3,1\} < \hbar\omega_c$

**Fig 22:** $d = \sqrt{\frac{15\pi\Phi_o}{4B}}$

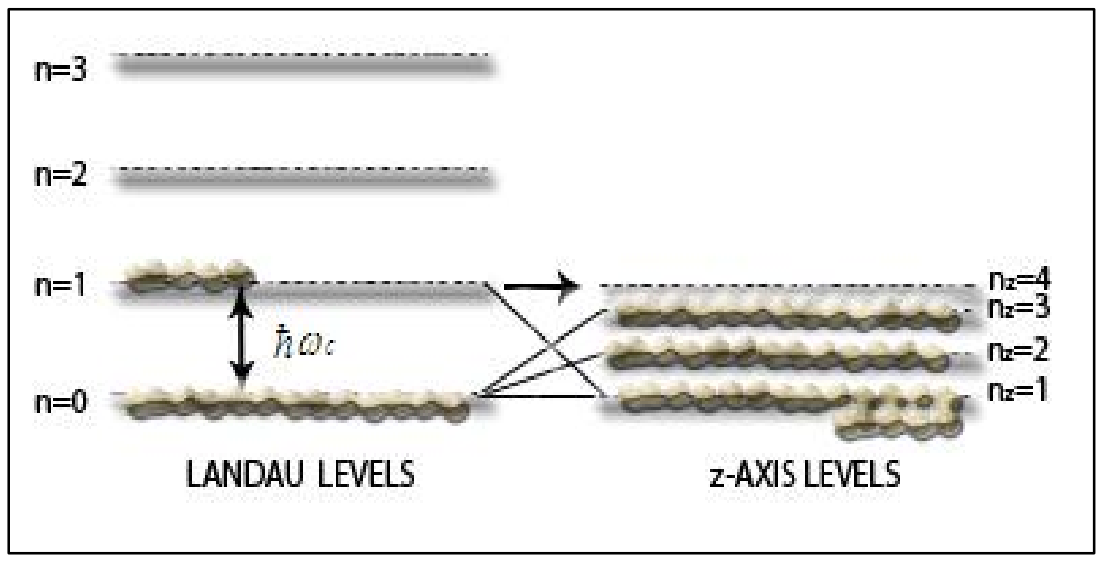

The equality $\Delta E_z\{4,1\} = \hbar\omega_c$ is satisfied

**Fig 23:** $d > \sqrt{\frac{15\pi\Phi_o}{4B}}$

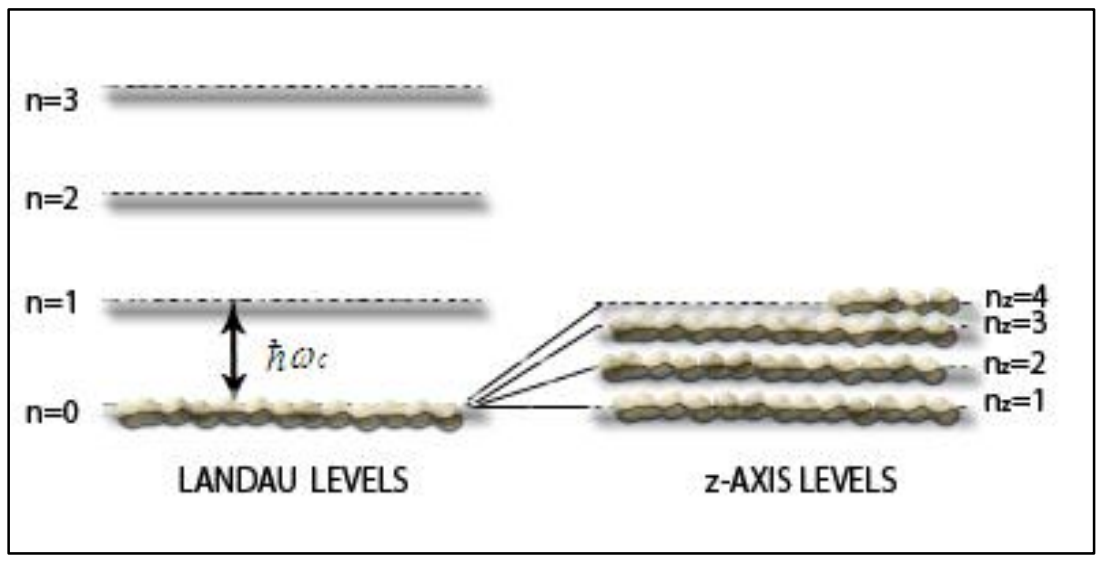

Only the lowest LL states are occupied (combined with all 4 $z$-axis levels), because: $\Delta E_z\{4,1\} < \hbar\omega_c$

From the above examples one should note that sometimes, in intermediate steps, the energetics involve competitions between LLs and QW levels that are not necessarily the lowest possible**;** observe e.g. Fig. 18, where in the competition of energy-differences it was the levels $n_z$=2 and $n_z$=3 that were involved, and not the lowest $n_z$=1 level. This is because the $n_z$=1 level has already been combined with both available LLs and there is no extra freedom for this level to be involved any more. (Note again that the energy comparisons in all the above figures are made only for energy-**differences** that stand side to side, and not for the absolute energy values**;** if we wanted the absolute spectrum, we would have to add the two contributions, and then we would have crossovers at the points of transitions ï it is actually in this form of crossover that we will detect possible effects of the above type later in Section 6 on topological insulators). More subtle behaviors in the energetic comparisons like this one we will also see in the examples that follow, and these give to the results a certain form of unpredictability**;** they are only determined by the system itself when the occupational procedure is run (under the energy criteria set up earlier and the Pauli principle). This leads to an interesting pattern of possible occupation scenarios (with corresponding *consequences on measurable quantities* that will be shown later below).

Let us finally present the results for the fifth window of $B$-values (that involves slightly more complex comparisons between LLs and QW levels), which is:

$$\frac{1}{10} n_A \Phi_o \le B \le \frac{1}{8} n_A \Phi_o$$

Since the number of windows of $d$-values turns out to be rather large (21), we will present this case only with figures, as we did before, but with no commentary. One should again observe that not all cases refer to comparisons between the lowest LL and QW levels.

**Fig 24:** $d \le \sqrt{\frac{3\pi\Phi_o}{4.4.B}}$

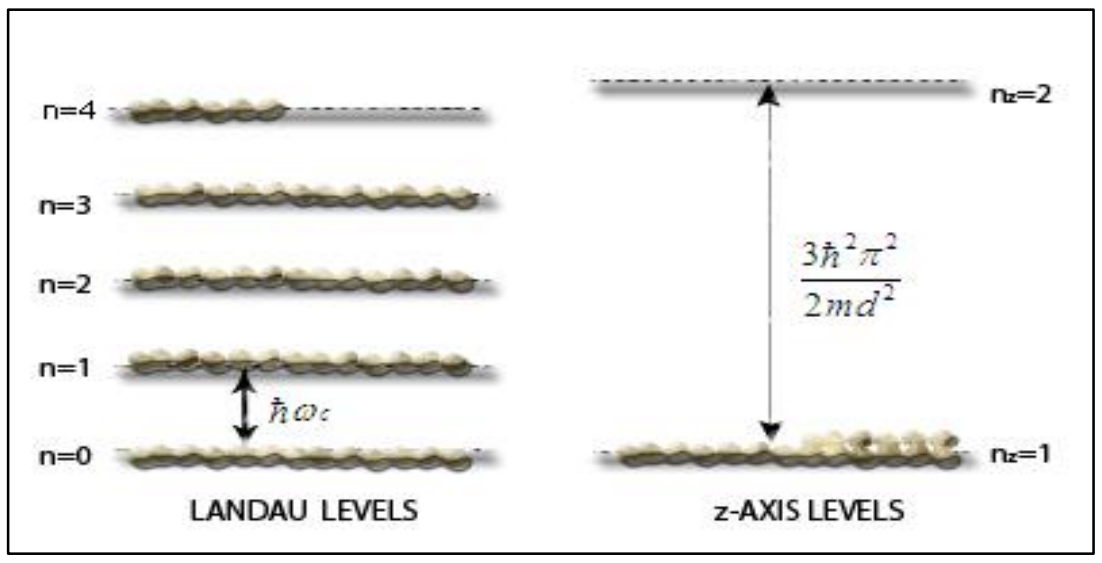

Only lowest QW is occupied because $\Delta E_z\{1, 2\} > 4\hbar\omega_c$

**Fig 25:** $d = \sqrt{\frac{3\pi\Phi_o}{4.4.B}}$

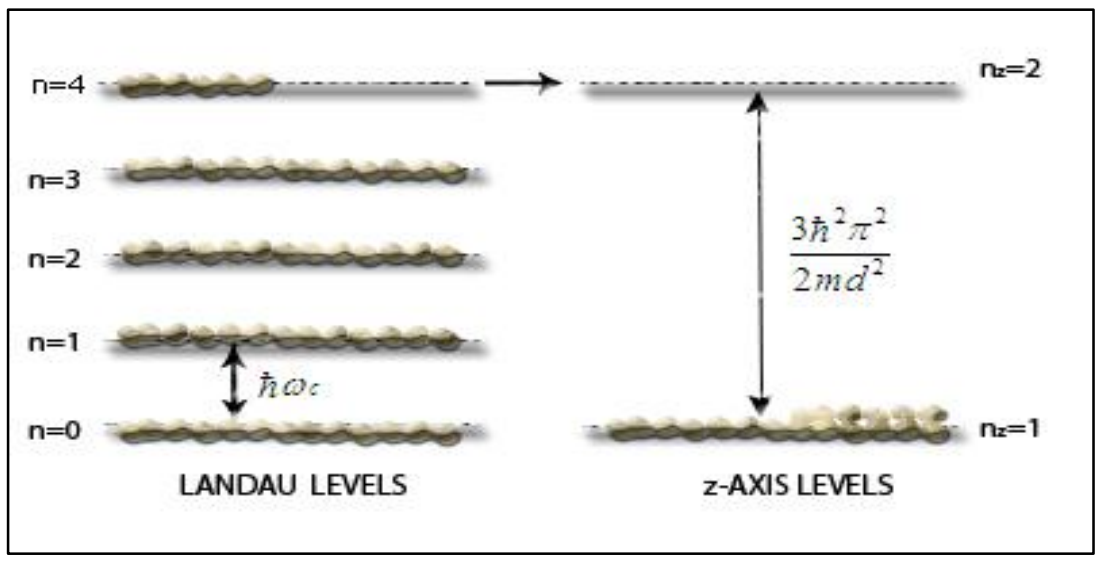

The equality $\Delta E_z\{1, 2\} = 4\hbar\omega_c$ is satisfied

**Fig 26:** $d \ge \sqrt{\frac{3\pi\Phi_o}{4.4.B}}$

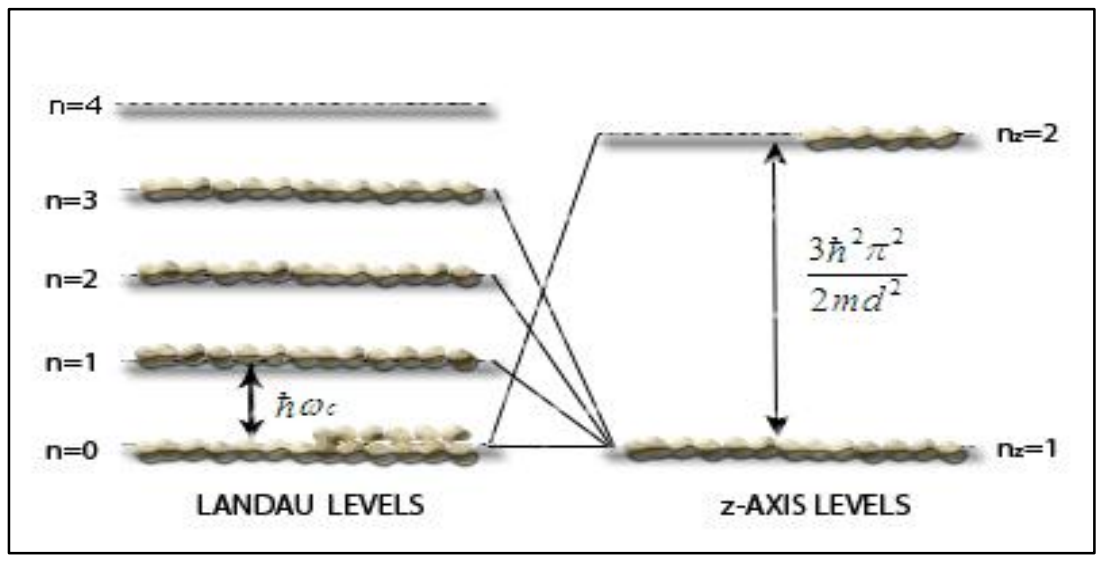

States $\{n=4,\ n_z=1\}$ are abandoned and states $\{n=0,\ n_z=2\}$ are occupied, because: $\Delta E_z\{1, 2\} < 4\hbar\omega_c$

**Fig 27:** $d = \sqrt{\frac{3\pi\Phi_o}{4.3.B}}$

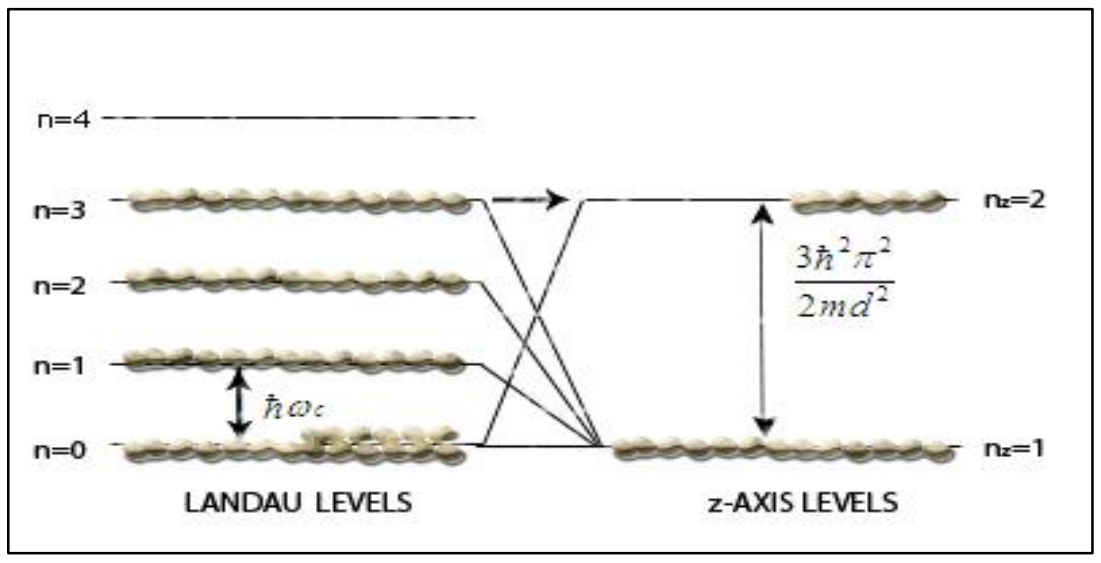

The equality $\Delta E_z\{1, 2\} = 3\hbar\omega_c$ is satisfied

**Fig 28:** $d \ge \sqrt{\frac{3\pi\Phi_o}{4.3.B}}$

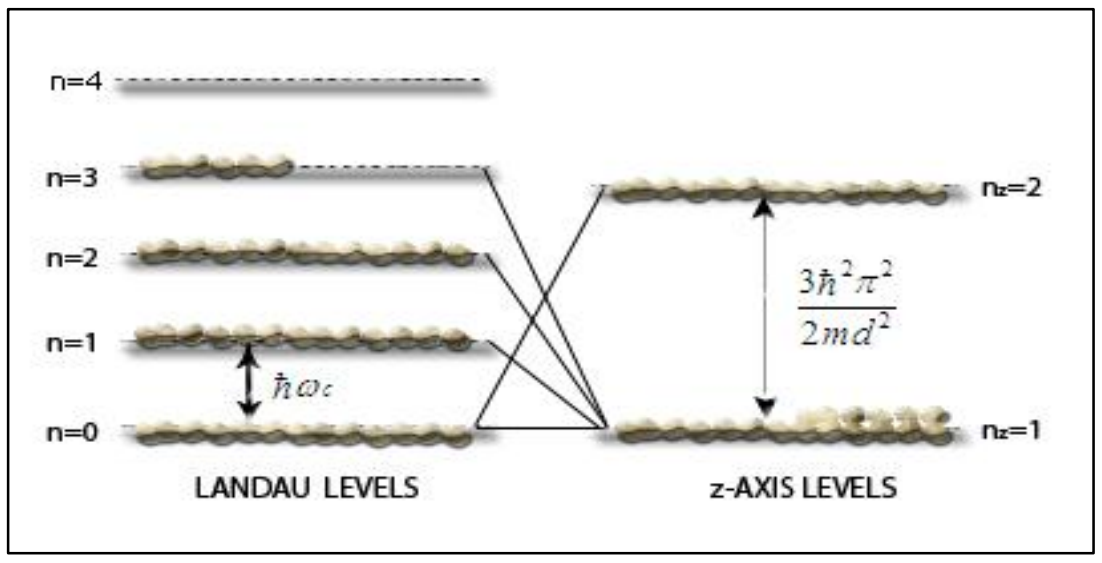

States $\{n=0,\ n_z=2\}$ are fully occupied, while $\{n=3,\ n_z=1\}$ are partially filled, because: $\Delta E_z\{1, 2\} < 3\hbar\omega_c$

**Fig 29:** $d = \sqrt{\frac{3\pi\Phi_o}{4.2.B}}$

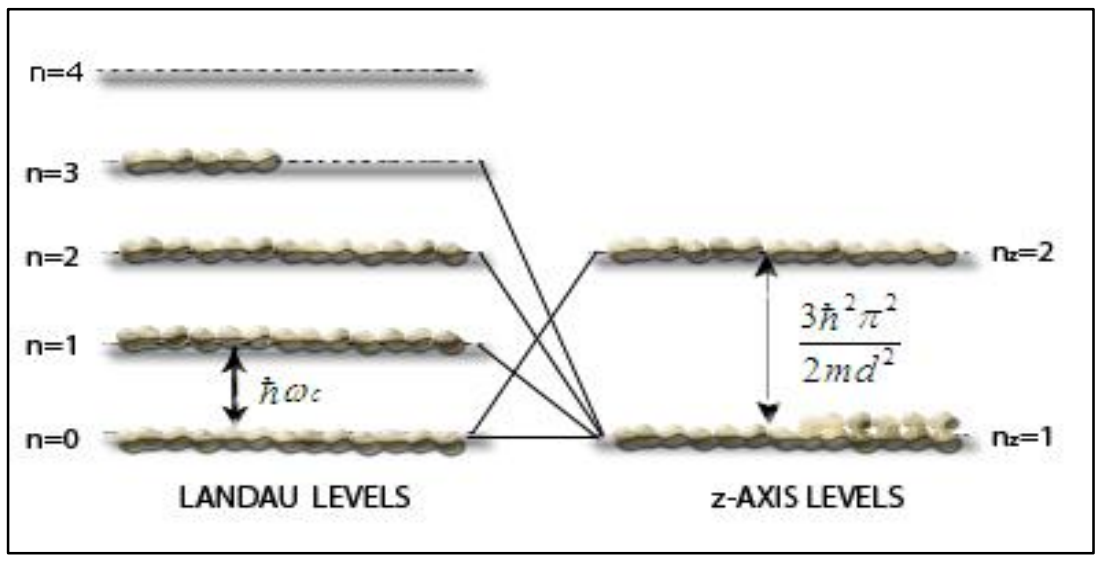

The equality $\Delta E_z\{1, 2\} = 2\hbar\omega_c$ is satisfied

**Fig 30:** $d \geq \sqrt{\dfrac{3\pi\Phi_o}{4.2.B}}$

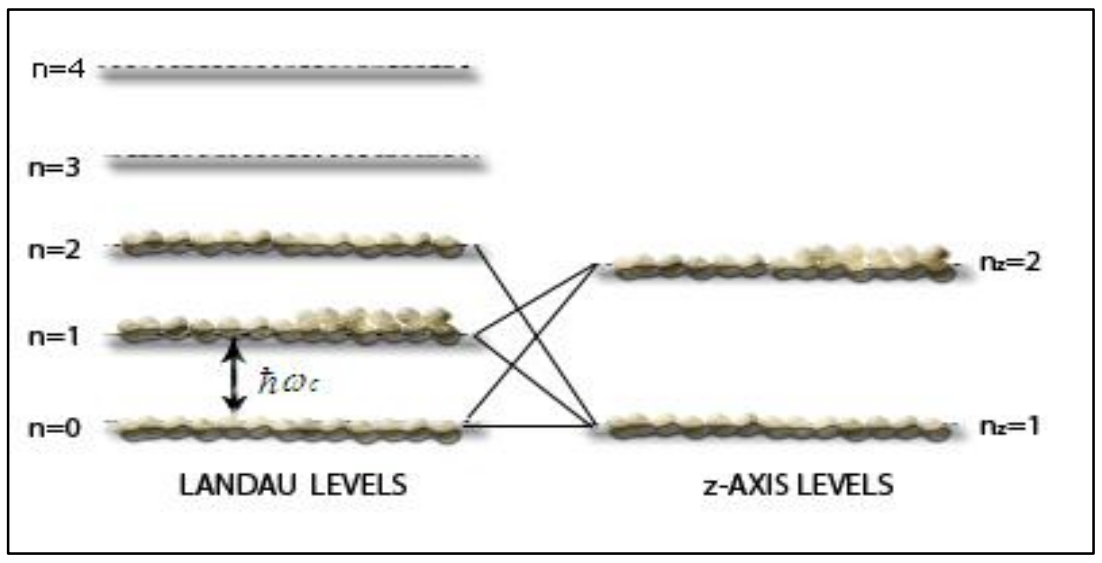

States {$n$=3, $n_z$=1} are abandoned and {$n$=1, $n_z$=2} are partially occupied, because: $\Delta E_z\{1,2\} < 2\hbar\omega_c$

**Fig 31:** $d = \sqrt{\dfrac{3\pi\Phi_o}{4B}}$

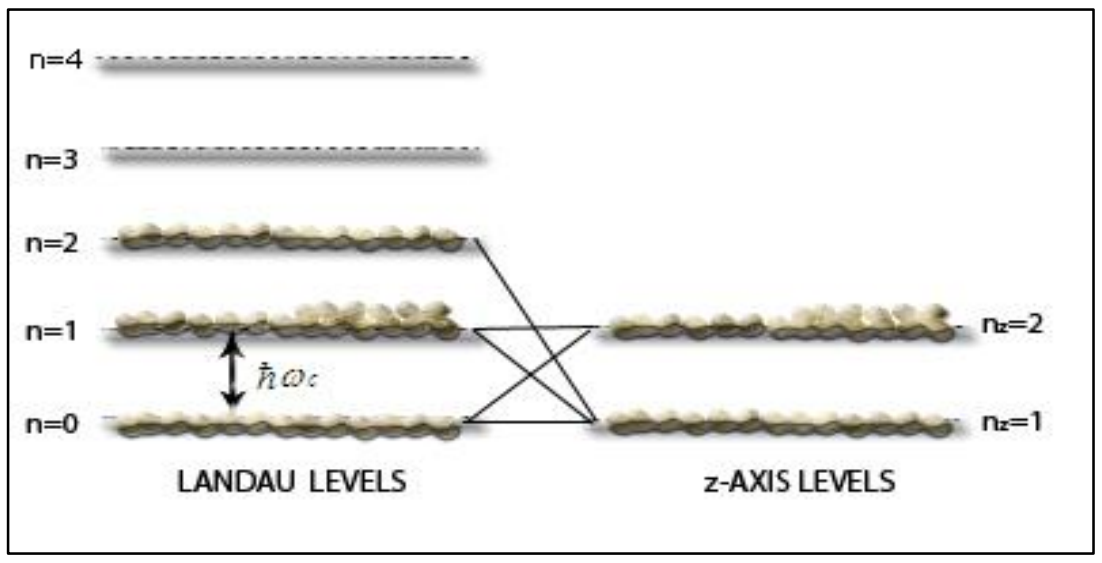

The equality $\Delta E_z\{1,2\} = \hbar\omega_c$ is satisfied

**Fig 32:** $d \geq \sqrt{\dfrac{3\pi\Phi_o}{4B}}$

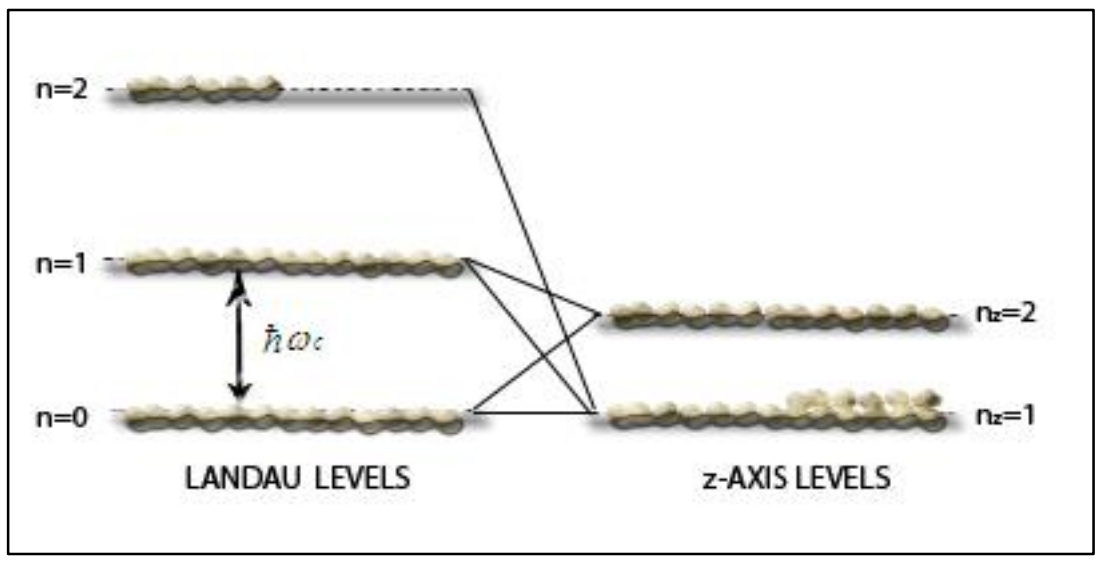

States {$n$=1, $n_z$=2} are fully occupied, while {$n$=2, $n_z$=1} are now partially filled, because: $\Delta E_z\{1,2\} < \hbar\omega_c$

**Fig 33:** $d = \sqrt{\dfrac{4\pi\Phi_o}{4B}}$

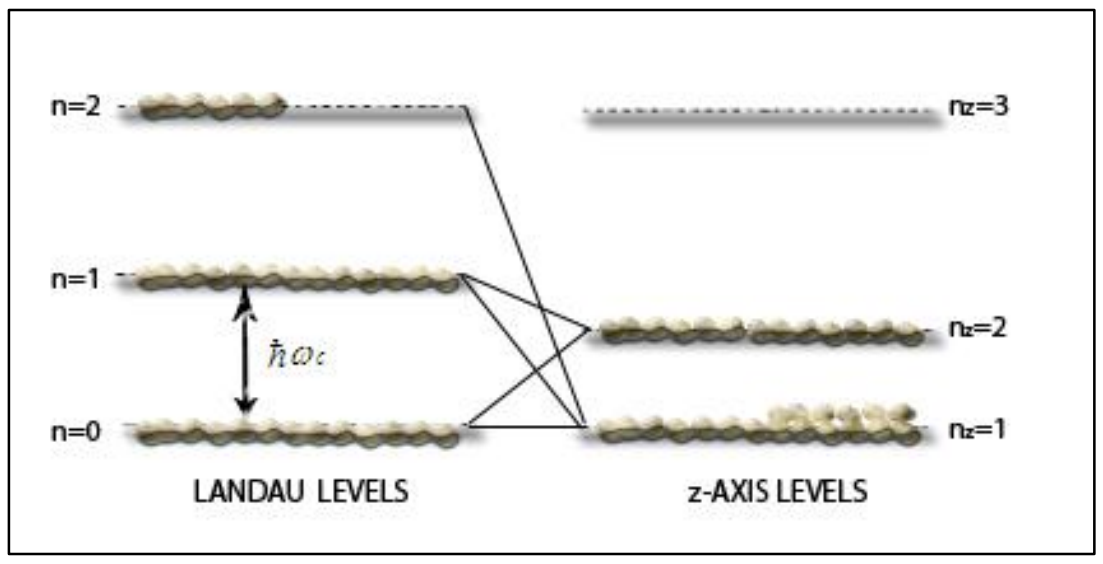

The equality $\Delta E_z\{3,1\} = 2\hbar\omega_c$ is satisfied

**Fig 34:** $d \geq \sqrt{\dfrac{4\pi\Phi_o}{4B}}$

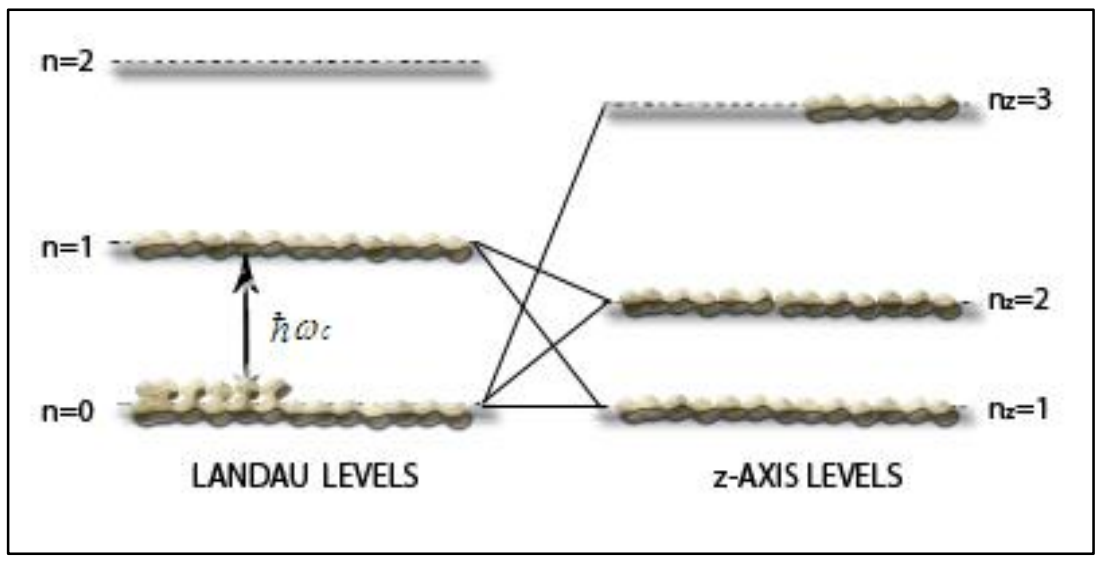

States {$n$=2, $n_z$=1} are abandoned, and {$n$=0, $n_z$=3} are now occupied, because: $\Delta E_z\{3,1\} < 2\hbar\omega_c$

**Fig 35:** $d = \sqrt{\dfrac{5\pi\Phi_o}{4B}}$

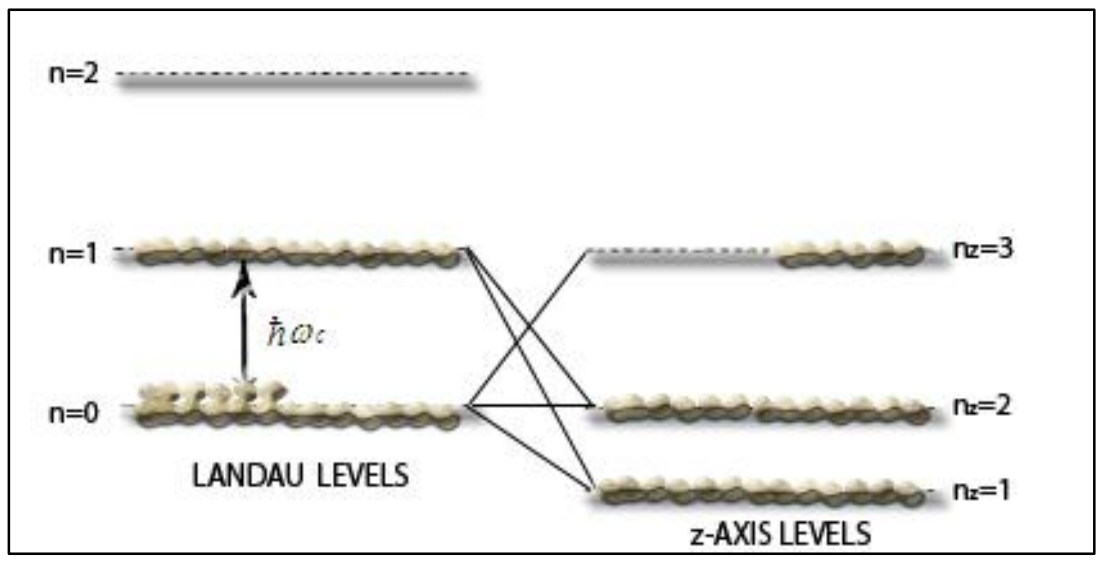

The equality $\Delta E_z\{3,2\} = \hbar\omega_c$ is satisfied.

**Fig 36:** $d \geq \sqrt{\frac{5\pi\Phi_o}{4B}}$

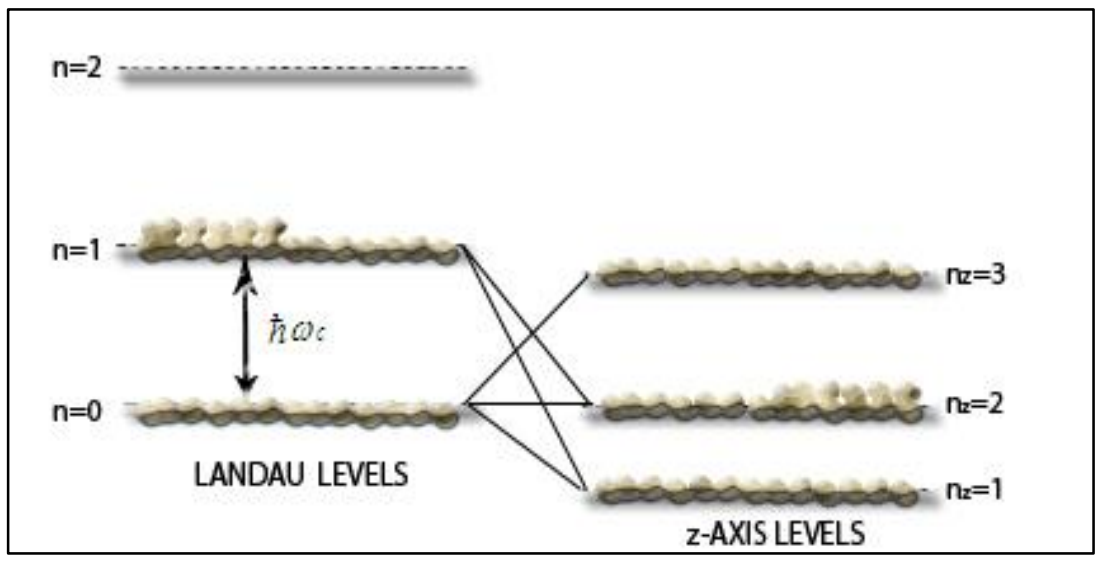

States {$n$=0, $n_z$=3} are fully occupied, while {$n$=1, $n_z$=2} are now partially filled, because: $\Delta E_z\{3,2\} < \hbar\omega_c$

**Fig 37:** $d = \sqrt{\frac{8\pi\Phi_o}{4B}}$

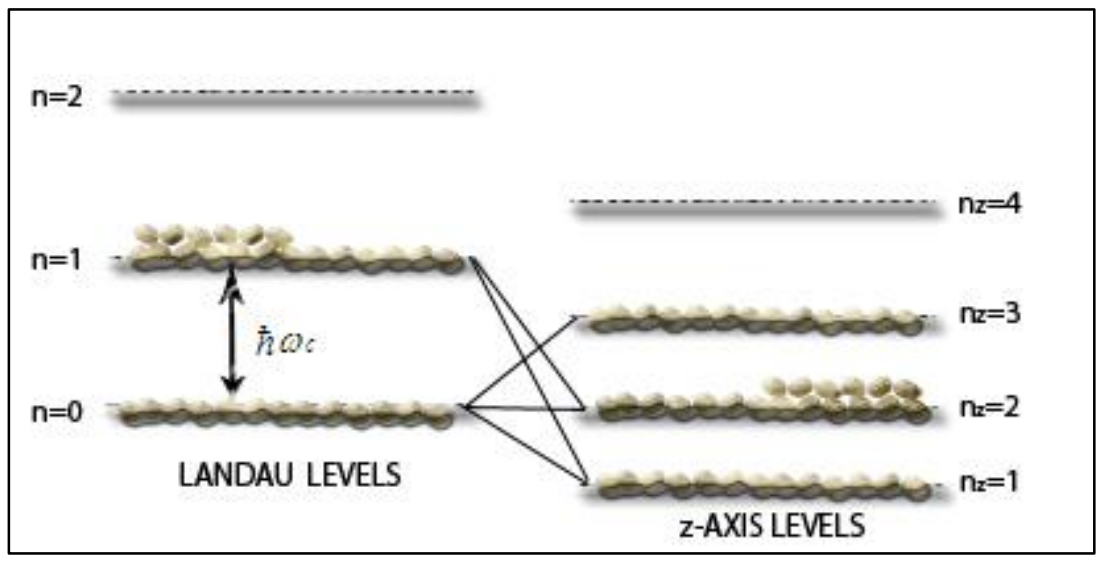

No further changes

**Fig 38:** $d \geq \sqrt{\frac{8\pi\Phi_o}{4B}}$

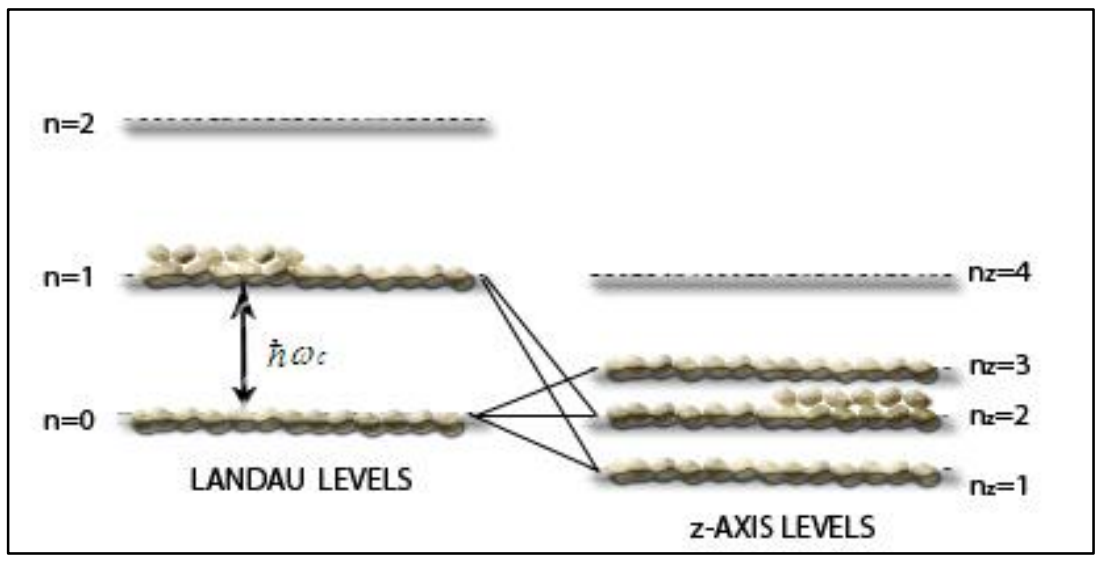

No further changes

**Fig 39:** $d = \sqrt{\frac{12\pi\Phi_o}{4B}}$

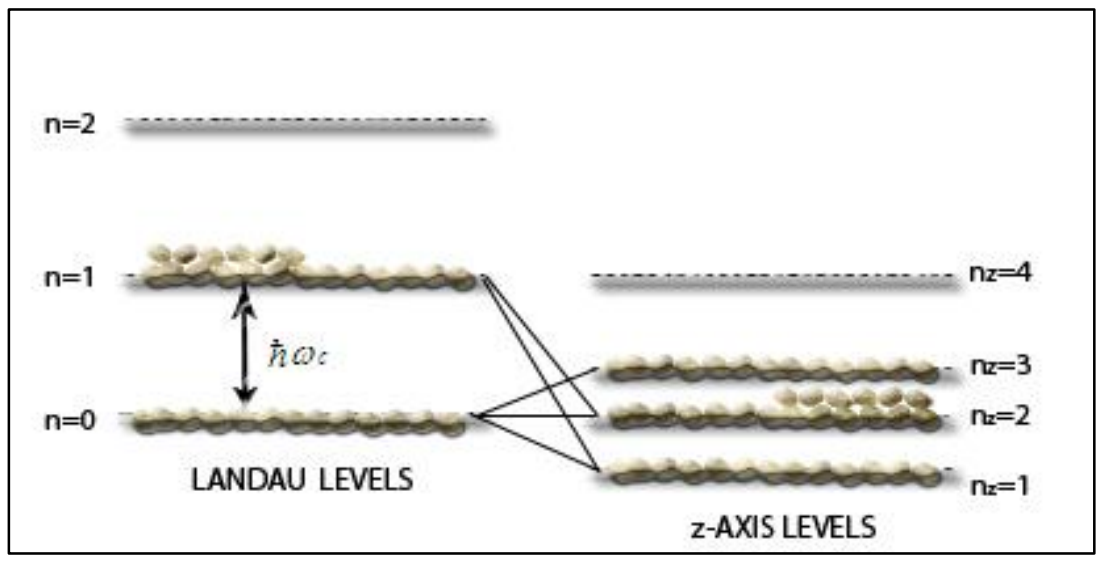

The equality $\Delta E_z\{4,2\} = \hbar\omega_c$ is satisfied

**Fig 40:** $d \geq \sqrt{\frac{12\pi\Phi_o}{4B}}$

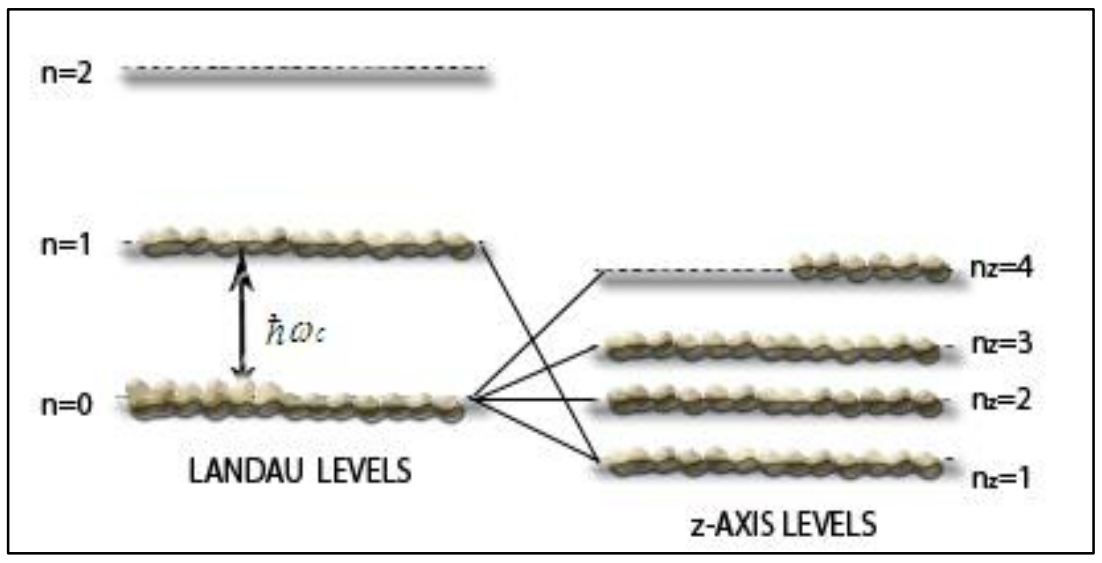

States {$n$=1, $n_z$=2} are abandoned, and {$n$=0, $n_z$=4} are partially occupied, because: $\Delta E_z\{4,2\} < \hbar\omega_c$

**Fig 41:** $d = \sqrt{\frac{15\pi\Phi_o}{4B}}$

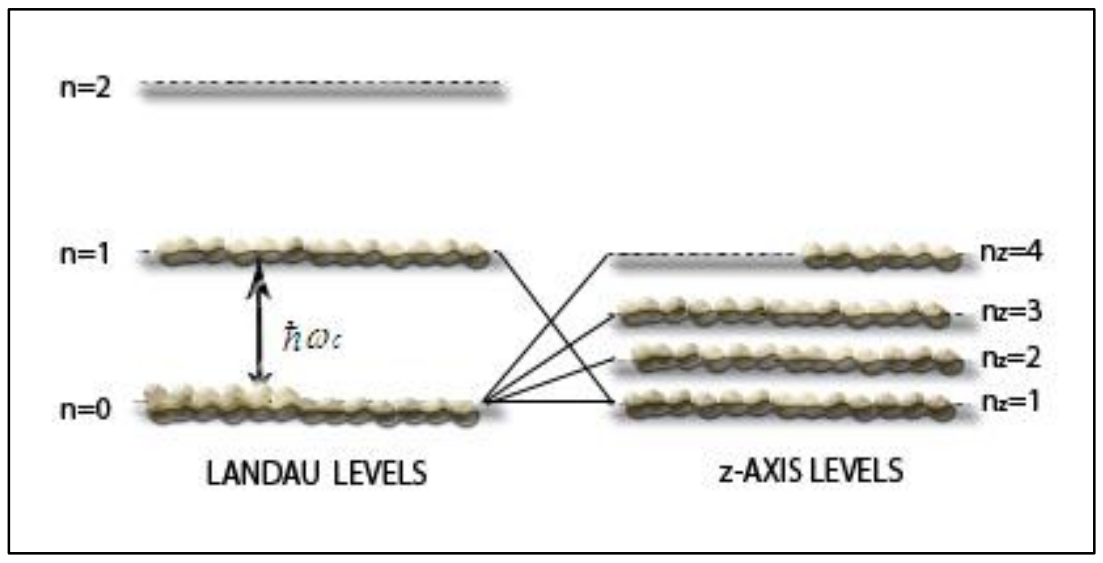

The equality $\Delta E_z\{4,1\} = \hbar\omega_c$ is satisfied

**Fig 42:** $d \geq \sqrt{\dfrac{15\pi\Phi_o}{4B}}$

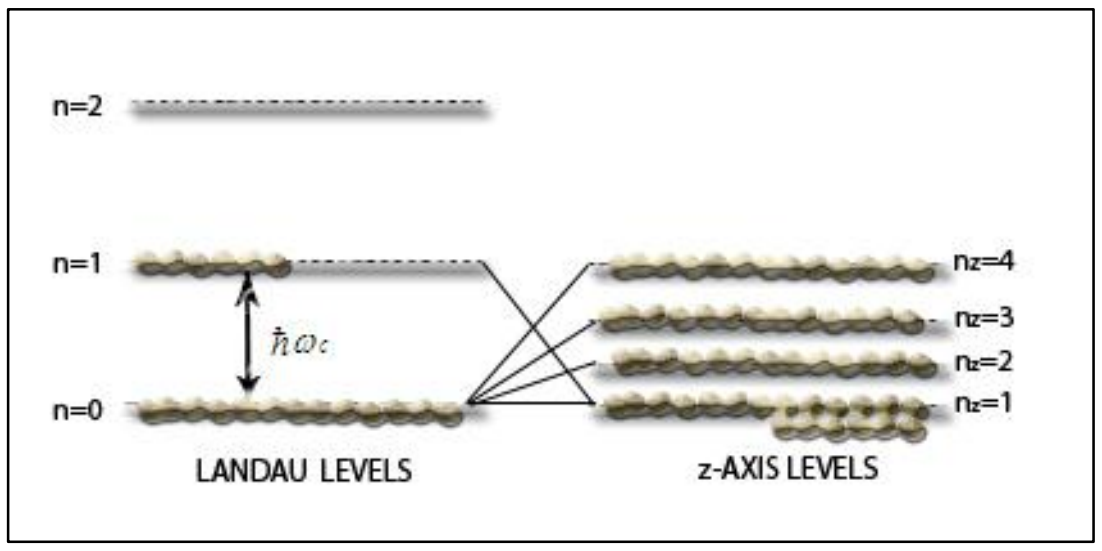

States $\{n=0,\ n_z=4\}$ are fully occupied, while $\{n=1,\ n_z=1\}$ are partially filled, because: $\Delta E_z\{4,1\} < \hbar\omega_c$

**Fig 43:** $d = \sqrt{\dfrac{24\pi\Phi_o}{4B}}$

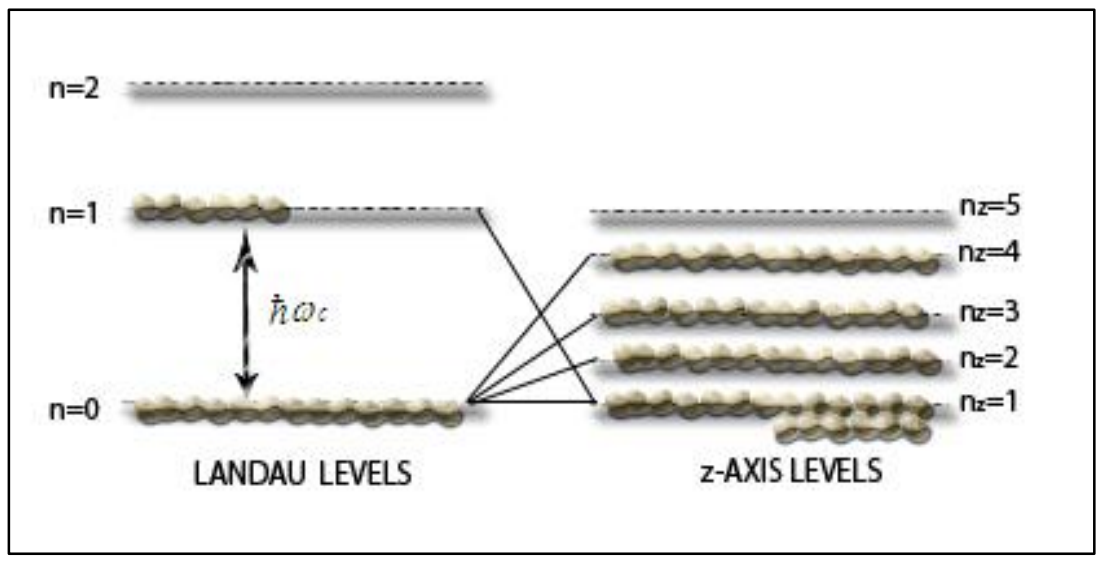

The equality $\Delta E_z\{5,1\} = \hbar\omega_c$ is satisfied

**Fig 44:** $d \geq \sqrt{\dfrac{24\pi\Phi_o}{4B}}$

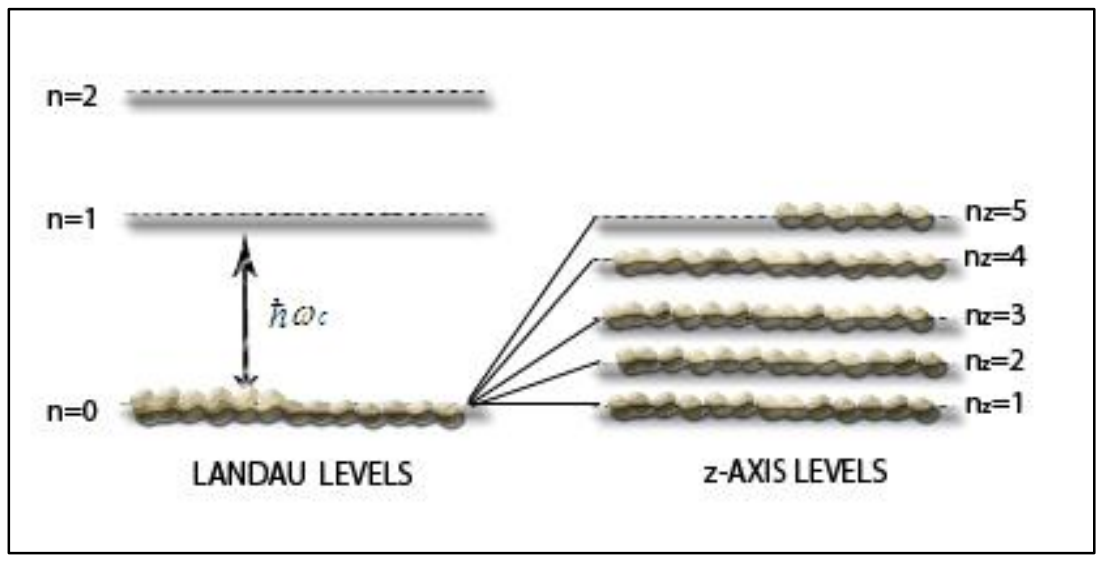

All distinct *z*-axis levels are occupied (combined with the lowest LL), because: $\Delta E_z\{5,1\} < \hbar\omega_c$

From these examples we can observe some well-defined patterns at the two ends of the procedure (i.e. of the range of variation of *B* and *d*), but we also observe a certain unpredictability that requires utmost care in the middle of the procedure (note for example in Fig.ôs 35-40 that comparisons have to necessarily involve higher QW levels). After the optimal scenarios are carefully found and run, for every window of *B* and *d* values, it is straightforward to write down analytically the total energy for each case. The most important information left is then to draw the graphs of the total energy, magnetization and susceptibility as functions of the thickness *d*, or magnetic field *B*, or both. (Once again, although fixed *B*-variations describe better the theoretical patterns, fixed *d* is the experimentally relevant case (which is also shown) ï the combined variation also being provided in 2D graphs later that demonstrates everything in a compact manner). In the figures below we take the areal density to be $n_A = 10^{16} m^{-2}$. We first plot the 1D graphs (i.e. with respect to one variable only, the 2nd held fixed), and later we present some 2D graphs (under combined variation of *B* and *d*). First we keep *B* fixed (and the reader should recall that, although the thickness *d* is treated as an independent variable, the windows of *d*-values (for which we have a particular analytical expression for the total energy E) *do* depend on *B).*

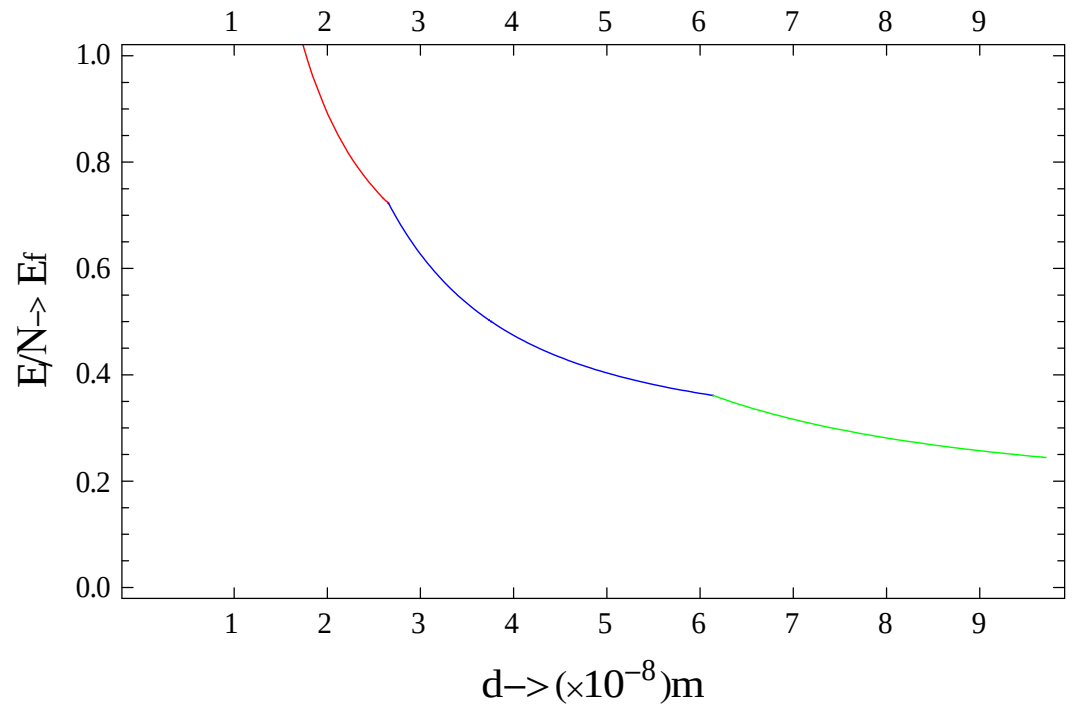

[ȷ ]

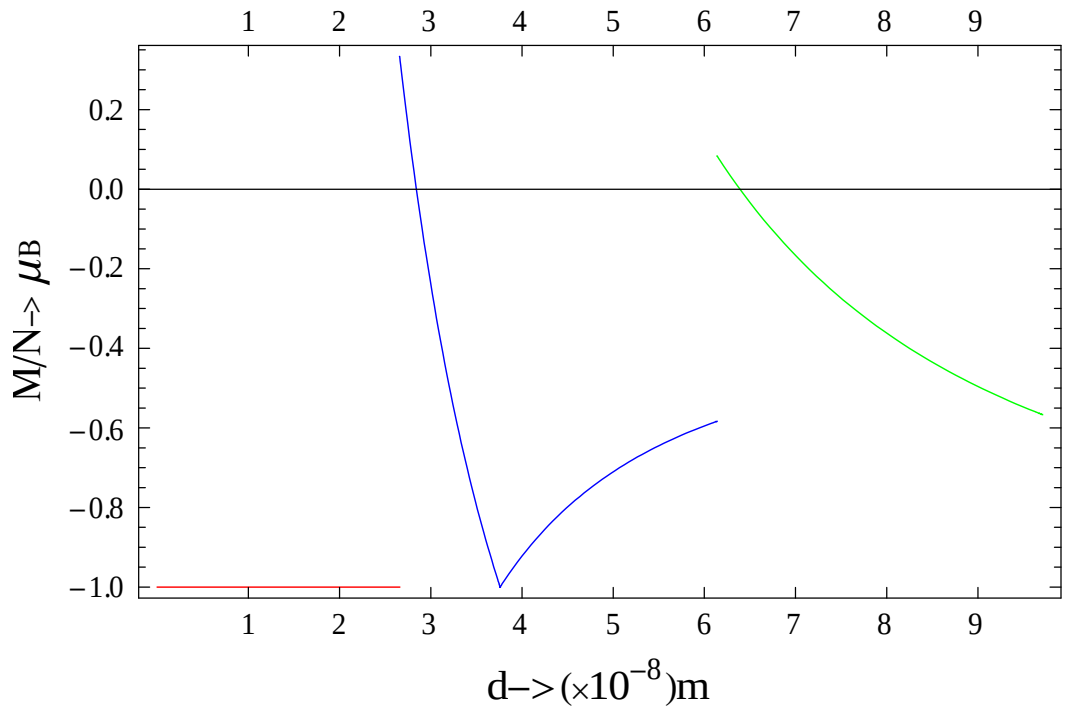

[ȸ]

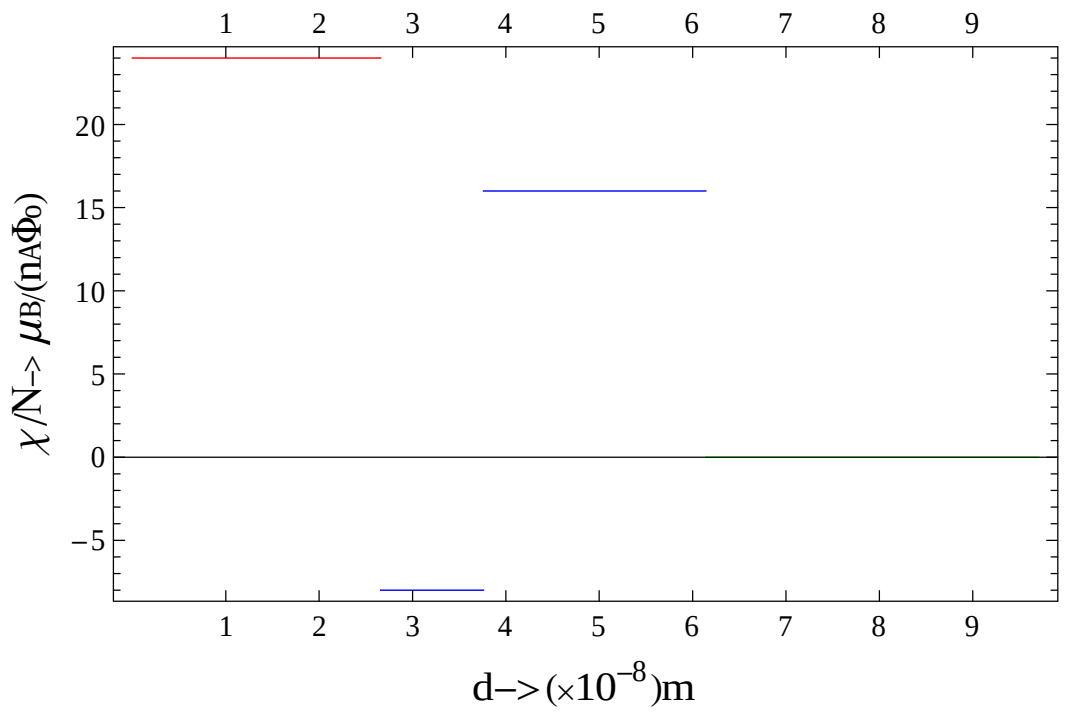

[C]

**Fig 45:** Graphs: A) Energy, ȸ) Magnetization, C) Susceptibility per electron as functions of thickness *d* when the magnetic field is 1/6nAū ɵ (hence we have complete LL filling). Different *d*-windows are presented with different color. Note that susceptibility can be negative (as opposed to the 2D case)

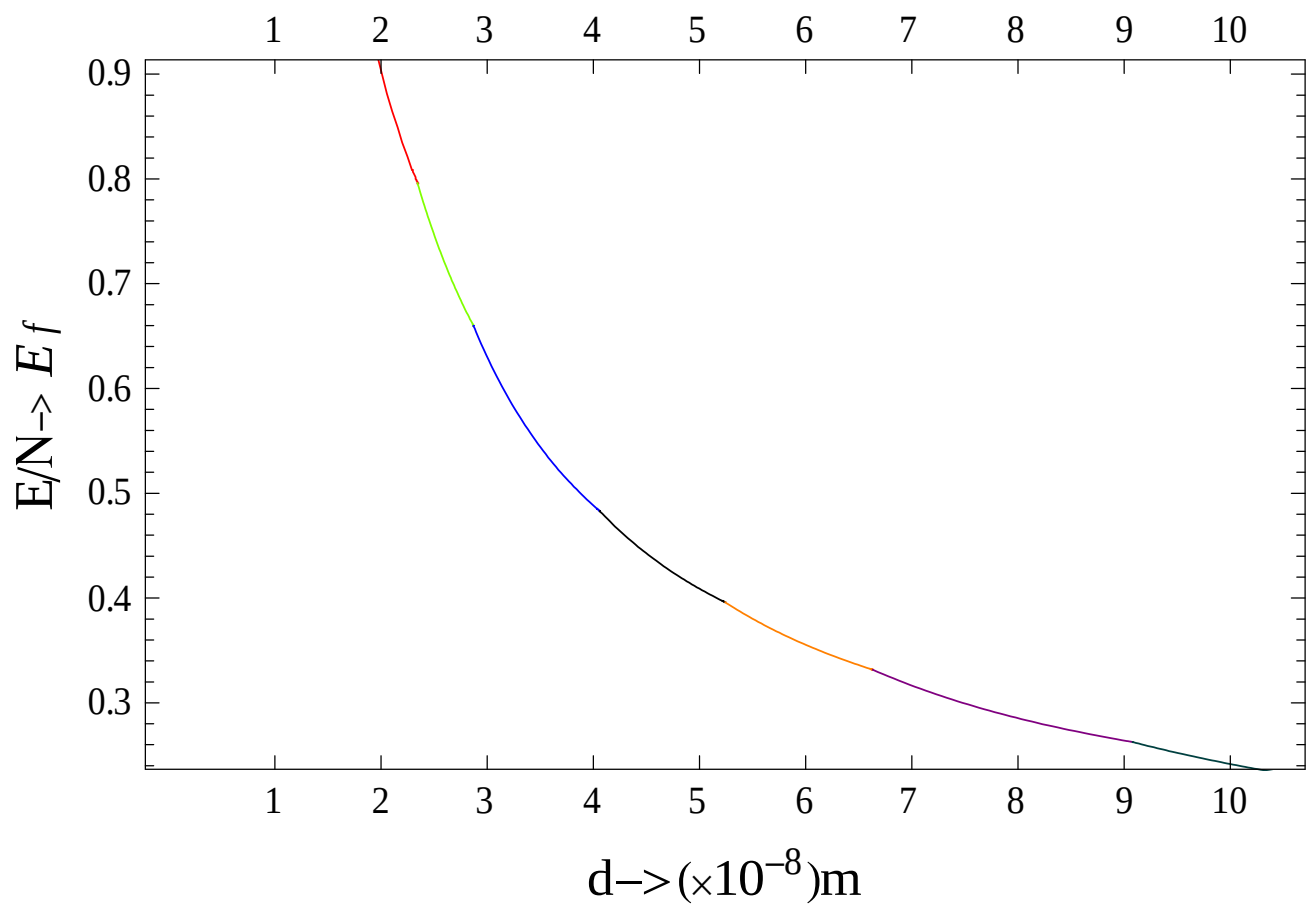

[D]

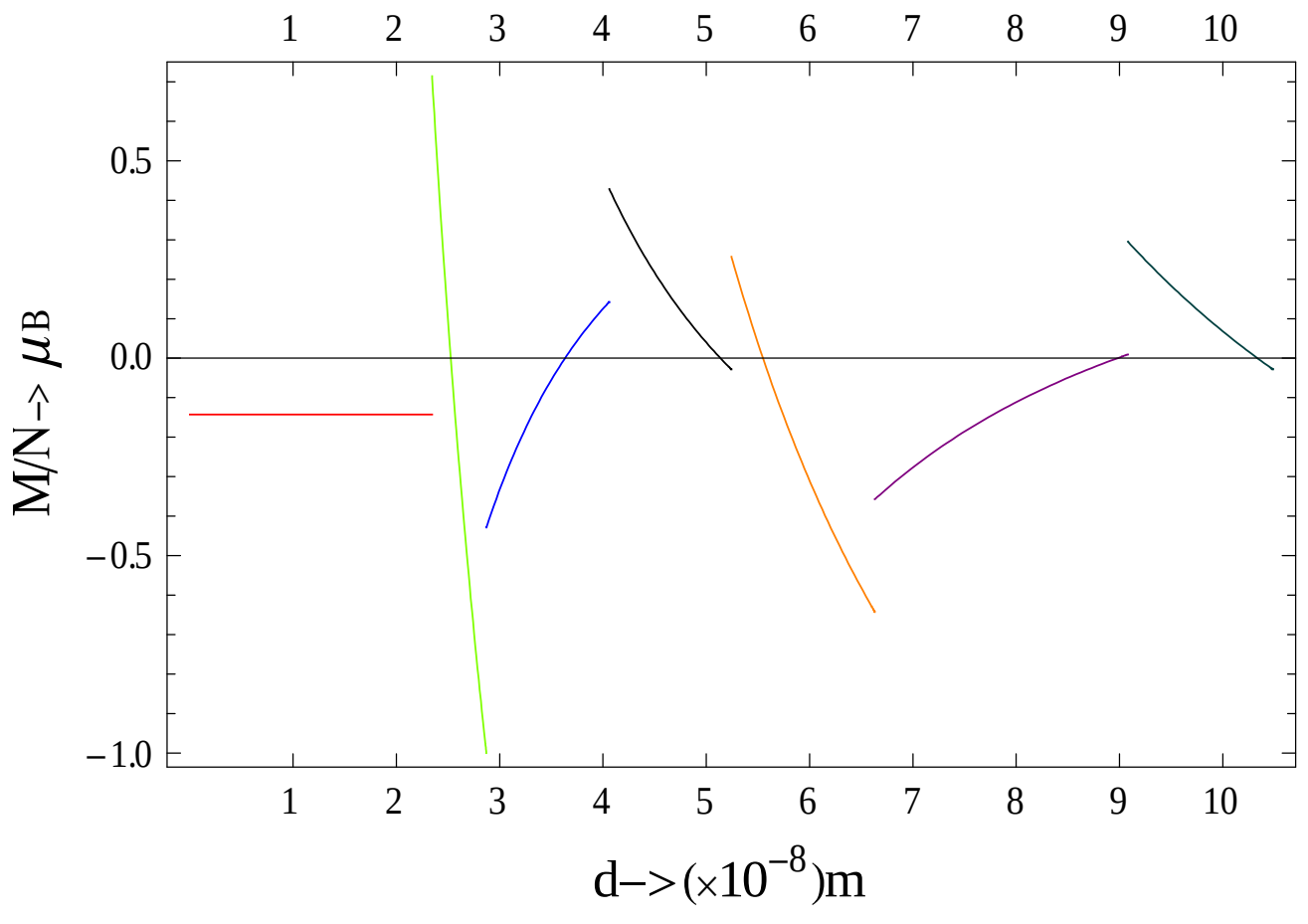

[Ⱥ]

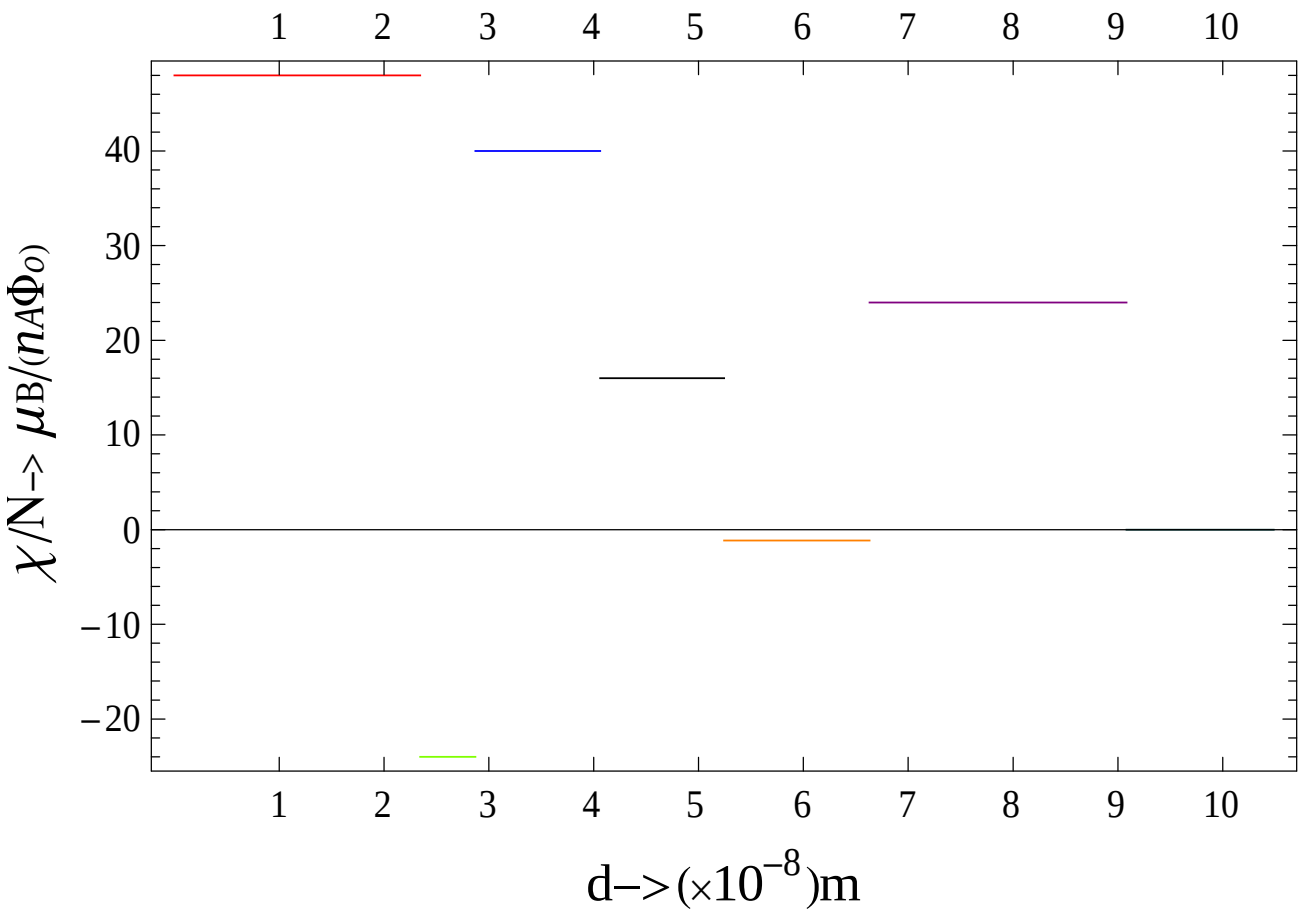

[Ȼ]

**Fig 46:** Graphs: D) Energy, Ⱥ) Magnetization, Z) Susceptibility per electron as functions of thickness *d* when the magnetic field is 1/7nAū ʘ (hence we now have partial LL filling). Although the total energy is everywhere continuous, note the discontinuities that take place in magnetization and magnetic susceptibility; the latter may have negative values, contrary to the 2D system that only gave positive values. Different *d*-windows are presented with different color (although it should be noted that all transitions shown here are ñinternal transitionsò, and this will affect later figures, when *B* will be varied (for fixed *d*), where each of these transitions will appear as ñinternal breakingsò, with branches that will have the same color).

We should note again that the new (internal) transitions found above correspond to incomplete LL filling, and one would be tempted to speculate that these might lead to interesting effects (pertinent to fractional fillings and the FQHE) *if interactions were included* (even turned on perturbatively)**;** however, let us make the choice to restrict ourselves to noninteracting particles for consistency of the approach. (After all we want to ultimately apply this line of reasoning to topological insulators (see Section 6) which are actually defined in a one-electron Physics picture).

Next, we present again 1D graphs but in the case where the thickness *d* is kept constant and magnetic field *B* is varied. Note the discontinuities in magnetization and magnetic susceptibility for some values of the magnetic field *B*. Furthermore, there are cases where magnetization may also have discontinuities in the interior of a *B*-window (see for example graph [M] at value of 1/*B* =15/$n_A$ū 0), an example of an ñinternal breakingò with both its branches shown with the same color**;** such breakings (that are actually phase transitions corresponding to partial LL filling, as we saw earlier) have not been noted in theoretical treatments in the past, and they are not in accordance with the dHvA effect).

**Thermodynamic quantities for** $d = \sqrt{\frac{30\pi}{n_A}}$ (97 *nm*)

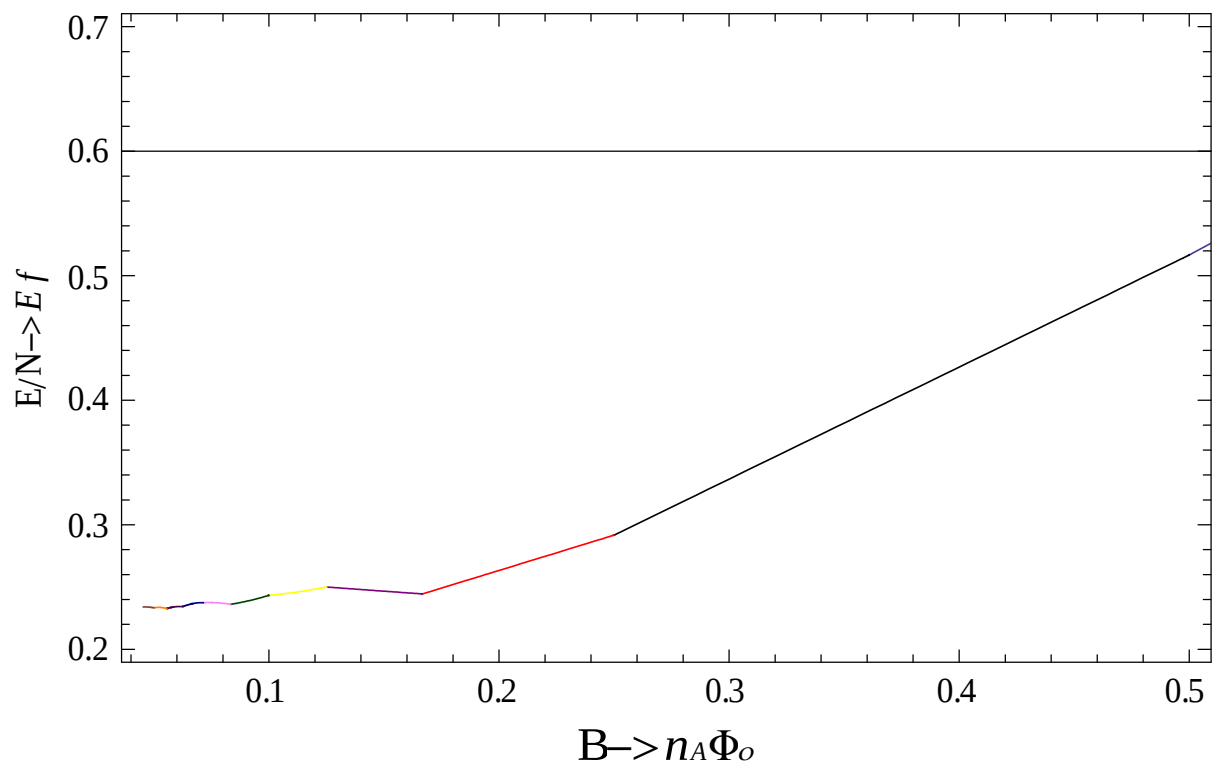

[Ɨ]

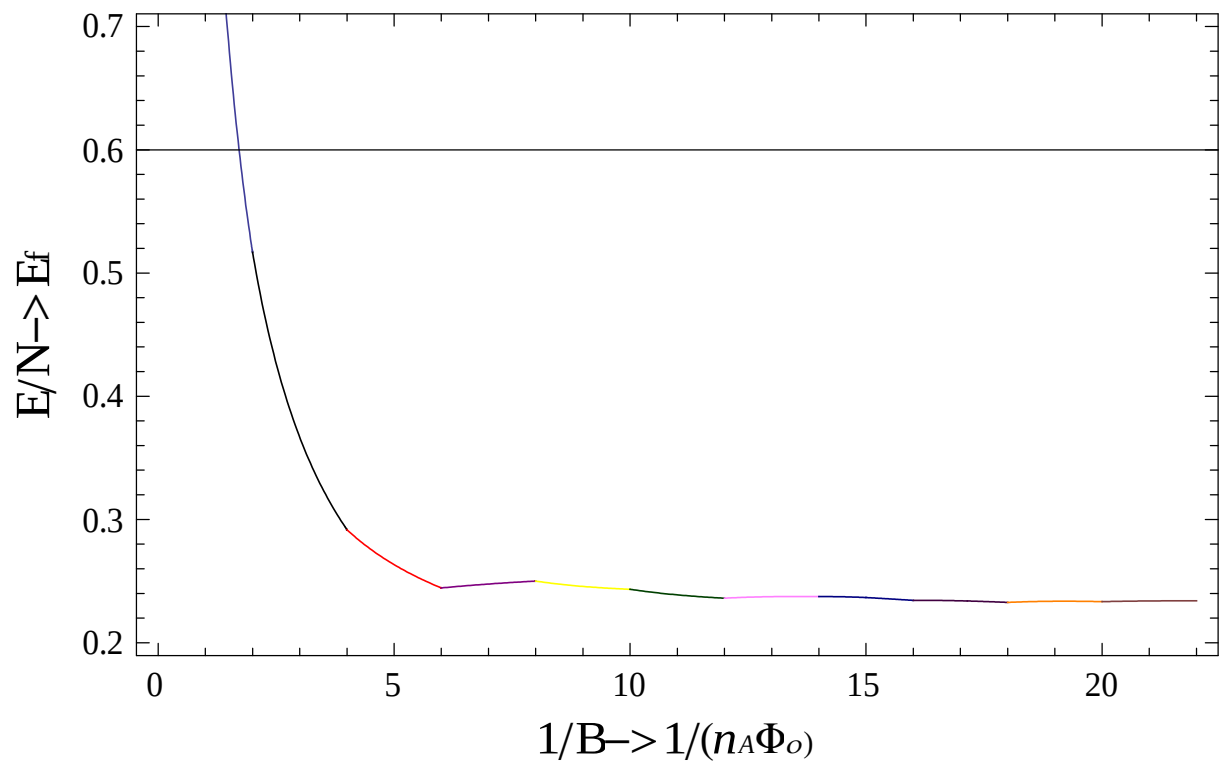

[Ʈ]

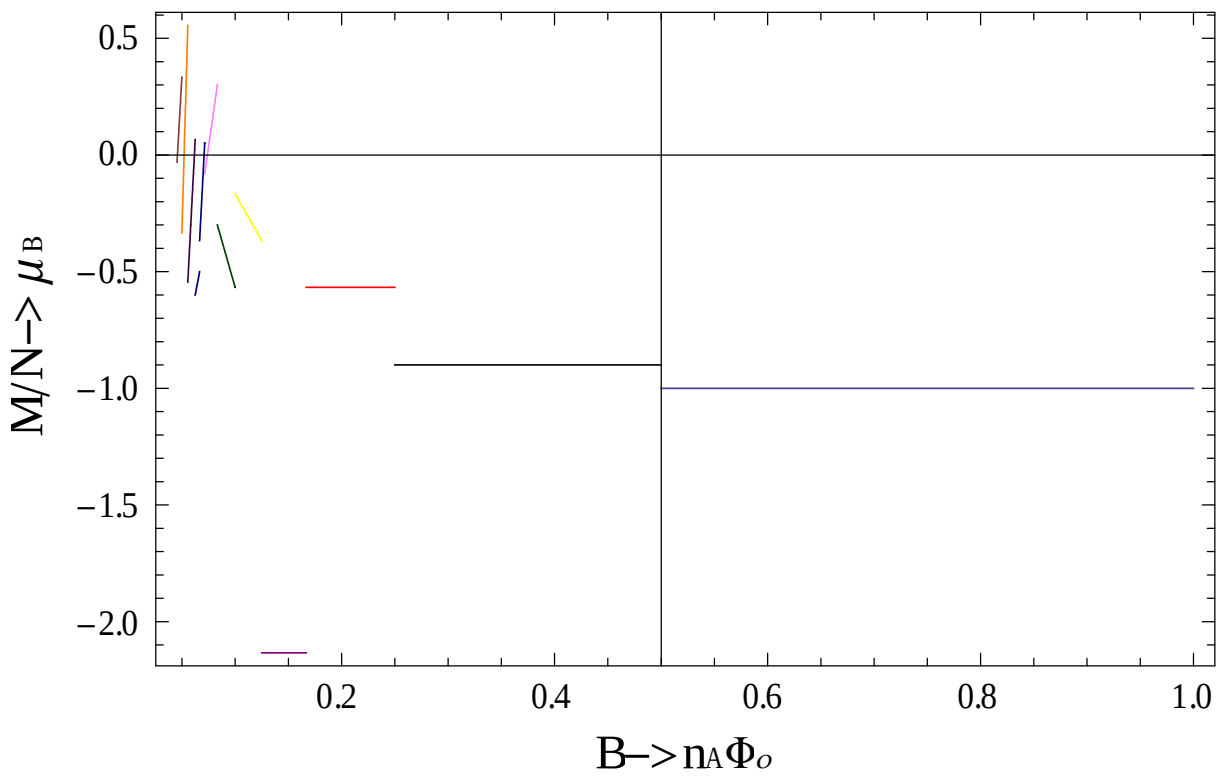

[L]

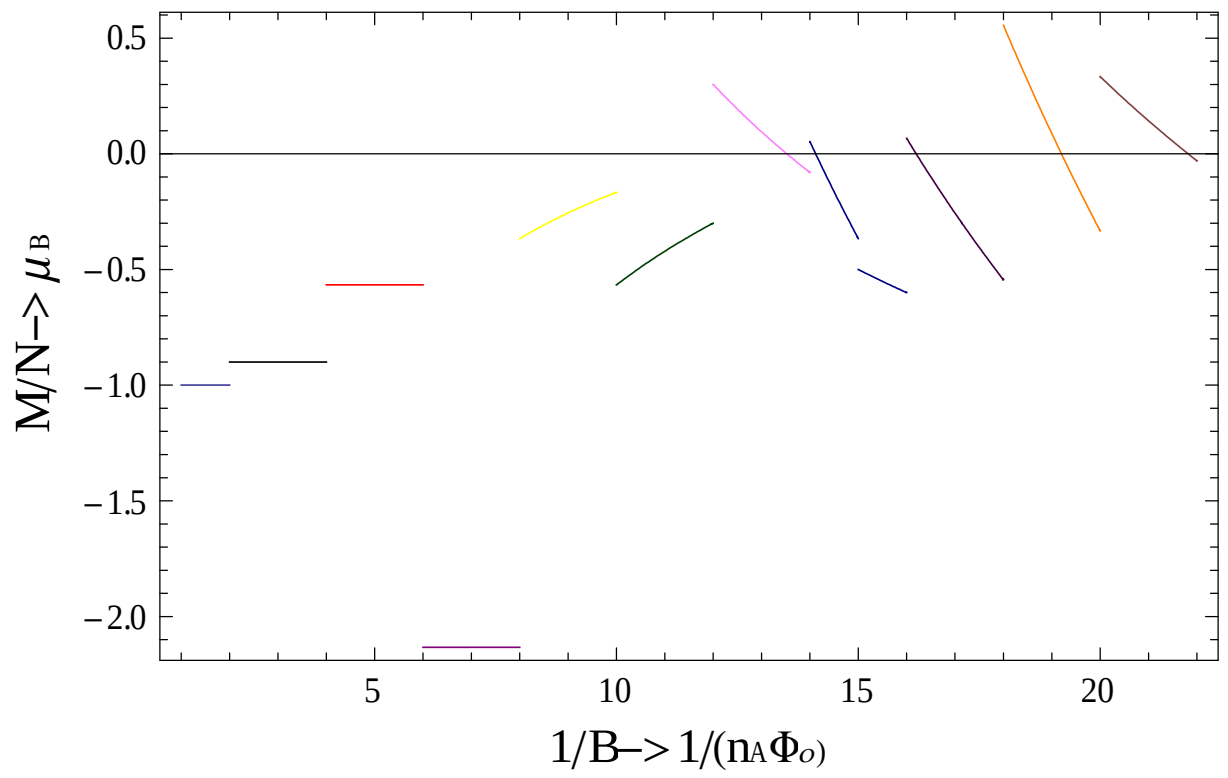

[ɀ]

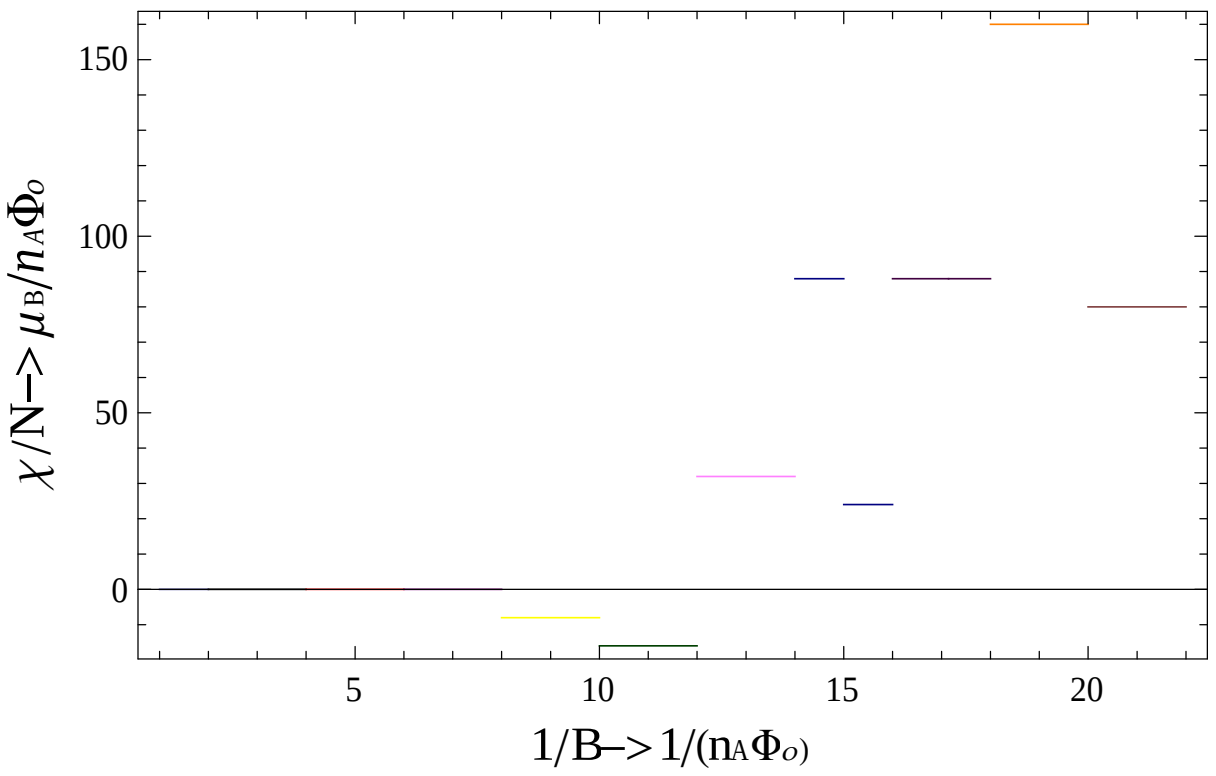

[ʔ]

**Fig 47:** Graphs: I) Energy, L) Magnetization per electron as functions of ȸ and Ʈ) Energy, ɀ ) Magnetization, ʔ) Susceptibility as functions of inverse *B*.

➢ **Thermodynamic quantities for:** $d = \sqrt{\dfrac{375\pi}{2n_A}}$ (242 $nm$)

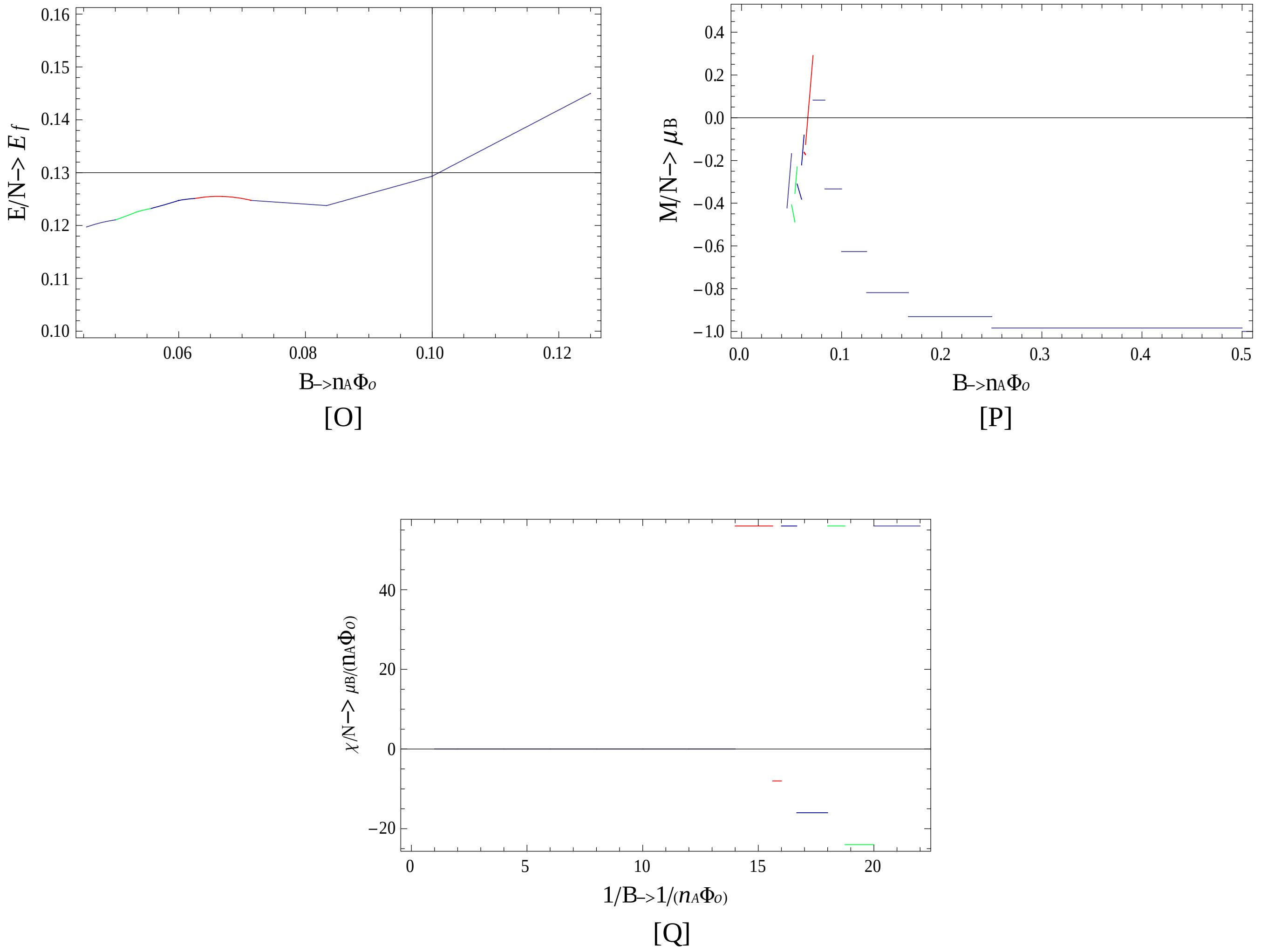

**Fig 48:** Graphs: O) Energy, P) Magnetization per electron as functions of $B$ and Q) Susceptibility per electron as function of inverse $B$. (Note internal transitions at 1/B ~ 15.5 and and also ~18.8 in units of $1/n_A\Phi_0$)

Some comments concerning these graphs are now in order. Energy is always (as in the case of 2D) a continuous function of the magnetic field, as expected on general physical grounds. Graph O shows the energy as a function of $B$ for somewhat large thickness (about 242 nm), which, as we shall see in the next section, looks as almost identical (in numerical values) to the energy that comes out for the case of full 3D space with periodic boundary conditions (see Section 5, although we will see that the energy for that system is perfectly smooth (continuous and differentiable), while here it still have cusps (the magnetization has discontinuities)). All thermodynamic quantities such as energy, magnetization and susceptibility converge to the corresponding full three dimensional quantities when the thickness is very large, signifying that boundary conditions (here a double rigid wall) don't actually matter when the space available to electrons is very large (at least for this conventional system).
While energy is a continuous function of $B$, magnetization and susceptibility on the other hand are not. With respect to the critical values of $B$, where all energy states are fully occupied (or fully empty), this is not a surprise. We could predict these discontinuities by examining the semiclassical dHvA effect, according to which magnetization and susceptibility are periodic functions of $1/B$, with period $2/n_A\Phi_o$. But this is not the only type of discontinuities here**;** from graphs [P] and [Q] one notices that there are cases where magnetization and susceptibility have discontinuities even *in the interior* of a dHvA window of *B*-values (see for example graph P and Q at values of 1/B ~ 15.5 and 18.8 in units of $1/n_A\bar{u}_0$). This is a new observation, a result not captured by other approaches, and demonstrates the nontrivial role that thickness *d* plays even in this simple problem.

In the following, we also present the corresponding 3D graphs of all thermodynamic properties, as functions of combined variations of both *B* and *d* for the first few windows of *B*-values.

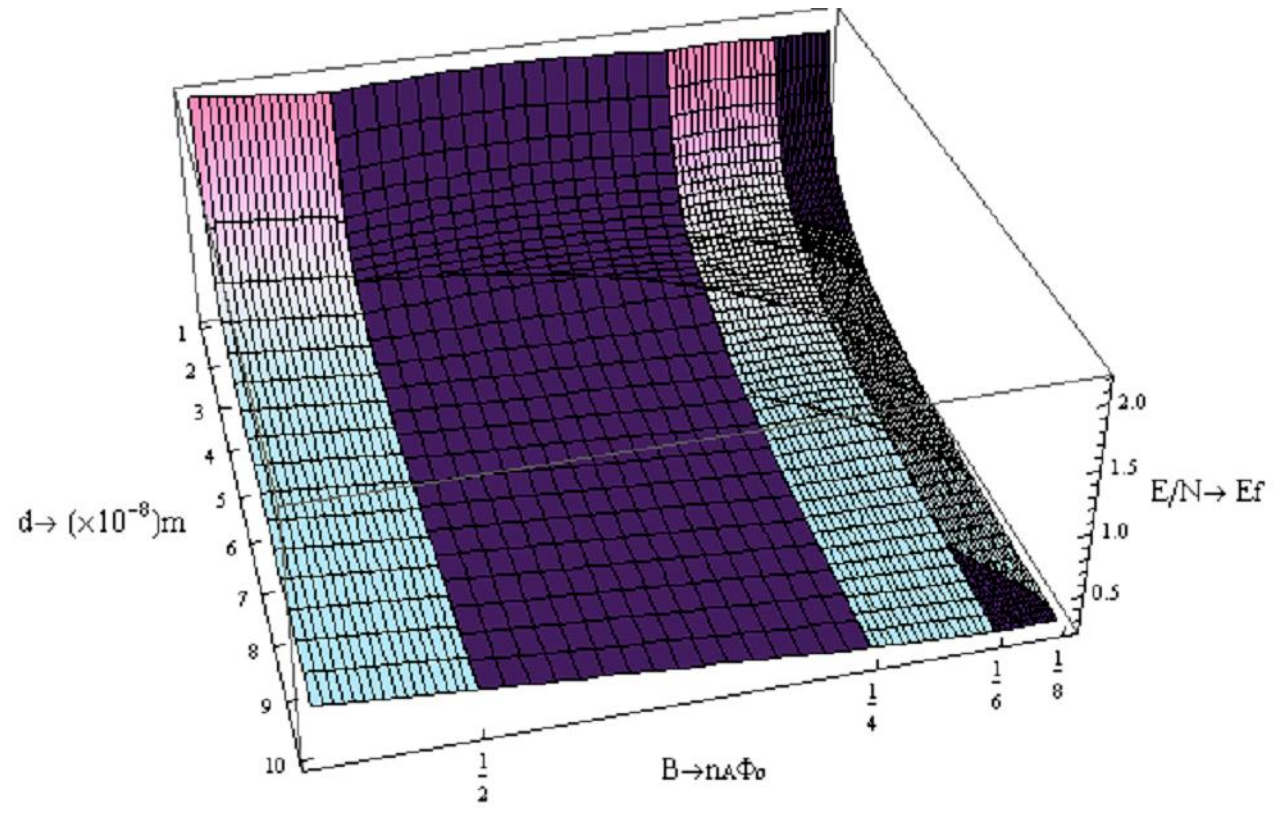

**Fig 51:** Total energy as a function of both *B* and *d*

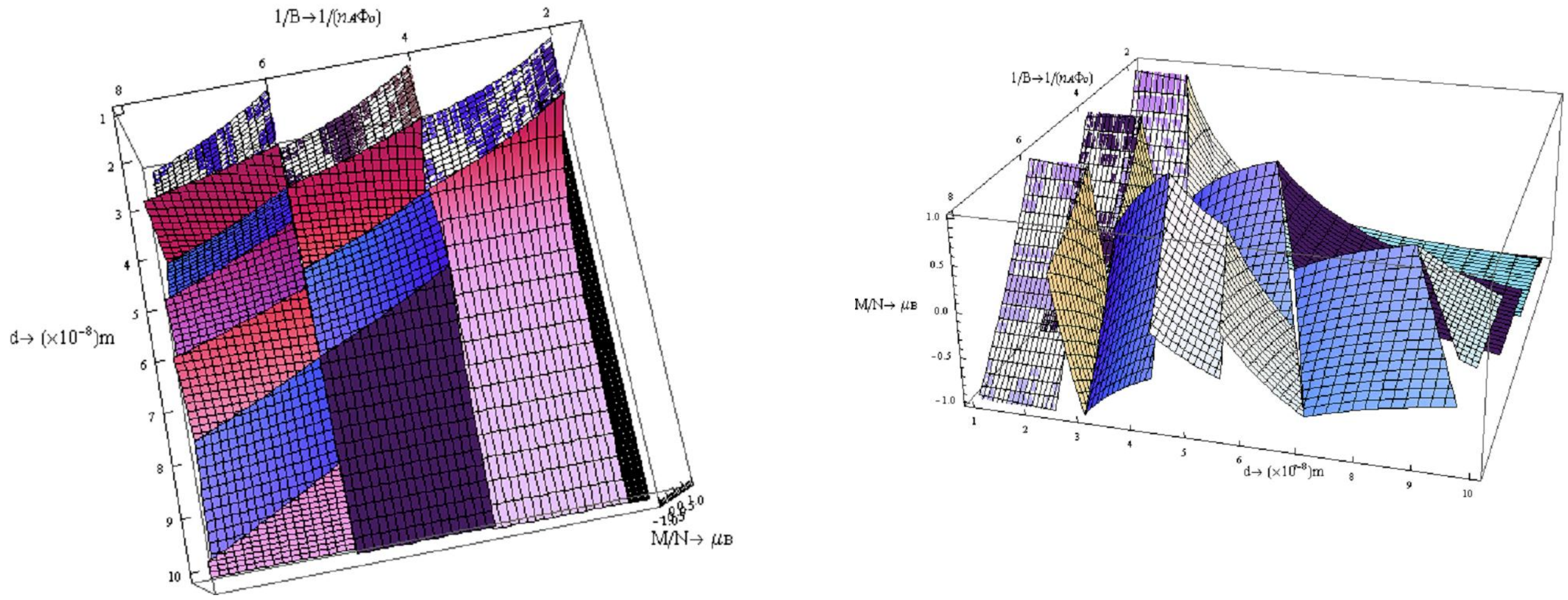

**Fig 52:** Magnetization as a function of both *B* and *d*.

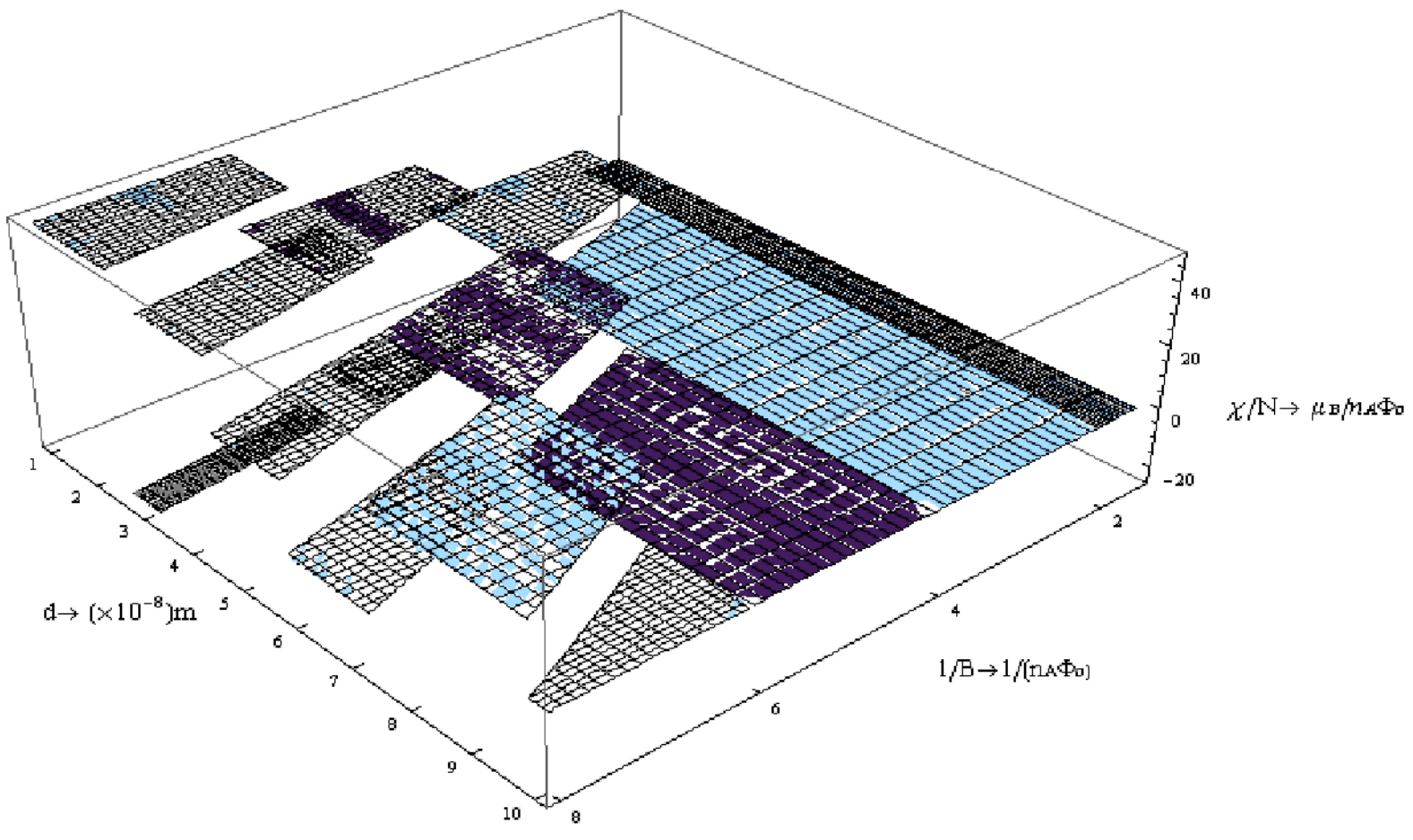

**Fig 53:** Magnetic Susceptibility as a function of *B* and *d*.

Let us summarize some observations concerning all these results. Unlike the energy, magnetization and susceptibility are strongly discontinuous, both in *B*- and *d*-axis (one should notice the oscillations along the *B*-axis when thickness is very small, where the system behaves as effectively being two dimensional). These rapid discontinuities create, for a very thin film, a sawtooth behavior similar to the case of two dimensions. We conclude that the system oscillates between paramagnetism and diamagnetism (hence it experiences phase transitions), since magnetization also changes its slope during the increase of *B* or *d*. As thickness increases, the single particle energetic configurations change, in a manner that is not predictable a priori, and this has consequences. There are new, qualitatively different from 2D, transitions occuring: magnetization can be discontinuous even in the interior of a *B*-window (when the highest LL is incompletely filled), something that violates the standard periodicities given by dHvA effect (that are always related to complete filling in 2D). This comes out of the energy interplays between LL and QW levels and, as already emphasized, it occurs in patterns that not easily predictable. These patterns are analysed more closely (analytically) in the Appendix. But we can quickly give here some further quantitative observation**:** Magnetization and susceptibility may have several discontinuities for arbitrary values of *B* and *d*, and we have noted that in each *B*ï window there are exactly $\rho(\rho-1)/2+1$ *d*-windows (with $\rho$ being the total number of combined states involved). This means that magnetization and susceptibility have $\rho(\rho-1)/2$ discontinuities inside that window. For example, if *B* is equal to $(n_A\Phi_o)/5$, then $\rho=3$ and magnetization will have discontinuities at 24 *nm*, 34 *nm* and 56 *nm* (for a full analytical discussion see Appendix).

**Relation to transport properties ï Hall conductivity**

The usual criterion for the existence of IQHE is that the Fermi energy must lie in a bulk energy gap (actually the well-known mobility gap created by disorder) and then chiral currents flow along the edges. This picture is valid for a planar two dimensional system, where no freedom in *z*-axis is present. One then wonders how this picture is modified in the case of our interface. How are the diamagnetic chiral currents generalized in the presence of a strongly quantized *z*-direction? A first thought is that these one dimensional current channels now become surface currents that move in opposite directions in the two opposite edge-surfaces of the interface. This may not be a bad picture, since from the energy spectrum (4.1) one notes that there is in fact no dispersion at all; the net velocity in *z* axis is zero. Including a confining potential in *x*-direction, surface diamagnetic currents are then created in two edge surfaces, while the net velocity in the bulk of the system vanishes. The net current is then zero. For a nonzero surface current one needs to shift the electrochemical potentials of the edges by a moderate amount, and this is directly achieved by applying a small in-plane electric field, which controls the number of edge surface states.

Now let us consider a clean sample and see when the Fermi energy lies in a gap: This condition is only met whenever the magnetic field is exactly of the form $\frac{1}{2\rho} n_A \Phi_o$, in the sense that there are no infinitesimally neighboring empty states for the electron to be scattered in. We expect that at these special values of *B* where all LLs are fully occupied the Hall conductivity will be quantized as in the usual 2D case in units of $e^2/h$ for every value of thickness *d*. This actually comes out of a semiclassical treatment of the problem, where the Hall conductivity is of the form:

$$\sigma_{\rm H} = \frac{n_V ec}{B} = \frac{n_A ec}{dB}, \tag{4.26}$$

where $n_V = n_A / d$ is the average volume density. Substituting in (4.26) the special values of *B* we have:

$$\sigma_{\rm H} = \frac{2\rho e^2}{hd} \tag{4.27}$$

a result that may apply to multilayered QHE systems [7]. An alternative way to obtain the above result is to use the analog of the mathematical relation (2.10) introduced in Section 2, that relates the discontinuities of orbital magnetization with the corresponding discontinuities of chemical potential at the critical values of *B*, namely

$$\sigma_{\rm H} = \frac{ec\Delta {\rm M}}{d\Delta\mu S} \tag{4.28}$$

Let us use (4.28) in an example in order to check the validity of (4.27) when $B = (1/2\rho) n_A \Phi_o$, namely, when $\mathcal{J}$ sets of combined states are fully occupied. Then the criterion of quantization of conductivity is fulfilled, because Fermi energy is in a gap. If $\mathcal{J}$=1 (only 1 combined state occupied) then from (4.8) we have for the energy:

$$\frac{E_1}{N} = E_f \left[ \frac{B}{n_A \Phi_o} + \left( \frac{\pi}{2n_A d^2} \right) \right], \text{ valid for every } d$$

and from (4.16):

$$\frac{E_2}{N} = E_f \left[ \left( \frac{B}{n_A \Phi_o} \right) - 6 \left( \frac{B}{n_A \Phi_o} \right) \left( \frac{\pi}{2n_A d^2} \right) + \left( \frac{4\pi}{2n_A d^2} \right) \right], \text{ valid for } d \geq \sqrt{\frac{3\pi\Phi_o}{4B}} = \sqrt{\frac{3\pi}{2n_A}}$$

The corresponding magnetizations are for this case:

$$\frac{M_1}{N} = -\mu_B$$

$$\frac{M_2}{N} = \mu_B \left[ -1 + 6 \left( \frac{\pi}{2n_A d^2} \right) \right],$$

with $\Delta M = M_2 - M_1 = 6N\mu_B \left( \frac{\pi}{2n_A d^2} \right)$.

The chemical potentials are (compare highest single-particle energies in Fig.ôs 1 and 3):

$$\mu_1 = \frac{\hbar\omega_c}{2} + \frac{\hbar^2\pi^2}{2md^2}$$

$$\mu_2 = \frac{\hbar\omega_c}{2} + \frac{4\hbar^2\pi^2}{2md^2}$$

with $\Delta\mu = \mu_2 - \mu_1 = \frac{3\hbar^2\pi^2}{2md^2} = 3\mu_B\Phi_o \frac{\pi}{2d^2}$

By then applying (4.28) we get:

$$\sigma_{\mathrm{H}} = \frac{ec\Delta\mathrm{M}}{d\Delta\mu S} = \frac{ec6N\mu_B\left(\frac{\pi}{2n_Ad^2}\right)}{d3\mu_B\Phi_o\frac{\pi}{2d^2}S} = 2\frac{e^2}{dh} \tag{4.29}$$

in full accordance with (4.27) for $\mathcal{J}$=1.

Another example is when $\mathcal{J}$=2, and again for complete filling $B$ must be:

$$B = \frac{1}{4}n_A\Phi_o\,,$$

Let us also consider the case when $d$ lies in the following window:

$$d \le \sqrt{\frac{3\pi\Phi_o}{4B}} = \sqrt{\frac{3\pi}{n_A}}$$

The discontinuity of $M$ is connected with the two neighboring energies, one shown in (4.15) and the other could i.e. be the second expression in Table 3 (the choice is of course made with respect to thickness, so that the two ranges match):

$$\frac{E_1}{N} = E_f\left[-4\left(\frac{B}{n_A\Phi_o}\right)^2 + 3\left(\frac{B}{n_A\Phi_o}\right) + \left(\frac{\pi}{2n_Ad^2}\right)\right]$$

$$\frac{E_2}{N} = E_f\left[4\left(\frac{B}{n_A\Phi_o}\right)^2 + \left(\frac{B}{n_A\Phi_o}\right) - 12\left(\frac{B}{n_A\Phi_o}\right)\left(\frac{\pi}{2n_Ad^2}\right) + \left(\frac{4\pi}{2n_Ad^2}\right)\right]$$

The corresponding magnetizations are respectively:

$$\frac{M_1}{N} = \mu_B\left[8\left(\frac{B}{n_A\Phi_o}\right) - 3\right]$$

$$\frac{M_2}{N} = \mu_B\left[-8\left(\frac{B}{n_A\Phi_o}\right) - 1 + 12\left(\frac{\pi}{2n_Ad^2}\right)\right]$$

So,

$$\Delta\mathrm{M} = M_2 - M_1 = N\mu_B\left[-8\left(\frac{B}{n_A\Phi_o}\right) - 1 + 12\left(\frac{\pi}{2n_Ad^2}\right) - 8\left(\frac{B}{n_A\Phi_o}\right) + 3\right] = N\mu_B\left[-16\left(\frac{B}{n_A\Phi_o}\right) + 12\left(\frac{\pi}{2n_Ad^2}\right) + 2\right]$$

For $B = \frac{1}{4}n_A\Phi_o$,

$$\Delta\mathrm{M} = N\mu_B\left[12\left(\frac{\pi}{2n_Ad^2}\right) - 2\right]$$

Now, we have for the chemical potentials (compare Fig.ôs 2 and 6):

$$\mu_1 = 3\frac{\hbar\omega_c}{2} + \frac{\hbar^2\pi^2}{2md^2}$$

$$\mu_2 = \frac{\hbar\omega_c}{2} + \frac{4\hbar^2\pi^2}{2md^2}$$

$$\Rightarrow \Delta\mu = \mu_2 - \mu_1 = \frac{\hbar\omega_c}{2} + \frac{4\hbar^2\pi^2}{2md^2} - 3\frac{\hbar\omega_c}{2} - \frac{\hbar^2\pi^2}{2md^2} = -\hbar\omega_c + \frac{3\hbar^2\pi^2}{2md^2},$$

$$\Delta\mu = -\frac{2\hbar eB}{2mc} + \frac{c}{e}\frac{3\hbar^2\pi^2 e}{2md^2c} = \mu_B\left(-2B + \frac{3\hbar\pi^2 c}{d^2 e}\right) \xrightarrow{B=\frac{1}{4}n_A\Phi_o} = \frac{\mu_B\Phi_o n_A}{4}\left(-2 + \frac{12\pi}{2n_A d^2}\right)$$

Substituting then in (4.28) we have:

$$\sigma_{\rm H} = \frac{ec\Delta \mathrm{M}}{d\Delta\mu S} = \frac{ec}{dS}\frac{N\mu_B\left[12\left(\frac{\pi}{2n_A d^2}\right) - 2\right]}{\frac{\mu_B\Phi_o n_A}{4}\left(-2 + \frac{12\pi}{2n_A d^2}\right)} = 4\frac{ecN}{dS\Phi_o n_A} = 4\frac{e^2}{dh}, \tag{4.30}$$

in full accordance with (4.27) with $\jmath$ =2. If $B = (1/6)n_A\Phi_o$ we would then find $\sigma_{\rm H} = 6e^2/dh$, *independently of the choice of thickness*, and so on. In conclusion, we see that if $B$ has the exact value needed for the combined sets of degenerate states to be fully occupied, the transverse conductivity is quantized, with universal values that are essentially the same as those of a 2D system (like in a multilayered QHE system [7]). The issue of the new transitions reported here for partial LL filling requires, as already noted, a closer investigation (as in this case we have the standard issue of the enormous degeneracy of the many-body states involved, and to draw conclusions on transport one has to include electron-electron interactions (see however Subsection 4.2 for a rather unconventional picture)).

### 4.1 Inclusion of Zeeman term

When the gyromagnetic ratio $g^*$ is nonvanishing, the previous results will be modified, and here we give a quick discussion of the general manner in which the presence of $g^*$ is expected to affect them. By including the Zeeman term in our model we have the following single particle energy spectrum:

$$\varepsilon_{n,k_z} = (n + \frac{1}{2})\hbar\omega_c^* + \frac{\hbar^2 k_z^2}{2m^*} \pm \frac{g^*}{2}\mu_B B, \tag{4.1.1}$$

where $g^*$ is for simplicity considered to be positive, $m^*$ is electronôs effective mass, $\mu_B = e\hbar/2mc$ is the Bohr magneton (with $m$ being electronôs vacuum mass) and $\omega_c^* = eB/m^*c$ is the effective cyclotron frequency. The wavenumber $k_z$ is still quantized in the following manner: $k_z = \pi n_z/d$, with $n_z = 1,2,3...$ We may write (4.1.1) in a more convenient form, namely

$$\varepsilon_{n,k_z} = (n + \frac{1}{2} \pm \frac{g^*}{4}\frac{m^*}{m})\hbar\omega_c^* + \frac{\hbar^2 k_z^2}{2m^*} \tag{4.1.2}$$

(that directly shows the well-known fact that, for the special case (of noninteracting electrons in vacuum) with $m^* = m$ and $g^* = 2$, the Zeeman splitting is exactly equal to the LL splitting).

For the purposes of our calculation, we will confine $g^*$ in the range:

$$0 \le g^* \le 2,$$

and will also assume m*<m. The effect of Zeeman coupling is to split all Landau levels in two sublevels, where electrons are being placed according to their spin orientation, namely spin up and spin down (see fig. 4.1.1).

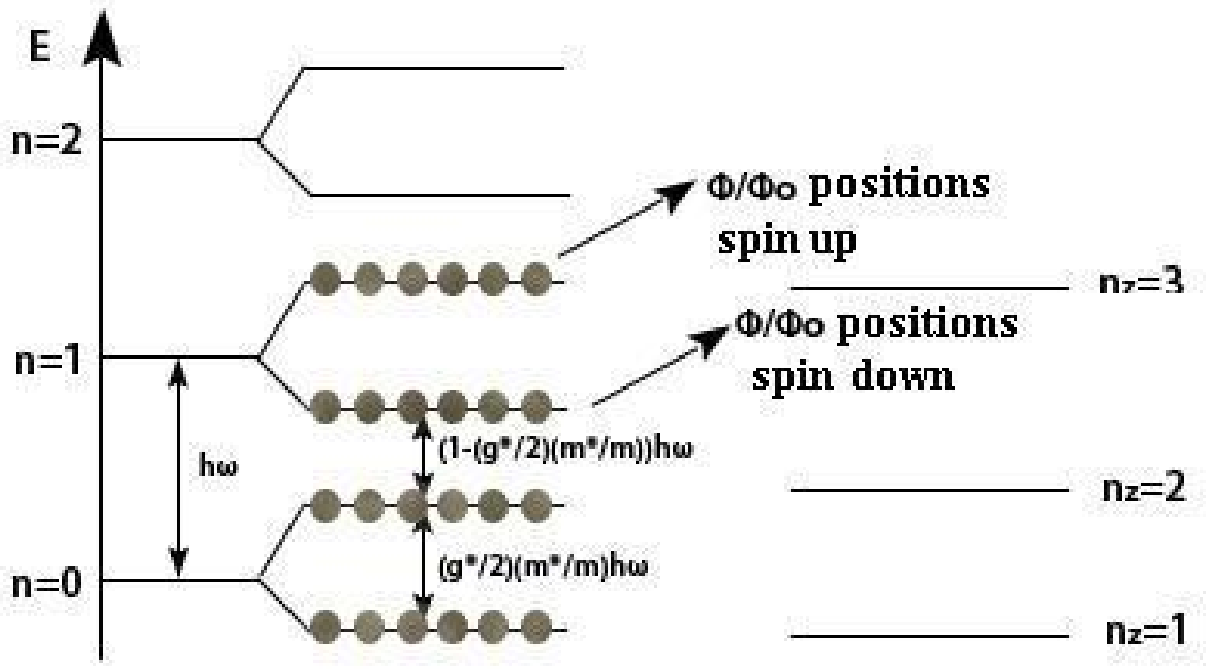

**Fig 4.1.1**. Energetic configuration in the presence of Zeeman splitting (we have set $\omega = \omega_c$ )

We now have a different background structure for the possible energetic competitions: First, the earlier degeneracy of each LL is partially lifted**;** each Zeeman sublevel contains ū /ū ɵ independent states (that can only accommodate electrons of a single spin). Second, gaps between different sublevels appear which depend on the gyromagnetic ratio and on the effective mass (while the original inter-LL gaps are still present and equal to $\hbar\omega_c$). Third, if we happen to have $m^* = m$ and $g^* = 2$ then Zeeman splitting coincides with LL spacing and nearby sublevels fall on top of each other, doubling therefore their degeneracy to 2ū /ū ɵ as before (except the lowest zero-energy state that remains with a degeneracy ū /ū ɵ). In what follows we will denote each set of degenerate quantum states with:

$$\{(n, X), n_z\}, \qquad \text{where ј} = \uparrow \text{ or } \downarrow$$

(with a fourth quantum number *l* (that counts each sublevel degeneracy) omitted ï since this will naturally be accounted for in the occupation procedure as earlier).

Let us now examine the first *B*-window that naturally comes up for this problem, namely:

$$B \geq n_A \Phi_o \,, \tag{4.1.3}$$

where $n_A = N / S$ is always the constant areal density. It is now clear that for such a *B*, only the lowest sublevel is occupied (combined with $n_z = 1$), namely $\{n = 0 \downarrow, n_z = 1\}$, with total energy given by:

$$E = N\varepsilon\{n = 0 \downarrow, n_z = 1\} \,, \tag{4.1.4}$$

or in units of the effective Fermi energy:

$$\frac{E}{N} = E_f{}^* \left[ \frac{B}{n_A \Phi_o} \left( 1 - \frac{g^*}{2} \frac{m^*}{m} \right) + \frac{\pi}{2 n_A d^2} \right] \tag{4.1.5}$$

The above result describes a completely polarized state, where all spins are parallel in a direction opposite to *B*. For $g^* = 0$ it coincides with (4.8) as it should (while for $g^*$=2 and $m^*$=*m* there remains only the *z*-term, due to the zero-energy of planar motion in this case).

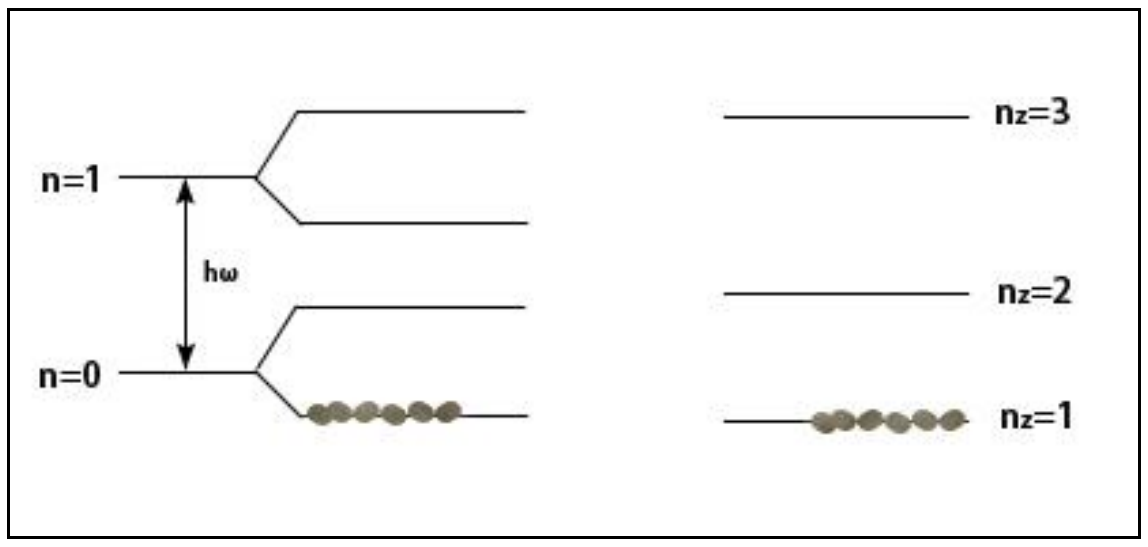

**Fig 4.1.2:** When *B* lies in the first window, all electrons fall into a completely polarized state

This first *B*-window leads therefore to a total energy linear in *B*. Let us now lower *B* so that we lie in the second *B*-window which is:

$$\frac{1}{2} n_A \Phi_o \leq B \leq n_A \Phi_o \qquad (4.1.6)$$

Here, due to Pauli principle, we are forced to accommodate the extra *N*-ū /ū ɞ electrons into another sublevel. This requires care since we have to take into account the finite thickness *d*, which will decide for us the proper occupation scenario. We have three options: we can just change LL index and move to *n*=2, or we can change Zeeman sublevel and so reverse *N*-ū /ū ɞ electronsô spin, or we can change only QW level and restart with the lowest possible values of all the remaining quantum numbers. Let us examine which option is the most favorite one.

At first, we should immediately note that changing LL index, will cost more energy than changing sublevel (reversing spins). So we are really left with two options. The choice between them is thickness-dependent. It can be made by examining when the two remaining options (for an extra single electron) become equal in energy, which will immediately determine the transition between the two scenarios, namely:

$$\varepsilon \underbrace{\left\{ n=0\uparrow, n_z=1 \right\}}_{\text{Change spin sublevel}} = \varepsilon \underbrace{\left\{ n=0\downarrow, n_z=2 \right\}}_{\text{Change QW sublevel}}$$

$$\left(\frac{1}{2}+\frac{g^*}{4}\frac{m^*}{m}\right)\hbar\omega_c^* + \frac{\hbar^2\pi^2}{2m^*d^2} = \left(\frac{1}{2}-\frac{g^*}{4}\frac{m^*}{m}\right)\hbar\omega_c^* + \frac{4\hbar^2\pi^2}{2m^*d^2} \Rightarrow d_{\text{crit}} = \sqrt{\frac{3\pi\Phi_o}{2g^*B}\frac{m}{m^*}} \qquad (4.1.7)$$

From this we can infer the following: When the thickness *d* is lower than (4.1.7), (namely, when QW gaps are large enough) it is favourable to place the extra electron in the next available spin-up sublevel that lies in the *n*=0 LL, and keep it in the QW level $n_z$=1**;** and when the thickness is larger than (4.1.7) it must go to $n_z$=2 by keeping its spin down in the same sublevel without violating Pauli principle (see fig. 4.1.3 and 4.1.4). If we substitute in (4.1.7) the largest value of *B* (namely $B = n_A\Phi_o$), then we find a new criterion for 2-dimensionality (for $d \leq d_{\min}$) , which depends strongly on $g^*$:

$$d_{\min} = \sqrt{\frac{3\pi}{2n_A}\left(\frac{m}{m^*g^*}\right)} \qquad (4.1.8)$$

This is of course different from (3.6) (note that if we set $g^* = 0$ it tends to infinity) as it describes spinï related Physics (we are always in the lowest LL, unlike the situation of the previous subsection). Let us then determine the new energies:

$$E = \frac{\Phi}{\Phi_o}\varepsilon\left\{n=0\downarrow, n_z=1\right\} + \left(N-\frac{\Phi}{\Phi_o}\right)\varepsilon\left\{n=0\uparrow, n_z=1\right\}$$

$$\Rightarrow \qquad \frac{E}{N} = E_f^*\left[-g^*\frac{m^*}{m}\left(\frac{B}{n_A\Phi_o}\right)^2 + \frac{B}{n_A\Phi_o}\left(1+\frac{g^*}{2}\frac{m^*}{m}\right) + \frac{\pi}{2n_Ad^2}\right], \quad d \leq d_{\text{crit}} \qquad (4.1.9)$$

$$E=\frac{\Phi}{\Phi_o}\varepsilon\left\{n=0\downarrow,n_z=1\right\}+\left(N-\frac{\Phi}{\Phi_o}\right)\varepsilon\left\{n=0\downarrow,n_z=2\right\}$$

$$\Rightarrow\qquad \frac{E}{N}=E_f{}^*\left[\frac{B}{n_A\Phi_o}\left(1-\frac{g^*}{2}\frac{m^*}{m}\right)-3\left(\frac{B}{n_A\Phi_o}\right)\left(\frac{\pi}{2n_Ad^2}\right)+\frac{4\pi}{2n_Ad^2}\right],\quad d\geq d_{\text{crit}} \qquad (4.1.10)$$

Note that if we set $g^*=0$ and $m^*=m$ we get:

$$\frac{E}{N}=E_f\left[\frac{B}{n_A\Phi_o}+\frac{\pi}{2n_Ad^2}\right]\ \text{valid for any } d, \qquad (4.1.11)$$

which coincides with (4.8) as it should.

Fig. 4.1.3 $d\leq d_{\text{crit}}$

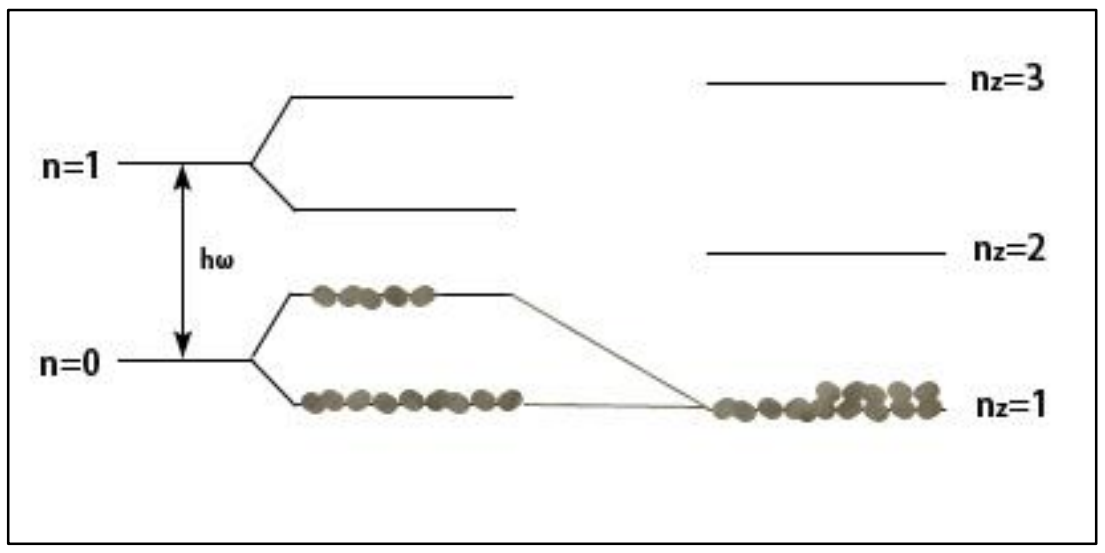

Fig 4.1.4 $d\geq d_{\text{crit}}$

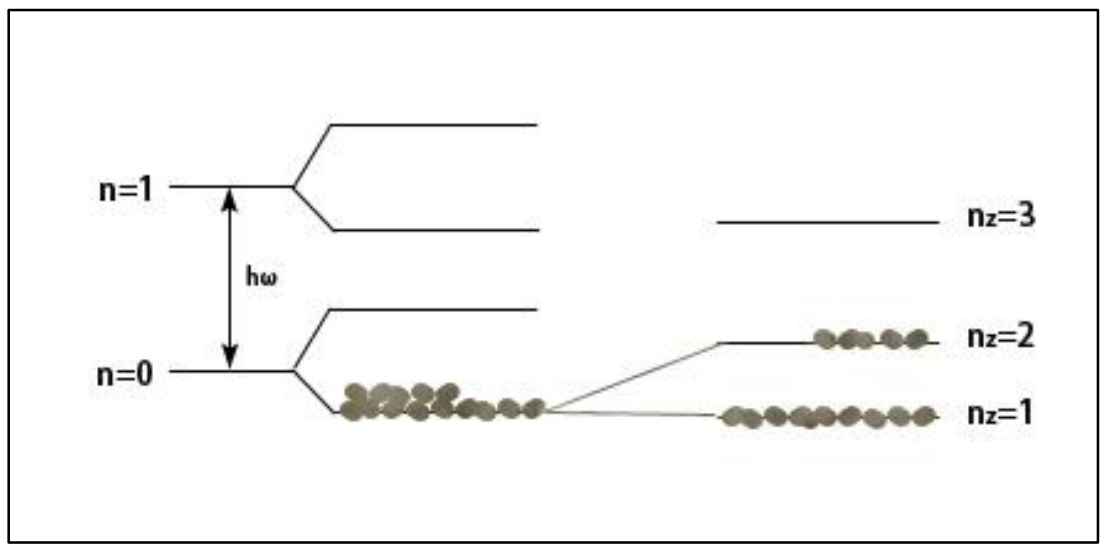

Following a similar line of reasoning for all *B*-windows, we can find all energetically optimal configurations (that are now richer in transitions compared to the ones in the previous subsection) but we will not show any further examples. In the following, we will first present one dimensional figures based on the above example, as well as the corresponding 2D ones (with combined variation of variables). The reader may compare them with those of previous subsection to see the differences.

In the figures we always use the following values of $g^*$, $m^*$, and $n_A$:

$$g^*=0.8,\ m^*=m,\ \text{and}\quad n_A=10^{16}\,el/m^2$$

(4.1.12)

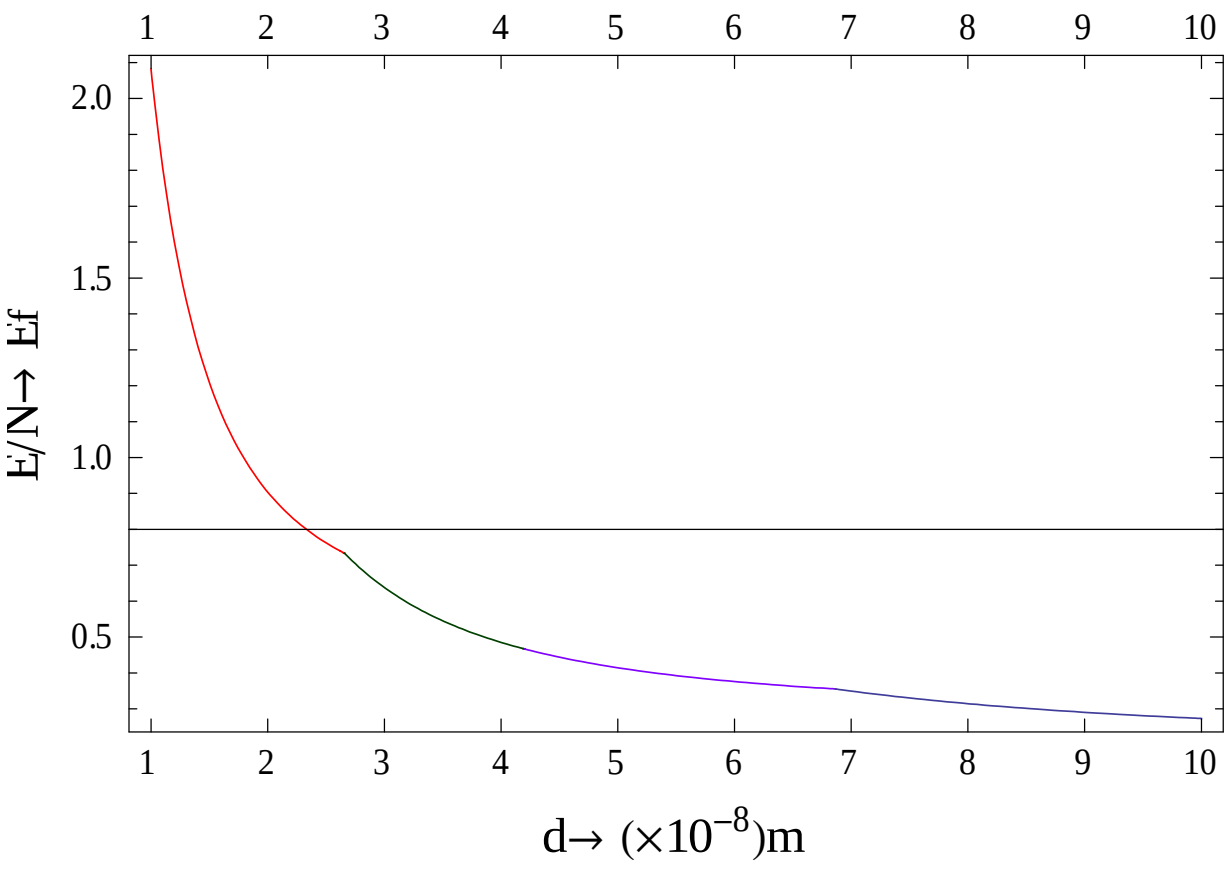

**[A]**

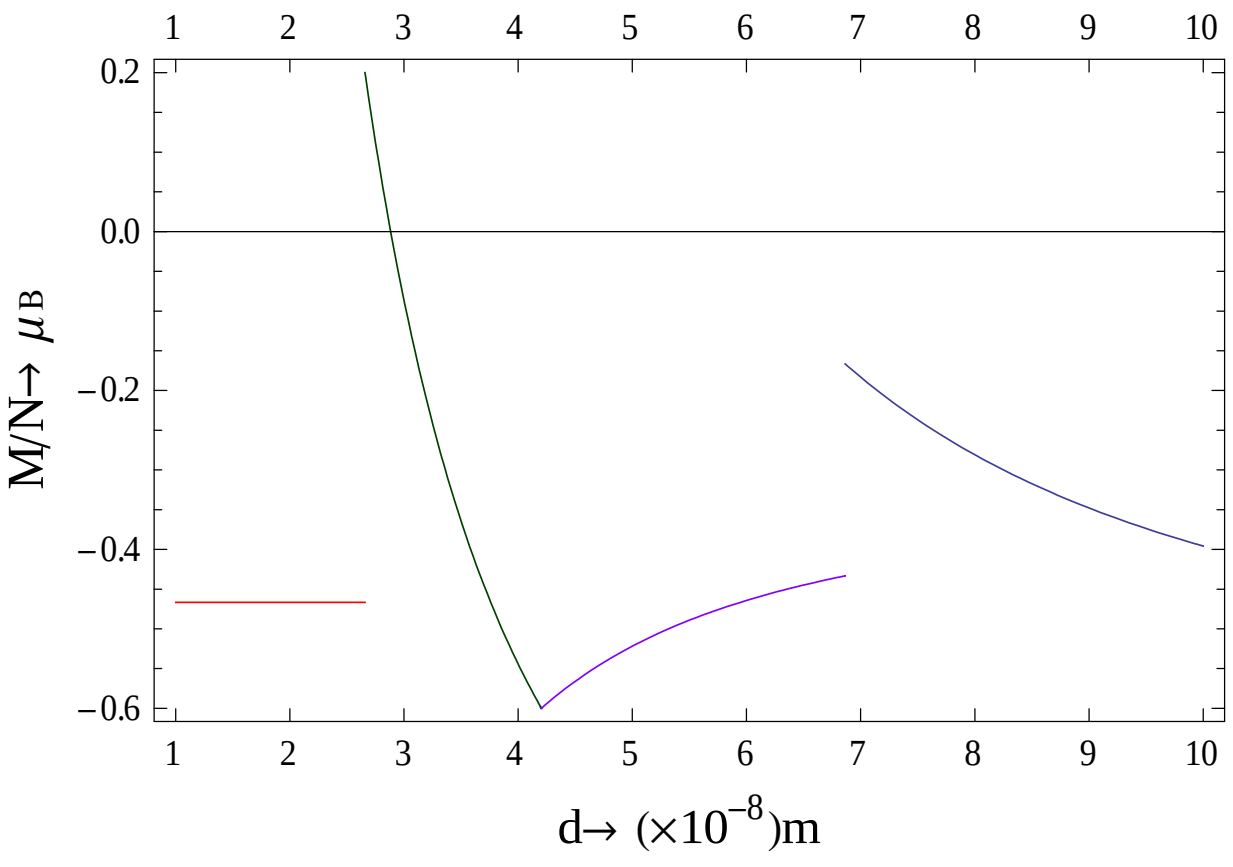

**[B]**

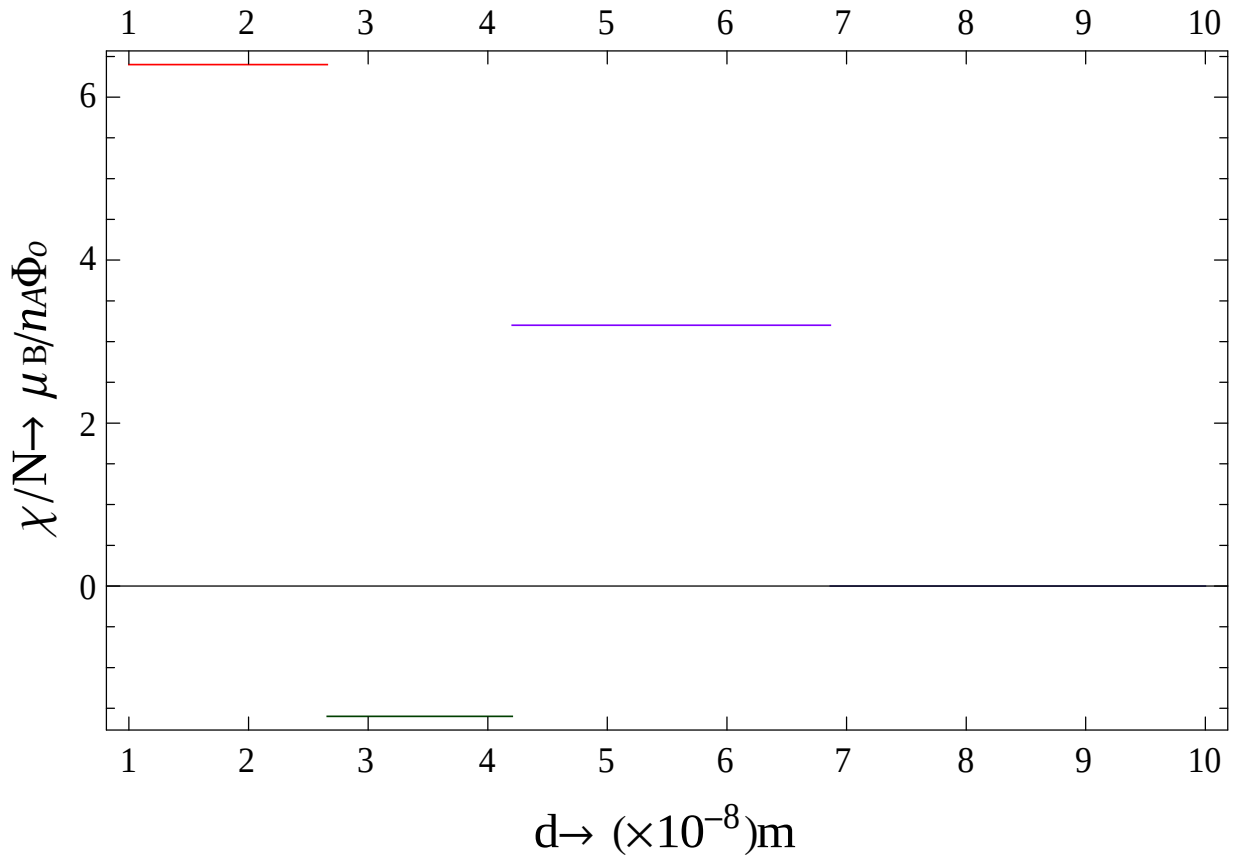

**[C]**

**Fig 4.1.5:** Graphs [A] Energy, [B] Magnetization, [C] Susceptibility as functions of thickness $d$ for ***b***=$1/3\ n_A\Phi_o$ .

These results appear to have a resemblance with the corresponding ones that we saw earlier in the main part of this Section. The main difference is that the magnetization discontinuities now occur more frequently.

We should note here that inclusion of such Zeeman splitting later in the full 3D problem will give rise in certain cases to pronounced local minima in energy (see end of Section 5).

Below we also provide the corresponding 2D graph for the magnetization:

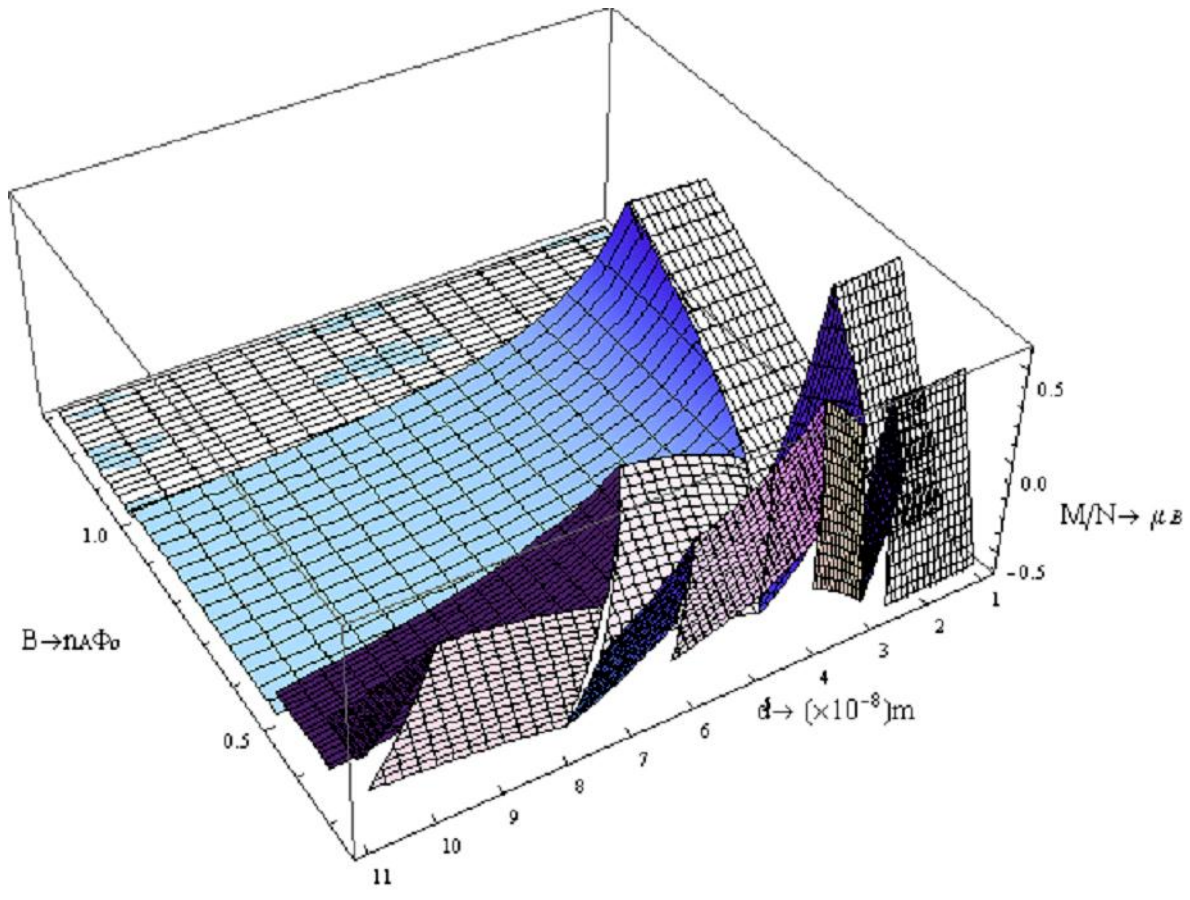

Fig 4.1.6 Magnetization as function of both $B$ and $d$.

[The corresponding ñtransport-relatedò discussion may have a relation to interesting ñspin-Physicsò at the edges (because of the above Zeeman-induced spin-asymmetry)**;** this is especially so if the system were folded into an Aharonov-Bohm cylinder (with nonzero thickness), because of Berryôs phase effects on the opposite spins that effectively feel an inhomogeneous magnetic field (due to the nonvanishing curvature). However, this is an interesting issue that deserves a separate study.]

**4.2 Inclusion of electronï electron interactions: Composite Fermions**

Although we found new transitions that correspond to partial LL filling, and, as already noted, we would be tempted to speculate that these might have something to do with fractional fillings and the FQHE if interactions were included, we have for consistency made the choice to restrict ourselves to noninteracting particles in the largest part of this article. However, in this Section we will make an exception, and consider for awhile interacting electrons, since the same line of reasoning and the same general approach that was followed so far can also be followed for the so-called Λ-levels (that are the Landau Levels corresponding to a system of noninteracting Composite Fermions (CFs), the IQHE of CFs (i.e. with completely filled Λ-levels) corresponding, as is well-known, to the FQHE of the original strongly interacting electron system).
Indeed, dealing with interactions between electrons inside a magnetic field is an extremely challenging problem already in 2D [24], and it is even more so in the presence of a finite thickness, like in the systems of our interest.

The picture of Composite Fermions (CFs) in 2D was devised and developed by Jain [25], and in this each electron is, loosely speaking, attached to 2*p* flux quanta (*p*=integer) in order to create a CF (more rigorously there is a Chern-Simons transformation [24] that maps through a many-body Aharonov-Bohm transformation the strongly interacting system of electrons to almost noninteracting CFs). The CF method has been very successful in describing electron states in two dimensions which are in FQHE state with very high accuracy. In our case, we have an extra *z*-direction, and we are allowed in principle to use Jainôs method, as in our conventional system the planar and *z*-motions are decoupled and the (Chern-Simons) transformation performed to give the CFs only affects the 2D motion. (Note, however, that this may not be a good model for topologically nontrivial systems).

We remind here the reader that in the approximation of noninteracting CFs the energy spectrum of each CF is given by

$$\varepsilon_{n,k_z} = \hbar\left|\omega_c^*\right|(n+1/2) + \frac{\hbar^2 k_z^2}{2m^*}$$

where $\omega_c^* = \frac{eB}{m^* c}$ (with m* being the effective mass, which also depends on $B$ ï see below), with $B^* = B - 2p\Phi_o n_A$ being the well-known effective magnetic field felt by the CFs (*p* being the earlier mentioned integer), and with the last term being the thickness-related contribution with again $k_z = \frac{\pi n_z}{d}$, $n_z = 1, 2, 3....$ (for rigid boundary conditions as earlier).
Note that the same quantization condition is valid for $k_z$, since it is not affected by the CF (or Chern-Simons) transformation. Let us choose as an example the integer *p* to be unity (meaning that 2 flux quanta are attached to each electron). Physics is now controlled by the effective magnetic field $B^*$, which determines the orbital 2D motion of CFôs on the *xy* plane. The degeneracy of Λ-levels now depends only on $B^*$, and the noninteracting CFs will have to be properly accommodated in the available Λ-levels.

Following the same method as earlier, we start from strong magnetic fields such as $\frac{1}{2} n_A \Phi_o \le B^* \le \infty$, so that only the lowest Λ-level is occupied. Simultaneously (by reversing the above with respect to *B* and with *p*=1), the real magnetic field *B* lies in the range

$$\frac{5}{2} n_A \Phi_o \le B \le \infty \tag{4.2.1}$$

The total energy for this window is trivial:

$$E = N\varepsilon\{n = 0, n_z = 1\} \tag{4.2.2}$$

$$\Rightarrow \frac{E}{N} = \frac{\hbar e B^*}{2m^* c} + \frac{\hbar^2 \pi^2}{2m^* d^2} \qquad (4.2.3)$$

(with *N* the number of CFs which is obviously the same as the number of electrons). We can choose to write the energy in units of 2D Fermi energy defined with the bare electronic mass *m*, in which case we have

$$\frac{E}{N} = E_f \left[ \left( \frac{B^*}{n_A \Phi_o} \right) \frac{m}{m^*} + \left( \frac{\pi}{2 n_A d^2} \right) \frac{m}{m^*} \right], \qquad (4.2.4)$$

an expression valid, in the range (4.2.1), for every thickness *d*. The next $B^*$-window is naturally the following:

$$\frac{1}{4} n_A \Phi_o \le B^* \le \frac{1}{2} n_A \Phi_o \qquad (4.2.5)$$

Here we have two possible types of states to place the extra (namely ʔ -2ū */ū ⊙) CFs into: $\{n = 0, n_z = 2\}$ and $\{n = 1, n_z = 1\}$. The system will choose the minimum energy state in a way that depends strongly on the critical *d*-value determined by:

$$\varepsilon\{n = 0, n_z = 2\} = \varepsilon\{n = 1, n_z = 1\}$$

$$\frac{3\hbar\omega_c^*}{2} + \frac{\hbar^2\pi^2}{2m^* d^2} = \frac{\hbar\omega_c^*}{2} + \frac{4\hbar^2\pi^2}{2m^* d^2} \Rightarrow d_{\text{crit}} = \sqrt{\frac{3\pi\Phi_o}{4B^*}} \qquad (4.2.6)$$

(which is actually equal to (4.4) with *B* replaced by *B**; this occurs more generally also in the energy results that follow, and it is the universality mentioned earlier (a type of law of corresponding states)). If thickness *d* is smaller than (4.2.6), the extra CFs occupy $\{n = 1, n_z = 1\}$ states, while if *d* is larger than (4.2.6) then $\{n = 0, n_z = 2\}$ states are preferred to be occupied. The total energy is then, for each case

$$E = 2\frac{\Phi^*}{\Phi_o}\varepsilon\{n = 0, n_z = 1\} + \left( N - 2\frac{\Phi^*}{\Phi_o} \right) \varepsilon\{n = 1, n_z = 1\}$$

$$\Rightarrow \frac{E}{N} = E_f \left[ -4\left( \frac{B^*}{n_A \Phi_o} \right)^2 \frac{m}{m^*} + 3\left( \frac{B^*}{n_A \Phi_o} \right) \frac{m}{m^*} + \left( \frac{\pi}{2 n_A d^2} \right) \frac{m}{m^*} \right], \; d \le d_{\text{crit}} \qquad (4.2.7)$$

$$E = 2\frac{\Phi^*}{\Phi_o}\varepsilon\{n = 0, n_z = 1\} + \left( N - 2\frac{\Phi^*}{\Phi_o} \right) \varepsilon\{n = 0, n_z = 2\}$$

$$\Rightarrow \frac{E}{N} = E_f \left[ \left( \frac{B^*}{n_A \Phi_o} \right) \frac{m}{m^*} - 6\left( \frac{B^*}{n_A \Phi_o} \right) \left( \frac{\pi}{2 n_A d^2} \right) \frac{m}{m^*} + \left( \frac{4\pi}{2 n_A d^2} \right) \frac{m}{m^*} \right], \; d \ge d_{\text{crit}} \qquad (4.2.8)$$

For the figures shown below we use the following input:

$$\frac{m}{m^*} = \frac{1.95}{\sqrt{\frac{B^*}{n_A \Phi_o} + 2}} \quad , \quad n_A = 10^{16}\, {}^{el}\!/_{m^2}$$

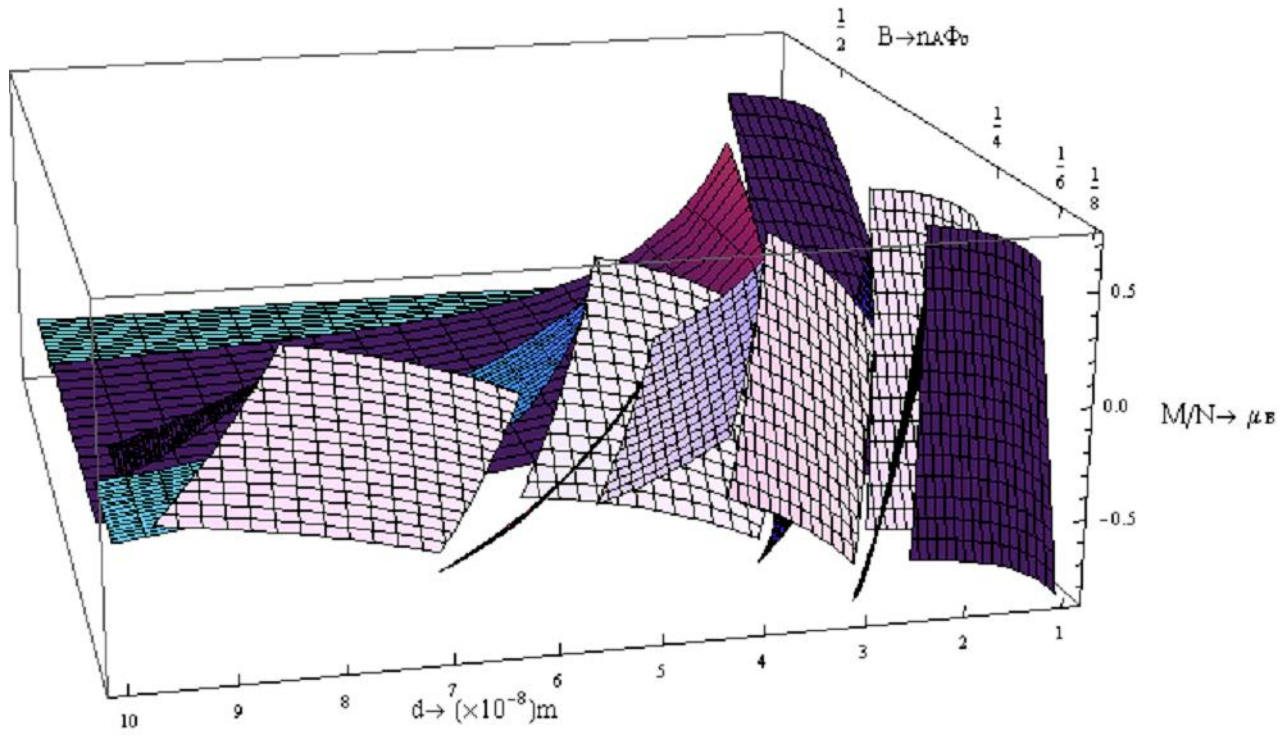

**Fig. 4.2.1**: Magnetization as function of *B* and *d*

Note that global magnetization as function of both *B* and *d* is considerably different from the one given in Fig. 52. This is because interactions have a further significant role on thermodynamic properties through the *B*-dependent mass given above. Below we also give a comparison between 1D graphs of magnetization for CFs (left) and for noninteracting electrons (right), for *p*=1 (and for corresponding states).

We can see some qualitative differences between the two systems which, however, deserve a closer investigation (especially with respect to the noninteracting CFs approximation). Similarly, the issue of ñinternal transitionsò (at partial ş -level filling) for CFs is well beyond the scope of the present article).

**Interacting system** **Noninteracting system**

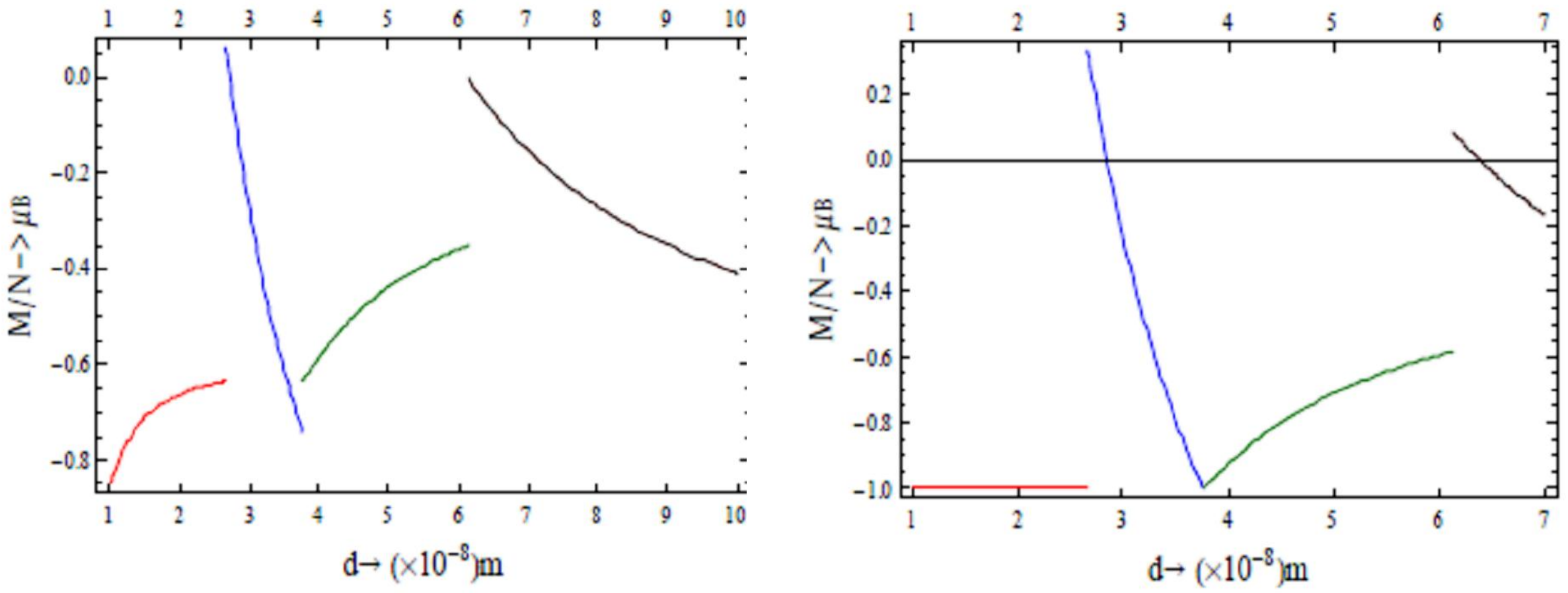

**A** **B**

**Figure 4.2.2: A):** Magnetization per Composite Fermion as a function of width *d* for $B^* = (1/6)nA\Phi_o$ or $B = (13/6)nA\Phi_o$. **B):** Magnetization per electron as a function of width *d* for $B = (1/6)nA\Phi_o$.

## 5. Electron gas inside a magnetic field in full 3D space

Let us now for comparison deal with the case of noninteracting electrons in full 3D space (with periodic boundary conditions imposed along the direction of the field). Although one might expect that things will now be getting more complicated, especially in the presence of a homogeneous magnetic field, this problem is actually more tractable than the earlier one of the finite-thickness interface (and amenable to closed form solutions for the thermodynamic functions). The basic origin of the simplification is the fact that $k_z$ is now a quasicontinuous variable. In fact, the quantum mechanical problem of 3D electron gas at zero temperature in a magnetic field has been studied many years ago [27], with direct use of the grandcanonical potential $\boldsymbol{q}$, with generalized Riemann functions (or Hurwitz zeta functions) appearing in the results at low temperatures (see also [28]). Our aim in this Section is to determine **exactly** the energetics of the ground state (of noninteracting electrons) by using a very different, simpler and more physical method of energy interplays, when the electron system occupies combined Landau Levels with (now quasicontinuous) *z*-axis levels, having always in mind the minimum total energy requirement at T=0. It will turn out (not surprisingly) that all thermodynamic properties (e.g. Energy, Magnetization and Susceptibility) will be determined analytically, in terms of imaginary parts of Hurwitz zeta functions. However, it will also turn out from these exact solutions that we can determine the exact quantal manner in which the semiclassical dHvA periodicity is violated, which could be relevant for certain 3D solid state systems (but of very low density as we shall see). (In this respect, our method is superior compared to earlier semiclassical approaches that cannot detect these violations).

We start by writing again the single particle energies that emerge from solution of the Schrodinger equation in space with cubic geometry (with length *L*), by now imposing periodic boundary conditions in the *z*-direction (the direction of the applied magnetic field). The single-particle energy spectrum consists then of the Landau Levels that describe the motion in *xy*-plane plus a (nonrelativistic) kinetic term (free wave) in *z*-direction, namely

$$\varepsilon_n = \hbar\omega_c\left(n+\frac{1}{2}\right)+\frac{\hbar^2 k_z{}^2}{2m}, \quad k_z = 2\pi\frac{n_z}{L}, \quad n_z = 0,\pm1,\pm2..., \quad n \text{ a nonnegative integer} \tag{5.1}$$

with *L* assumed macroscopic. We now have quasicontinuous $k_z$ (since $L \rightarrow \infty$) and strong quantization in the *xy*-plane. Let us first study the Pauli principle-respecting occupational procedure: electrons first occupy the lowest LL (for $n_z$=0) and then start building a 1D Fermi line segment along $k_z$ in *k*-space (with $k_z$ now taking also negative values). But this cannot go on for ever (even at T=0). There comes a point when the length of the segment (essentially the Fermi wavenumber $k_f$ in *z*-direction) is so large that it is no longer energetically favorable to continue this procedure of occupations; it may be preferable for the extra electron to be excited to the next LL and start building a new Fermi segment from the beginning (notice, without violating the Pauli principle). We can therefore have a Fermi segment corresponding to any occupied LL (but how many such segments we have will depend on the values of *B* and the electronic volume density $n_V$). Since the above method is different from the usual semiclassical treatment, let us first work out the simplest examples.

Let us first consider the case of extremely strong *B* (in a range to be determined below) so that all electrons are frozen in the lowest LL (*n*=0) and they form only a single Fermi segment (extending in *k*-space from ï $kf_1$ to $+kf_1$). The maximum $kf_1$ will occur when, energetically speaking (in the above spirit) another $k_{f2}$ (associated with the *n*=1 LL) is just about to form, and this will occur when

$$\hbar\omega_c = \frac{\hbar^2 k_{f1}^2}{2m} \tag{5.2}$$

and then with the standard substitution of a sum (over the quasicontinuous $k_z$) with an integral (in the limit of infinite *L*) we can determine $kf_1$ as follows:

$$N = \frac{2\Phi}{\Phi_o}\sum_{\vec{k}:occupied} 1 \rightarrow \frac{2\Phi}{\Phi_o}\frac{L}{2\pi}\int_{-kf_1}^{kf_1} dk\,, \tag{5.3}$$

from which it turns out that $kf_1 = \pi^2 l_B^2 n_V$ (where $l_B = \sqrt{\hbar c / eB}$ is the magnetic length and $n_V = N/V$ is the volume density, always for spinful electrons ï note that in astrophysical applications there is usually an extra factor of 2 involved [26]).

By then using (5.2), (5.3) and $\omega_c = \dfrac{eB}{mc}$, we obtain:

$$n_{crit1} = \left(\frac{16}{\pi}\right)^{1/2} \left(\frac{B}{\Phi_o}\right)^{3/2} \tag{5.4}$$

which for fixed $B$ gives the critical density below which we have the above assumed case of only a single LL participating in the occupational process. More interesting, however, is the case of fixed $n_V$ . Then (5.4) gives the critical magnetic field

$$B_{crit1} = \left(\frac{\pi}{16}\right)^{1/3} n_V^{2/3} \Phi_o \tag{5.5}$$

in the sense that, it is only for $B > B_{crit1}$ that we have the above scenario (of only a single LL being involved). In that case the total energy is

$$E = \underbrace{\left[\left(\frac{2\Phi}{\Phi_o}\right) \times \frac{L_z}{2\pi} \int_{-k_f}^{k_f} dk_z\right]}_{N} \times \frac{\hbar\omega_c}{2} + \left(\frac{2\Phi}{\Phi_o}\right) \times \frac{L_z}{2\pi} \frac{\hbar^2}{2m} \int_{-k_f}^{k_f} k_z^2 dk_z$$

$\Rightarrow E = N\dfrac{\hbar\omega_c}{2} + \dfrac{1}{3}N\dfrac{\hbar^2 k_f^2}{2m}$ or, in units of 3D Fermi energy ( $E_f = \dfrac{\hbar^2 \kappa_f^{\,2}}{2m}$ and $\kappa_f = (3\pi^2 n_V)^{2/3}$ ):

$$\frac{E}{N} = \frac{E_f}{(3\pi^2)^{2/3}} \left[ 2\pi \left( \frac{B}{n_V^{2/3}\Phi_o} \right) + \frac{\pi^2}{12} \left( \frac{\Phi_o n_V^{2/3}}{B} \right)^2 \right] \tag{5.6}$$

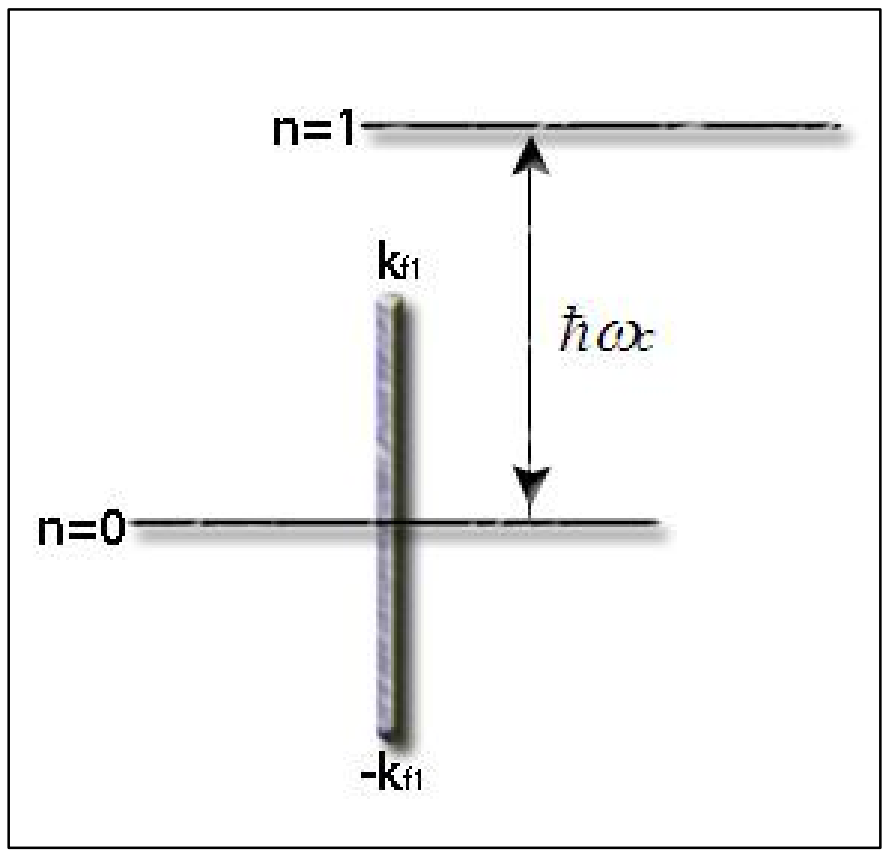

**Fig. 5.1:** Only one Fermi segment is created when $B$>$B_{crit1}$

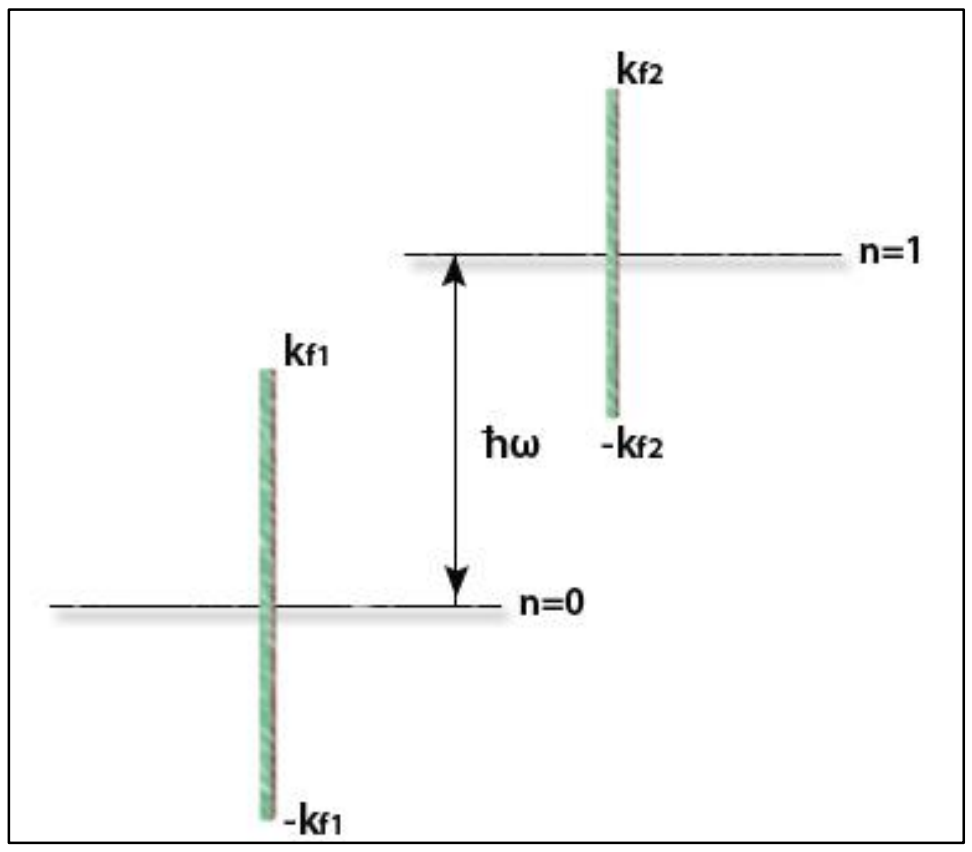

**Fig. 5.2:** Two Fermi segments are created when $B$<$B_{crit1}$ (and also when $B$>$B_{crit2}$ (see (5.10))

If we now drop $B$ to a value slightly lower than $B_{crit1}$, the lowest LL cannot accommodate all electrons, and the next LL (for $n$=1) will have to be used. We then start having a second Fermi line segment forming (extending from -$k_{f2}$ to +$k_{f2}$) associated with the $n$=1 LL, while we simultaneously have also a first Fermi segment (with a $k_{f1}$, always associated with the $n$=0 LL) that now increases in size as we keep placing more electrons**;** but now the actual manner in which we place the remaining electrons in the 2 LLs is back and forth in both of them in a way that the ñFermi heightò of the segment associated with $n$=0 and the one associated with $n$=1 will both keep increasing and will at every point (for every density) be related with each other through (the ñequilibrium relationò)

$$\frac{\hbar\omega_c}{2}+\frac{\hbar^2 k_{f1}^2}{2m}=\frac{3\hbar\omega_c}{2}+\frac{\hbar^2 k_{f2}^2}{2m} \tag{5.7}$$

where $kf_1=\pi^2 l_B^2 n_1$, $kf_2=\pi^2 l_B^2 n_2$ (coming out from an argument exactly like the one in (5.3), but now with partial densities). Eq. (5.7) will determine, for any given volume density $n_V$, the proper (energetically favorable) partition ($n_1, n_2$) of the total density between the 2 LLs involved. (5.7) again reflects the fact that the extra electron at every point of the occupational procedure must have the same single-particle energy for either of the two options (or scenarios). If (5.7) were not satisfied, and one side were larger than the other, it would indicate that the occupational procedure followed up to that point was not the energetically lowest. (Compare the above ñequilibrium conditionò with the one that was implemented in (3.5) of Section 3, or (4.3) of Section 4, when only two QW levels were occupied. Note that, although the cases are different, they are along a similar spirit).

From (5.7) and the expressions for $kf_1$, $kf_2$, we obtain the optimal density partition in the two LLs (by also utilizing $n_V=n_1+n_2$), the final result being

$$\left.\begin{aligned} n_1&=\frac{n_V}{2}\left(1+\frac{B^3}{B_{cr1}^3}\right)\\ n_2&=\frac{n_V}{2}\left(1-\frac{B^3}{B_{cr1}^3}\right)\end{aligned}\right\} \tag{5.8}$$

where $B_{crit1}$ is given by (5.5). The above partition of density is valid only for $B<B_{crit1}$. (Regarding the lowest value of $B$ allowed, i.e. the complete range of $B$-values where (5.8) is valid, see further below). Note how the full three-dimensionality and the extra presence of the magnetic field has modified the earlier found partition (3.9).

The total energy in the above case will be determined by

$$E=N_1\frac{\hbar\omega_c}{2}+\frac{1}{3}N_1\frac{\hbar^2k_{f1}^2}{2m}+N_2\frac{3\hbar\omega_c}{2}+\frac{1}{3}N_2\frac{\hbar^2k_{f2}^2}{2m}$$

$$\Rightarrow\frac{E}{N}=\frac{E_f}{(3\pi^2)^{2/3}}\left[4\pi\left(\frac{B}{n_V^{2/3}\Phi_o}\right)+\frac{\pi^2}{48}\left(\frac{\Phi_o n_V^{2/3}}{B}\right)^2-16\left(\frac{B}{n_V^{2/3}\Phi_o}\right)^4\right] \tag{5.9}$$

The lower value of the range of $B$ can then be determined by considering the next case (namely, when a 3rd Fermi segment (of LL index $n$=2) is about to form) for which we have the equilibrium condition $\frac{\hbar\omega_c}{2}+\frac{\hbar^2k_{f1}^2}{2m}=\frac{5\hbar\omega_c}{2}$ or, equivalently, $\frac{3\hbar\omega_c}{2}+\frac{\hbar^2k_{f2}^2}{2m}=\frac{5\hbar\omega_c}{2}$, and turns out to be

$$B_{crit2}=(3-2\sqrt{2})^{1/3}B_{crit1} \tag{5.10}$$

After the two examples, to proceed further with the most general case requires a little more mathematical sophistication. In the most general case, in every $i^{th}$ LL, electrons build a 1D Fermi segment that defines a Fermi wavevector, $k_{f_i}$, where the index $i$ (defined by $i=n+1$, n is a LL index) runs over all occupied LLs, and has positive integer values. When the magnetic field is a constant $B$, let us say that we know that the system occupies in general $k$ LLs ($k\geq 1$) and creates $k$ 1D Fermi segments in $z$-axis (and then $i$ runs from 1 to $k$). The associated $k_{f_i}$ ôs must be determined as in the example shown above, namely,

$$N_i = \frac{2\Phi}{\Phi_o}\sum_{\vec{k}:occupied} 1 \rightarrow \frac{2\Phi}{\Phi_o}\frac{L}{2\pi}\int_{-kf_i}^{kf_i} dk \Rightarrow kf_i = \pi^2 l_B^2 n_i\,, \tag{5.11}$$

where $n_i = N_i / V$ is the partial volume density (corresponding to the $i^{th}$ LL ($i$=1,2,3é .$k$)). A similar line of reasoning as that of Section 2 must then be followed: The last electrons on the ends of any of the $k$ 1D Fermi segments must have equal single particle energies, that is, in the spirit of Section 2, œequilibriumô is satisfied (otherwise, we would have transitions and rearrangements between the states so that equilibrium is recovered, to assure that the energetically optimal occupational procedure has been followed).

The appropriate mathematical expression for the equilibrium is then:

$$\frac{\hbar^2 k_{f1}^2}{2m} + \frac{\hbar\omega_c}{2} = \frac{\hbar^2 k_{f2}^2}{2m} + \frac{3\hbar\omega_c}{2} = ... = \frac{\hbar^2 k_{fk}^2}{2m} + \hbar\omega_c(k - \frac{1}{2}) \tag{5.12}$$

From the above condition, we can determine (with the use of (5.11)) in the general case (i.e. for any $k$) the proper partition of all 1D densities in each LL:

$$n_i^2 = n_1^2 - (i-1)\frac{B^3}{\left(\frac{\pi}{16}\right)\Phi_0^3}\,, \tag{5.13}$$

while it also holds that

$$n_V = \sum_{i=1}^{k} n_i\,, \tag{5.14}$$

where the index $i$ runs from 1 to $k$ and $n_V$ denotes the total (global) volume density of electrons. This is a system of $k$ equations with $k$ unknown variables and can be solved analytically. We will return to this solution soon. Let us first think of the appropriate values of magnetic field $B$ which force the system to occupy exactly $k$ LLs: from the equilibrium condition (5.12) we can find a critical value of $B$ as a function of $n_1$. When $B$ is exactly equal to this critical value, electrons start the occupation of $k$+1th LL. But this is rather easy to describe; it occurs when the 1D Fermi segment at $k$+1th LL is just about to be formed, namely

$$\frac{\hbar^2 k_{f1}^2}{2m} = k\hbar\omega_c\,, \tag{5.15}$$

so that $B$ is just:

$$B_{crit}^3(k) = \frac{1}{k}\left(\frac{\pi}{16}\right)\Phi_0^3 n_1^2 \tag{5.16}$$

(by following steps similar to the ones followed to derive (5.5)) ï but note that now $n_1$ also depends on $B$).

The same line of reasoning gives the other critical value of $B$, which makes electrons start the occupation of the $k$th LL:

$$B_{crit}^3(k-1) = \frac{1}{k-1}\left(\frac{\pi}{16}\right)\Phi_0^3 n_1^2\,, \tag{5.17}$$

and all the previous conditions are correct only in the case that the magnetic field varies in the following window:

$$B_{crit}(k-1) \geq B \geq B_{crit}(k) \tag{5.18}$$

(this being true for $k$>1; for $k$=1 we only have $B \geq B_{crit}(1)$ ).

We remind here the reader that the first linear density $n_1$ also depends on the magnetic field, and must be calculated analytically. Now, writing (5.13) in a more convenient form we obtain:

$$n_i = n_V \frac{\sqrt{\frac{B^3_{crit1}}{B^3}\left(\frac{n_1}{n_V}\right)^2 - (i-1)}}{\sqrt{\frac{B^3_{crit1}}{B^3}}} \tag{5.19}$$

where we set $B^3_{crit1} = \left(\frac{\pi}{16}\right)\Phi_0^3 n_V{}^2$ that was found to be the first critical value of *B* (see (5.5)). It is also convenient to define a quantity (a filling factor-like quantity):

$$a_1 = \frac{B^3_{crit1}}{B^3}\left(\frac{n_1}{n_V}\right)^2 \tag{5.20}$$

It is then easy to observe that when $B = B_{crit}(k)$ (see eq. (5.16)), $a_1 = k$ and when $B = B_{crit}(k-1)$ (see eq. (5.17)), $a_1 = k-1$, so it must hold that:

$$\boxed{B_{crit}(k-1) \geq B \geq B_{crit}(k) \Rightarrow k-1 \leq a_1 \leq k} \tag{5.21}$$

When *B* lies on a critical value, then $a_1$ is an integer (*k* or *k-1* accordingly), otherwise it must be a fractional (more generally irrational) real number. Eq. (5.19) then becomes:

$$n_i = n_V \frac{\sqrt{a_1 - i + 1}}{\sqrt{\frac{B^3_{crit1}}{B^3}}} \tag{5.22}$$

that is, the coefficient of $n_V$ is just the percentage of density which corresponds to the $(i)_{th}$ LL. Now, by using (5.14) we have:

$$\sqrt{\frac{B^3_{crit1}}{B^3}} = \sum_{i=1}^{k}\sqrt{a_1 - i + 1}$$

or

$$\sqrt{\frac{B^3_{crit1}}{B^3}} = \sqrt{a_1} + \sqrt{a_1 - 1} + \sqrt{a_1 - 2} + \ldots + \sqrt{a_1 - (k-1)} \tag{5.23}$$

Unfortunately, it does not seem possible to solve the above equation with respect to $a_1$. But one observes that (5.23) can be written with use of generalized Riemann or Hurwitz zeta functions (defined by $\zeta(s,a) = \sum_{i=0}^{\infty} 1/(i+a)^s$ ) as follows:

$$\sqrt{\frac{B^3_{crit1}}{B^3}} = -i\left[\zeta\left(-\tfrac{1}{2}, -a_1\right) - \zeta\left(-\tfrac{1}{2}, k-a_1\right)\right], \tag{5.24}$$

where *i* is the imaginary unit and *k* the number of occupied LLs and $k-1 \leq a_1 \leq k$, $0 \leq k - a_1 \leq 1$. So the difference of Hurwitz zeta functions must be a pure complex number:

$$\mathrm{Re}\left\{\zeta\left(-\tfrac{1}{2}, -a_1\right)\right\} = \mathrm{Re}\left\{\zeta\left(-\tfrac{1}{2}, k-a_1\right)\right\} \forall a_1 \tag{5.25}$$

$$\mathrm{Im}\left\{\zeta\left(-\tfrac{1}{2}, k-a_1\right)\right\} = 0, \text{ because } k - a_1 \geq 0 \tag{5.26}$$

Finally, we find that

$$\zeta\left(-\tfrac{1}{2},-a_1\right)-\zeta\left(-\tfrac{1}{2},k-a_1\right)=i\,\mathrm{Im}\left\{\zeta\left(-\tfrac{1}{2},-a_1\right)\right\} \tag{5.27}$$

$$\boxed{\mathrm{Im}\left\{\zeta\left(-\tfrac{1}{2},-a_1\right)\right\}=\sqrt{\frac{B_{crit1}^3}{B^3}}} \tag{5.28}$$

This is the key for the solution of this problem. It is only the imaginary part of Hurwitz zeta functions has physical meaning. With the help of (5.28), (5.16) and (5.17) we can then write down analytical expressions for the critical values of *B* that do not depend on $n_1$:

$$B_{crit}(k)=\frac{B_{crit1}}{\left(\mathrm{Im}\left\{\zeta\left(-\tfrac{1}{2},-k\right)\right\}\right)^{2/3}} \tag{5.29}$$

$$B_{crit}(k-1)=\frac{B_{crit1}}{\left(\mathrm{Im}\left\{\zeta\left(-\tfrac{1}{2},-(k-1)\right)\right\}\right)^{2/3}} \tag{5.30}$$

As a consistency test we can check that the above reproduce the earlier results (5.5) and (5.10) (see (5.34) below for the imaginary part of Hurwitz Zeta functions). For *k*=1, then

$B_{crit}(1)=\dfrac{B_{crit1}}{\left(\mathrm{Im}\left\{\zeta\left(-\tfrac{1}{2},-1\right)\right\}\right)^{2/3}}=\dfrac{B_{crit1}}{1}=B_{crit1}$ which is (5.5), and for *k*=2 we have

$B_{crit}(2)=\dfrac{B_{crit1}}{\left(\mathrm{Im}\left\{\zeta\left(-\tfrac{1}{2},-2\right)\right\}\right)^{2/3}}=\dfrac{1}{\left(1+\sqrt{2}\right)^{2/3}}B_{crit1}=\left(3-2\sqrt{2}\right)^{1/3}B_{crit1}$, which is (5.10)

It is also interesting to check what the differences of neighboring inverse $B_{crit}$ôs are, and relate their behavioral pattern to the standard period of the de Haas-van Alphen effect. It is true that in weak magnetic fields (hence large values of *k*) the system starts behaving semiclassically (then the segment sizes will come from cuts of Landau tubes inside a Fermi sphere), and then we expect an oscillating period similar to that of dHvA effect. Having calculated the critical values of *B* analytically, we have the ability to check the period directly, without any approximations. Indeed, the semiclassical dHvA period is:

$$\delta\left(\frac{1}{B}\right)=\frac{4\pi}{\left(3\pi^2\right)^{2/3}}\frac{1}{n^{2/3}\Phi_o}=\frac{\left(\frac{4}{9}\right)^{1/3}}{B_{crit1}}=\frac{0.76314}{B_{crit1}} \tag{5.31}$$

The difference of inverse *B* that we have found is (from (5.29) and (5.30)):

$$\delta\left(\frac{1}{B}\right)=\frac{1}{B_{crit}(k)}-\frac{1}{B_{crit}(k-1)}=\frac{\left(\mathrm{Im}\left\{\zeta\left(-\tfrac{1}{2},-k\right)\right\}\right)^{2/3}-\left(\mathrm{Im}\left\{\zeta\left(-\tfrac{1}{2},-(k-1)\right)\right\}\right)^{2/3}}{B_{crit1}} \tag{5.32}$$

Note that when *k*=1, then $\delta(1/B)=1/B_{crit1}$, which deviates from (5.31) at about 31%, while if *k*=2, then $\delta(1/B)=0.7996/B_{crit1}$ which deviates from (5.31) at only 5%. Now, comparing (5.31) with (5.32), leads to the conclusion that the following must be proved:

$$\left(\mathrm{Im}\left\{\zeta\left(-\tfrac{1}{2},-k\right)\right\}\right)^{2/3}-\left(\mathrm{Im}\left\{\zeta\left(-\tfrac{1}{2},-(k-1)\right)\right\}\right)^{2/3}\approx\left(\frac{4}{9}\right)^{1/3} \tag{5.33}$$

Using the well known relations (which are true, because $k$ is an integer):

$$\operatorname{Im}\left\{\zeta\left(-\tfrac{1}{2},-k\right)\right\}=\sum_{j=1}^{k}\sqrt{j} \text{ and } \operatorname{Im}\left\{\zeta\left(-\tfrac{1}{2},-(k-1)\right)\right\}=\sum_{j=1}^{k-1}\sqrt{j}\ , \tag{5.34}$$

then the following must be proved:

$$\left(\sum_{j=1}^{k}\sqrt{j}\right)^{2/3}-\left(\sum_{j=1}^{k-1}\sqrt{j}\right)^{2/3}\approx\left(\frac{4}{9}\right)^{1/3}=\left(\frac{2}{3}\right)^{2/3} \text{ when } k>>1 \tag{5.35}$$

For this let us momentarily think in a slightly different manner: instead of calculating $\delta\left(\frac{1}{B}\right)$, we can calculate $\delta\left(\frac{1}{B}\right)^{3/2}$ and then relate it to $\delta\left(\frac{1}{B}\right)$: Using (5.34) and (5.32) we obtain:

$$\delta\left(\frac{1}{B}\right)^{3/2}=\frac{\sqrt{k}}{B_{crit1}^{3/2}} \tag{5.36}$$

Equivalently, we can write $\delta\left(\frac{1}{B}\right)$ as:

$$\delta\left(\frac{1}{B}\right)=\frac{2}{3}\delta\left(\frac{1}{B}\right)^{3/2}\sqrt{B_{crit}(k)}=\frac{2}{3}\frac{1}{B_{crit1}}\frac{\sqrt{k}}{\left(\operatorname{Im}\left\{\zeta\left(-\tfrac{1}{2},-k\right)\right\}\right)^{1/3}} \tag{5.37}$$

And now comes the approximation: for large $k$ (weak magnetic fields) we must expand the term $\frac{\operatorname{Im}\left\{\zeta\left(-\frac{1}{2},-k\right)\right\}}{\left(\sqrt{k}\right)^{3}}$ around $k=\infty$, to see that it is almost equal to 2/3, which is indeed true:

$$\frac{\operatorname{Im}\left\{\zeta\left(-\tfrac{1}{2},-k\right)\right\}}{\left(\sqrt{k}\right)^{3}}=\frac{2}{3}+\frac{1}{2k}+\left(\frac{1}{k}\right)^{3/2}\zeta\left(-\frac{1}{2}\right)+\frac{1}{24k^{2}}-\frac{1}{1920k^{4}}+\mathrm{O}\left\{\left(\frac{1}{k}\right)^{11/2}\right\}\approx\frac{2}{3} \tag{5.38}$$

So, the result is:

$$\delta\left(\frac{1}{B}\right)=\frac{2}{3}\left(\frac{3}{2}\right)^{1/3}\frac{1}{B_{crit1}}=\left(\frac{2}{3}\right)^{2/3}\frac{1}{B_{crit1}} \tag{5.39}$$

as anticipated (see (5.33)). The conclusion is that in not extremely strong magnetic fields (for many LLs occupied) the system rapidly converges to the semiclassical behavior. But this semiclassical dHvA period is violated at exceedingly strong magnetic fields.

Unfortunately, the magnetic fields needed to observe these extreme quantum effects are very large, and therefore we cannot see them in the laboratory. We can however effectively reduce them by lowering the value of electronic number density. For example consider (5.5) which gives the first critical value of $B$ (the largest of all critical values). Nowadays, we may achieve magnetic fields up to 40 Tesla, so

$$n_{\max}=\left(\frac{16}{\pi}\right)^{1/2}\left(\frac{40T}{\Phi_o}\right)^{3/2}=2.17\times10^{24}\,m^{-3}$$

This can be considered as the maximum number density of charge carriers that a material must have in order for our extreme quantum results (reflected in the dHvA violations) to be experimentally seen. (The above density is 4 orders of magnitude smaller than typical metallic densities).

The last important step for this Section is to calculate the total energy and magnetization. The energy is just a sum over all occupied LLs and $z$-axis levels:

$$E=\sum_{j=0}^{k-1}\left(N_{j+1}\hbar\omega(j+\tfrac{1}{2})+\frac{1}{3}N_{j+1}\frac{\hbar^2 kf_{j+1}^2}{2m}\right) \tag{5.40}$$

where $kf_{j+1}^2=\pi^4 l_B^4 n_{j+1}^2$ (from (5.11)), which leads to

$$E=\sum_{j=0}^{k-1}\left(\hbar\omega jN_{j+1}+\frac{\hbar\omega N_{j+1}}{2}+\frac{1}{3}\frac{\hbar^2\pi^4 l_B^4}{2m}N_{j+1}n_{j+1}^2\right) \tag{5.41}$$

Using $N_{j+1}=n_{j+1}V$ and substituting $n_j$ with its equal from eq. (5.13), we find:

$$E=\sum_{j=0}^{k-1}\left(\hbar\omega N_{j+1}\frac{B_{crit1}^3}{B^3}\frac{n_1^2}{n^2}+\frac{\hbar\omega N_{j+1}}{2}+\frac{1}{3}\frac{\hbar^2\pi^4 l_B^4}{2m}Vn_{j+1}^3-\hbar\omega V\frac{B_{crit1}^3}{B^3}\frac{n_{j+1}^3}{n^2}\right) \tag{5.42}$$

Now, observing that

$$\frac{\hbar^2\pi^4 l_B^4}{2m}=\hbar\omega\frac{B_{crit1}^3}{B^3 n^2} \tag{5.44}$$

the energy becomes:

$$E=\sum_{j=0}^{k-1}\left(\hbar\omega N_{j+1}\frac{B_{crit1}^3}{B^3}\frac{n_1^2}{n^2}+\frac{\hbar\omega N_{j+1}}{2}-\frac{2}{3}\hbar\omega V\frac{B_{crit1}^3}{B^3}\frac{n_{j+1}^3}{n^2}\right) \tag{5.45}$$

The first term gives:

$$\sum_{j=0}^{k-1}\hbar\omega N_{j+1}\frac{B_{crit1}^3}{B^3}\frac{n_1^2}{n^2}=N\hbar\omega\frac{B_{crit1}^3}{B^3}\frac{n_1^2}{n^2}=N\hbar\omega a_1 \tag{5.46}$$

The second term gives:

$$\sum_{j=0}^{k-1}N_{j+1}\frac{\hbar\omega}{2}=N\frac{\hbar\omega}{2}, \tag{5.47}$$

and the third term gives:

$$\frac{2}{3}\hbar\omega V\frac{B_{crit1}^3}{B^3}\sum_{j=0}^{k-1}\frac{n_{j+1}^3}{n^2}=\frac{2}{3}\hbar\omega N\left(\frac{B^3}{B_{crit1}^3}\right)^{1/2}\sum_{j=0}^{k-1}\left(a_1-j\right)^{3/2} \tag{5.48}$$

Now, the sum $\sum_{j=0}^{k-1}\left(a_1-j\right)^{3/2}$ is just a difference of two zeta functions of order -3/2:

$$\sum_{j=0}^{k-1}(a_1-j)^{3/2}=i\left[\zeta(-3/2,-a_1)-\zeta(-3/2,k-a_1)\right] \tag{5.49}$$

Energy must be a real quantity, so it must hold that:

$$\zeta(-3/2,-a_1)-\zeta(-3/2,k-a_1)=i\,\mathrm{Im}\{\zeta(-3/2,-a_1)\} \tag{5.50}$$

and

$$\sum_{j=0}^{k-1}(a_1-j)^{3/2}=-\mathrm{Im}\{\zeta(-3/2,-a_1)\} \tag{5.51}$$

Finally, the energy per electron is:

$$\boxed{E/N=\hbar\omega\left[1/2+a_1+\frac{2}{3}\left(\frac{B^3}{B_{crit1}^3}\right)^{1/2}\mathrm{Im}\{\zeta(-3/2,-a_1)\}\right]} \tag{5.52}$$

Once again, it does not seem possible to solve (5.52) with respect to $a_1$ (there is no analytic expression for the inverse function of imaginary part of Hurwitz zeta functions). But this is not quite necessary, since we can evaluate (5.52) numerically, and then find the roots of (5.52) for every value of *B*. If up to this point all calculations are correct, the derivative of energy with respect to $a_1$ must vanish, that is, energy is indeed minimal, and the correct density distributions are given by (5.52). Although tedious, it is straightforward to check this expectation. Indeed we have:

$$\frac{\partial(E/N)}{\partial a_1}=\hbar\omega_c\left[1-\frac{2}{3}\left(\frac{B^3}{B_{crit1}^3}\right)^{1/2}\frac{\partial}{\partial a_1}\left[\sum_{j=0}^{k-1}(a_1-j)^{3/2}\right]\right]=\hbar\omega_c\left[1-\left(\frac{B^3}{B_{crit1}^3}\right)^{1/2}\sum_{j=0}^{k-1}(a_1-j)^{1/2}\right]=\hbar\omega_c\left[1-\left(\frac{B^3}{B_{crit1}^3}\right)^{1/2}\left(\frac{B_{crit1}^3}{B^3}\right)^{1/2}\right]=0$$

At the critical values of *B* (the ones expressed by (5.29)), energy has a simple analytic form, namely

$$E/N=\frac{\hbar eB_{crit1}}{mc}\left[\frac{1}{\left(\mathrm{Im}\{\zeta(-1/2,-k)\}\right)^{2/3}}\right]\left[1/2+k+\frac{2}{3}\frac{\mathrm{Im}\{\zeta(-3/2,-k)\}}{\mathrm{Im}\{\zeta(-1/2,-k)\}}\right] \tag{5.53}$$

(where *k* comes out of invertion of (5.29), actually labeling the critical point). (Note the amusing fact that

$$\hbar eB_{crit1}/mc=(2/3)^{2/3}\frac{\hbar^2(3\pi^2 n_V)^{2/3}}{2m}=(2/3)^{2/3}E_f(3D),$$

where $E_f(3D)=\dfrac{\hbar^2(3\pi^2 n_V)^{2/3}}{2m}$ the 3D Fermi energy of electrons when no magnetic field is applied on the cube (something that could also be seen directly from (5.5) as well).

In the limit then $B \rightarrow 0$ (or $k\rightarrow\infty$) (5.53) can be shown to tend to $(3/5)E_f(3D)$. This can also be seen from fig. 5.3 below).

Finally, by taking the derivative of (5.52) we can also determine analytically the magnetization per electron using the relation:

$$M=-\frac{\partial E}{\partial B}$$

But we will need to know the derivative of $a_1$ with respect to *B*; this can be calculated from (5.19), and the result is:

$$\frac{da_1}{dB}=\frac{-3B^{-5/2}B_{crit}^{3/2}}{i\left[\zeta\left(1/2,-a_1\right)-\zeta\left(1/2,k-a_1\right)\right]}=\frac{3B^{-5/2}B_{crit}^{3/2}}{\mathrm{Im}\left\{\zeta\left(1/2,-a_1\right)\right\}}$$

$$\Rightarrow M/N=-\mu_{\mathrm{B}}\left[1+2a_1+\frac{10}{3}x^{3/2}\,\mathrm{Im}\left\{\zeta\left(-3/2,-a_1\right)\right\}\right], \tag{5.54}$$

and for the magnetic susceptibility the corresponding procedure gives

$$X/N=\frac{\mu_{\mathrm{B}}}{B_{crit1}}\left[9x^{-5/2}\frac{1}{\mathrm{Im}\left\{\zeta\left(1/2,-a_1\right)\right\}}-5x^{1/2}\,\mathrm{Im}\left\{\zeta\left(-3/2,-a_1\right)\right\}\right] \tag{5.55}$$

where $\mu_{\mathrm{B}}=\dfrac{\hbar e}{2mc}$ is the Bohr magneton and $x=\dfrac{B}{B_{crit1}}$.

The above solves exactly the problem of noninteracting electrons in a uniform magnetic field in full 3D space *by applying a procedure (of equilibrium relations) that is in a similar line of reasoning as in earlier Sections* (actually the central line of approach that has been introduced in this article, to be used as a common tool for quite disparate problems ï see also next Section). It should also be noted that the above results are the limiting behaviors of the previous quasi-2D interface, when thickness is getting exceedingly large (see derivations in the Appendix, and especially how the first critical field (5.5) comes out from the rather involved results of Section 4).
Earlier works that follow different methods either do not give results for the total energy [27] or they mostly deal with a Relativistic system [28], and are both considerably involved in mathematics that do not quite reflect the basic Physics of the problem (i.e. the basic physical processes that are involved in the formation of the proper Fermi segments).

Below the reader can find plots of all thermodynamic properties as functions of *B*. We should note again the continuity of energy and magnetization (but with the latter having cusps, leading to discontinuities and a highly nonlinear behavior of susceptibility, something that we did not witness in the quasi-2D results of Section 4 where susceptibility was always piecewise constant).

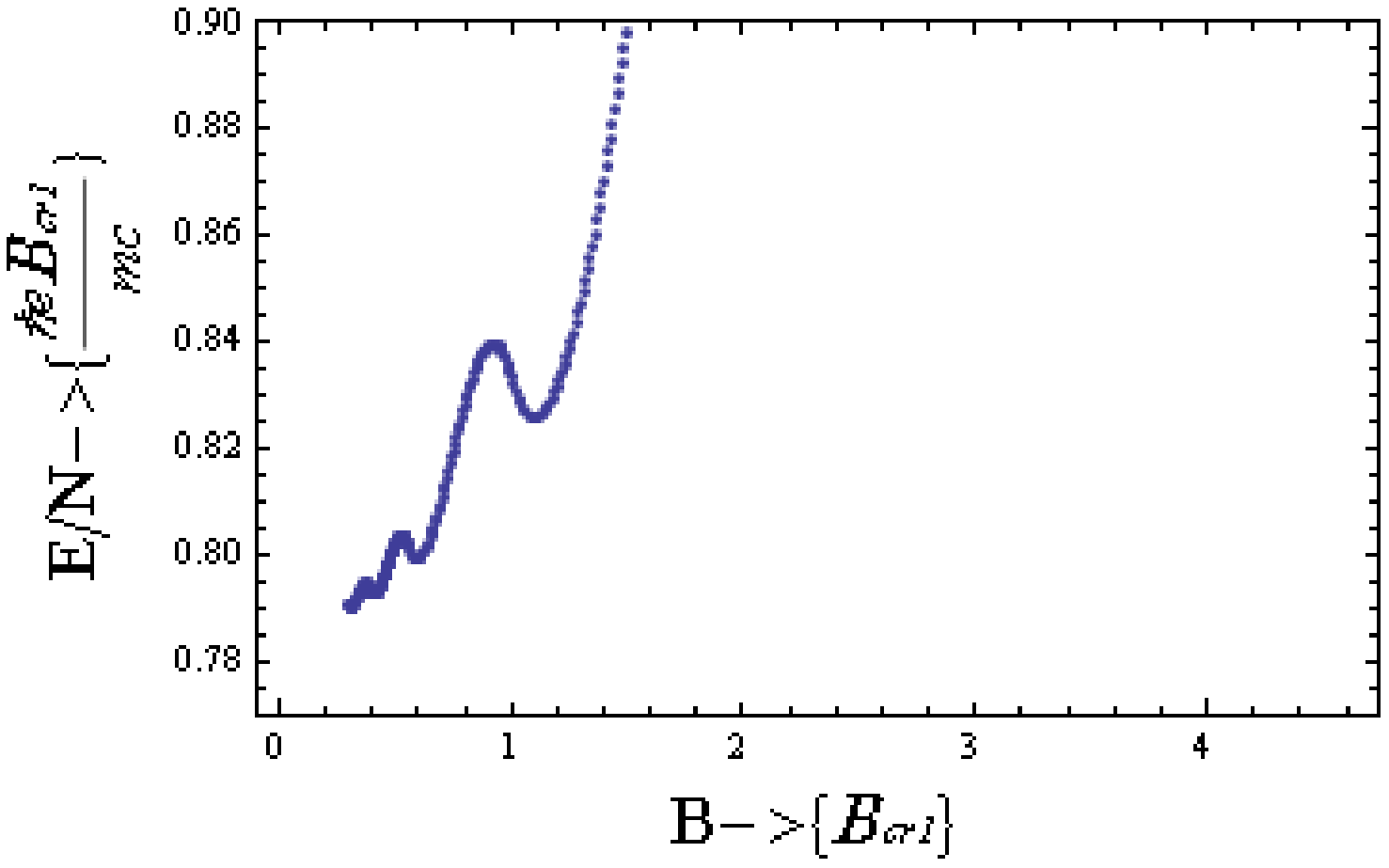

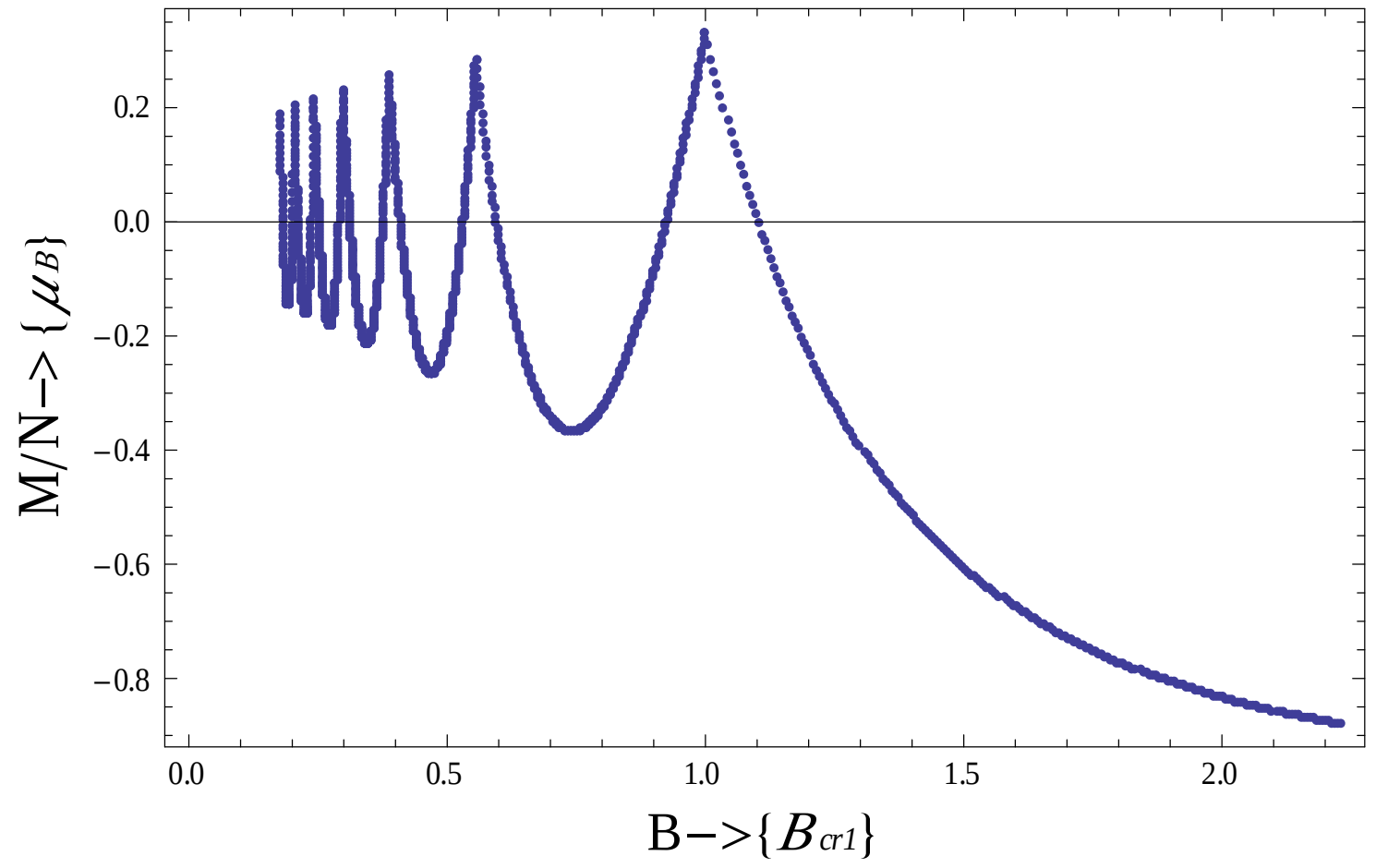

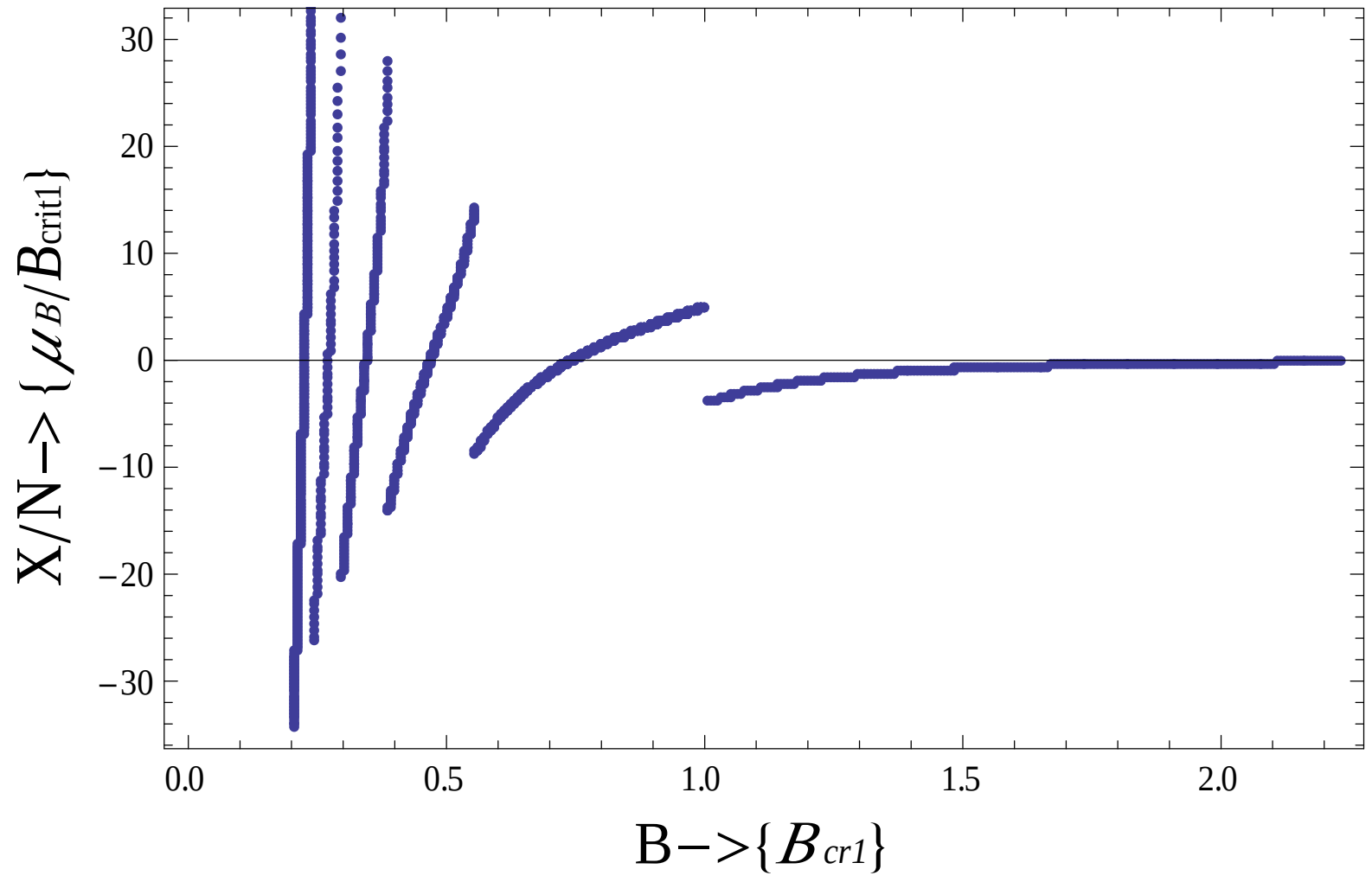

**Fig. 5.3:** Energy (in units of $\hbar e B_{crit1}/mc$), Magnetization (in units of $\mu_{\rm B}$) and Susceptibility (in units of $\mu_{\rm B}/B_{crit1}$) as functions of *B*. Susceptibility, apart from discontinuous, is highly nonlinear (compared to the quasi-2D cases of Section 4).

A final point must be made that concerns Zeeman coupling. Analytical solutions involving imaginary parts of Hurwitz zeta functions can also be obtained if the Zeeman term is included in the above calculations. The energy spectrum (5.1) is in that case modified as follows:

$$\varepsilon_{n,k_z} = (n+\frac{1}{2}\pm\frac{g^*}{4}\frac{m^*}{m})\hbar\omega_c^* + \frac{\hbar^2 k_z^2}{2m^*}$$

with $g^*$ the gyromagnetic ratio, $m^*$ the effective mass and $\omega_c^* = eB/m^*c$ the effective cyclotron frequency. Although the problem is also completely solvable, we here choose to only report the observation that, for sufficiently large $g^*$, we find a pronounced minimum in total energy as function of *B*. When i.e. the gyromagnetic ratio is $g^* = 1.5$ we obtain the behavior shown in Fig. 5.4. Such behaviors originate from the interplay of QW, Zeeman and LL Physics in the full 3D problem, and have not been reported earlier**;** as already noted in the Introduction, such minima may be important for the design of stable 3D quantum devices (in cases i.e. that the magnetic field can be self-consistently considered as self-generated).

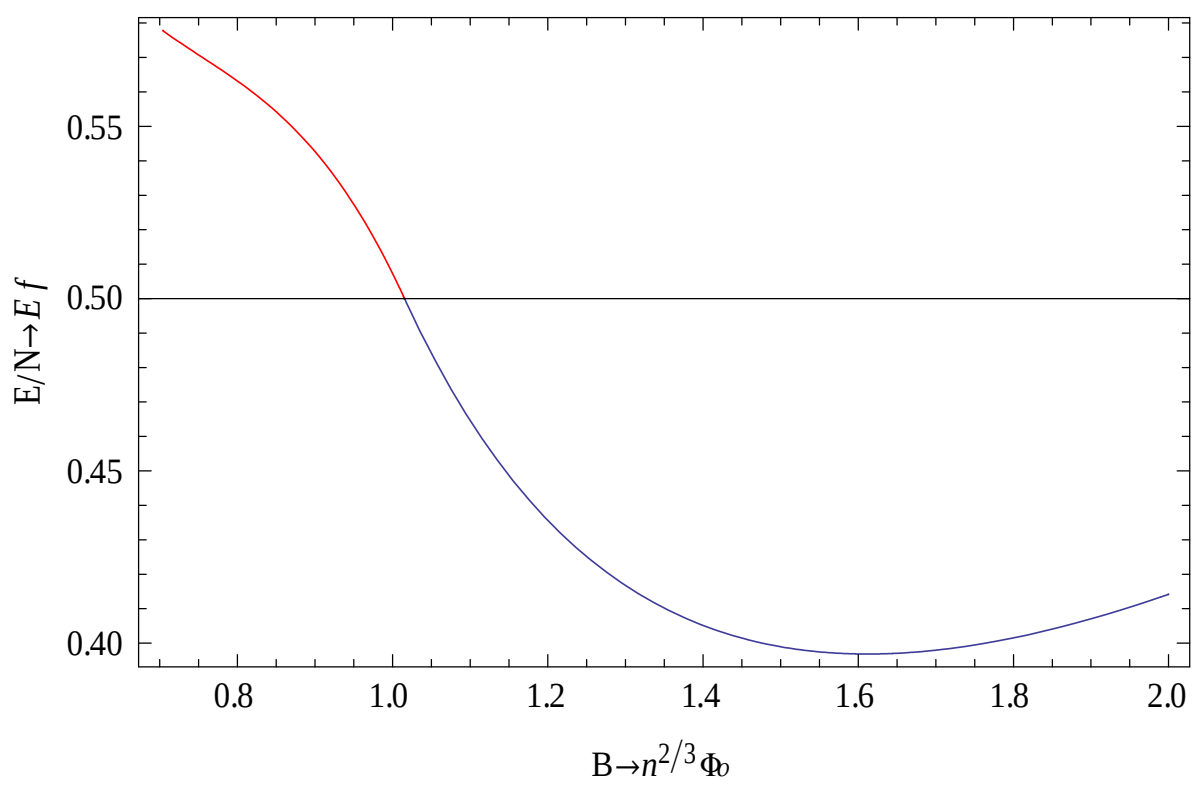

**Fig. 5.4:** Energy, as function of *B*, for $g^* = 1.5$ for the first two windows of *B* (plotted in different color). This minimum may be important in fabrication of small quantum devices.

### 6. Relevance and applicability to the dimensionality crossover in Topological Insulators

We have seen with an exact analytical solution and through a detailed energy interplay analysis that the finite thickness is not as innocent as widely believed or implied. Its presence does not merely provide just another variable and just another label to the wavefunctions and energy spectra (and this is basically due to the fact that the Pauli principle can be momentarily circumvented at every step, these steps forming a sequence that leads to interesting (and rather unpredictable) behaviors). The method that we have followed for determining ground state thermodynamic magnetic quantities such as the magnetization is not only exact but is also based on physically transparent arguments (on energy interplay and comparisons at the single-particle level, without the need of using the density of states). As a reward for this more physical approach, we have found, even for the above conventional systems, that special values of thickness induce certain ñinternal transitionsò (i.e. occurring at partial LL filling) that violate the standard de Haas-van Alphen periodicity, transitions that apparently have not been captured by other approaches. But as an equally important reward we should stress that, because of its simplicity and universality in its line of reasoning, the same method can also be applied to other more involved systems of current interest, such as 3D strong topological insulators (such as $Bi_2Se_3$) and its dimensionality crossover to 2D topological insulators (such as HgTe/CdTe wells).
To show this, we now briefly turn our attention to the well-known effective four-band model by Zhang et al. [29] that describes the low-energy behaviour of $Bi_2Se_3$. Such systems are described by a modified Dirac equation rather than the Schrodinger equation and this leads to very different wavefunctions (with nontrivial topological properties) and energy spectra, where the role of thickness is coupled to the 2D motion**;** however, the line of reasoning that we have developed and the general method that we have followed can still be applied in a similar manner. All one needs is essentially the one-particle spectrum which incorporates the effect of thickness (even though this effect may be strongly coupled to the 2D degrees of freedom). Indeed, even for the coupled problem, one could determine the single-particle energy for the lowest value of a thickness-related quantum number, then determine the same for the next higher value of this quantum number, and then study the comparison between the two energies ï looking for cases of *crossover* between the two that might occur not too far from the ũ-point ($k_x$=$k_y$=0). If there are also sufficient charge carriers that give a $k_F$ that is further away than the crossover point in *k*-space, this would be a strong indication that effects like the ones discussed above may also be present in these systems as well. We will carry out a quick calculation in the above spirit in what follows below but only in the thin-film limit (i.e. we will now have massive Dirac Fermions; this is even more relevant to our method as it has recently been found [30] that for thin films there is a gap opening and the Fermi level does not fall in the gap (hence surface carriers are present in the electron band [with an estimated areal density ~ $5*10^{16}$ $m^{-2}$])). And in this thin-film limit we will indeed find below theoretical evidence of a clear crossover close to the ũ-point (and inside the region of *k*-space where the Dirac equation is valid), with an estimated $k_f$ that is further away ï something that shows that for these more exotic systems a more careful study of effects like the ones presented in the present work is probably needed.

One can start with the effective model that describes the bulk states near the Γ-point for bulk $Bi_2Se_3$ [29]. The Hamiltonian is given by

$$H\left(\vec{k}\right) = \varepsilon_o\left(\vec{k}\right) I_{4\times4} + \begin{pmatrix} M\left(\vec{k}\right) & -iA_1\partial_z & 0 & A_2k_- \\ -iA_1\partial_z & -M\left(\vec{k}\right) & A_2k_- & 0 \\ 0 & A_2k_+ & M\left(\vec{k}\right) & iA_1\partial_z \\ A_2k_+ & 0 & iA_1\partial_z & -M\left(\vec{k}\right) \end{pmatrix} \tag{6.1}$$

In a basis $\left|p1_z^+,\uparrow\right\rangle$, $\left|p2_z^-,\uparrow\right\rangle$, $\left|p1_z^+,\downarrow\right\rangle$, $\left|p2_z^-,\downarrow\right\rangle$ where +(-) stands for even (odd) parity, with

$$\varepsilon_o\left(\vec{k}\right) = C - D_1\partial_z^2 + D_2k^2,\ M\left(\vec{k}\right) = M + B_1\partial_z^2 - B_2k^2,\ k_\pm = k_x \pm ik_y,\ k^2 = k_x^2 + k_y^2,$$

with the model parameters having values:

$M = 0.28eV$, $A_1 = 2.2eV\,\overset{o}{\mathrm{A}}$, $A_2 = 4.1eV\,\overset{o}{\mathrm{A}}$, $B_1 = 10eV\,\overset{o}{\mathrm{A}}{}^2$, $B_2 = 56.6eV\,\overset{o}{\mathrm{A}}{}^2$, $C = -0.0068eV$, $D_1 = 1.3eV\,\overset{o}{\mathrm{A}}{}^2$, $D_2 = 19.6eV\,\overset{o}{\mathrm{A}}{}^2$,

and with a 4-component trial wavefunction

$$\Psi = \Psi_\lambda e^{\lambda z} \tag{6.2}$$

(6.1) has been diagonalized [31] giving $\lambda_\alpha$ as functions of *E* and *k* [see eq. (5) of [31]]. By inverting them we obtain $E=E(\lambda_\alpha,k)$ and by focusing on the electron band we obtain

$$E_{el}^{\alpha=2}\left(\vec{k}\right) = -0.0068 + 19.6k^2 - \frac{12.8305\lambda_2^2}{\pi^2} - \sqrt{0.0784 - 14.886k^2 + 3203.56k^4 + \frac{7.5\lambda_2^2}{\pi^2} - \frac{11172.4k^2\lambda_2^2}{\pi^2} + 100\lambda_2^4} \tag{6.3}$$

Although then this problem must be treated numerically (for a self-consistent determination of E and λ's), we can immediately check the thin-film limit, where it is found [31] that $\lambda = i\, n_z \pi/d$; by plugging then this into (6.3) we obtain the single-particle spectrum for the electron band as a function of *k* for each $n_z$, namely

$$E_{el}^{\alpha=2}\left(\vec{k}\right) = -0.0068 + 19.6k^2 + \frac{12.8305n_z^2}{d^2} - \sqrt{0.0784 - 14.886k^2 + 3203.56k^4 - \frac{7.5n_z^2}{d^2} + \frac{11172.4k^2n_z^2}{d^2} + 100\frac{n_z^4}{d^4}} \tag{6.4}$$

By then using a value of $d$=10*nm* and plotting (6.4) for $n_z = 1$ and $n_z = 2$, we indeed find a crossover close to the Γ-point, as shown in Fig.'s 6.1-6.3.

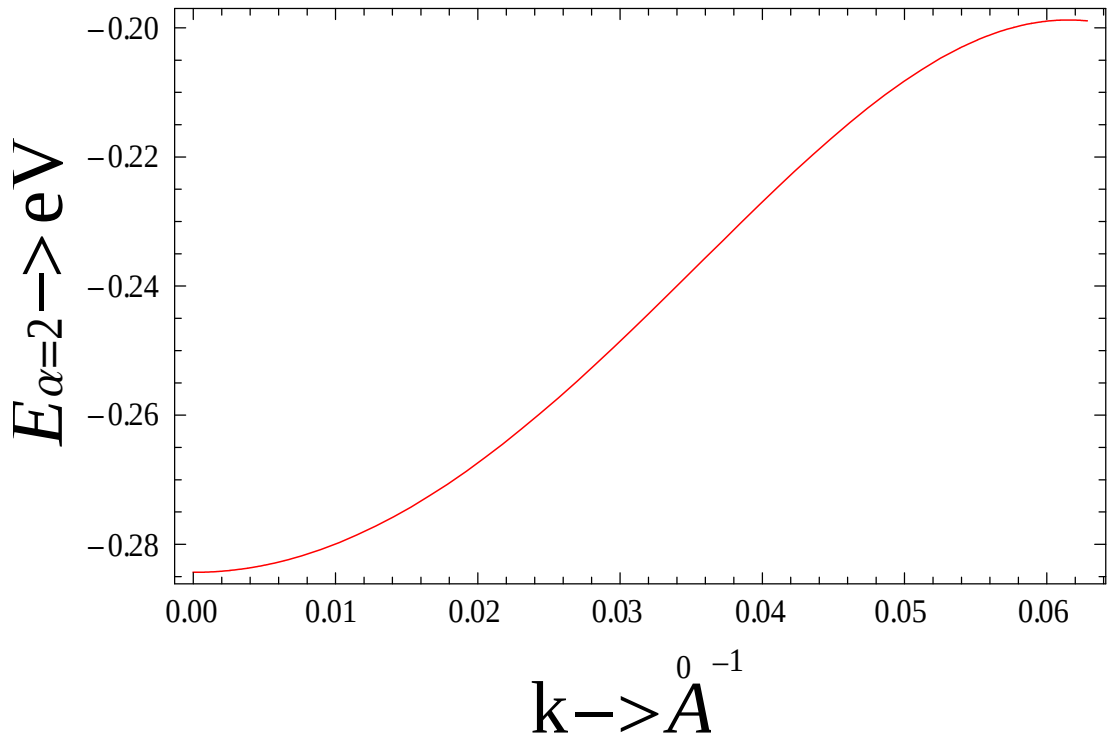

**Fig. 6.1:** $E_{el}^{\alpha=2}\left(\vec{k}\right)$ for $n_z=1$.

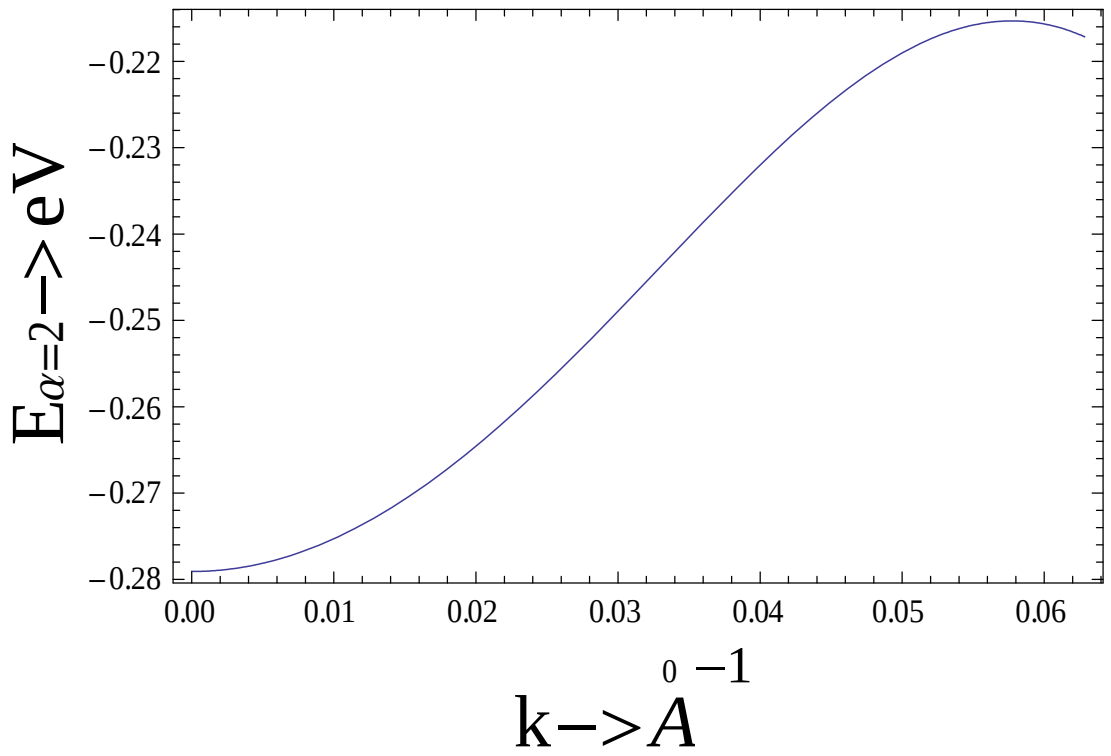

**Fig. 6.2:** $E_{el}^{\alpha=2}\left(\vec{k}\right)$ for $n_z=2$.

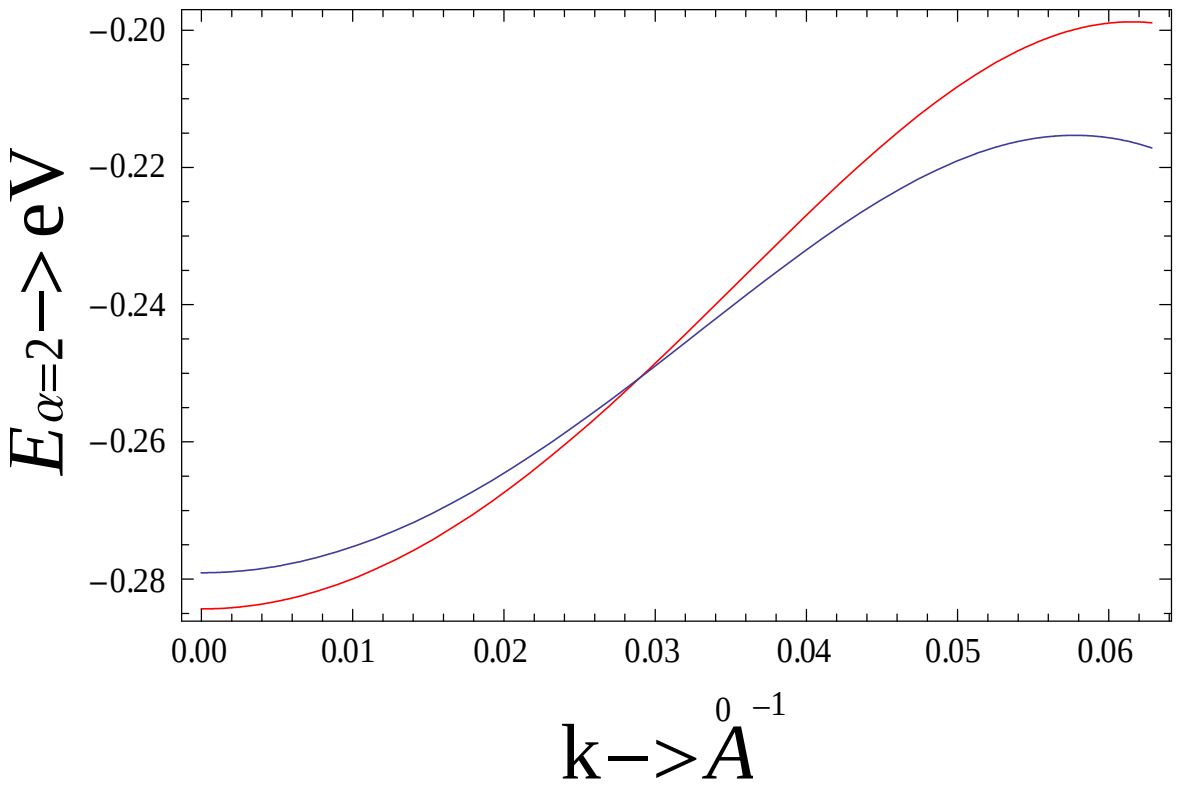

**Fig. 6.3:** $E_{el}^{\alpha=2}\left(\vec{k}\right)$ for $n_z=1$ and $n_z=2$ shown together

Moreover, note that, in analogy to Section 3, the occupational procedure is similar. For example, a Fermi wavevector for this band is now $k_F=\sqrt{\pi n_A}$ (with $n_A$ being the mean surface areal density) for a certain spin configuration. By then using the estimate of density given above, a value of $k_F=0.04\,\overset{o}{\mathrm{A}}^{-1}$ is obtained, that is further on the right of the crossover point in Fig. 6.3 (indicating that we have sufficient carriers that may exploit the crossover for internal transitions of the general type studied in this paper).
Independently, let us try to examine the first critical value of *d* where the first transition occurs, but now in this quasi-2D topological insulator (a generalization of (3.4)): we must now have

$$E_{el}^{\alpha=2}\left(k=k_F,n_z=1\right)=E_{el}^{\alpha=2}\left(k=0,n_z=2\right) \tag{6.5}$$

or equivalently

$$-0.0068+19.6k_F^2+\frac{12.8305}{d^2}-\sqrt{0.0784-14.886k_F^2+3203.56k_F^4-\frac{7.5}{d^2}+\frac{11172.4k_F^2}{d^2}+100\frac{1}{d^4}}=$$

$$-0.0068+\frac{12.8305\times 4}{d^2}-\sqrt{0.0784-\frac{7.5\times 4}{d^2}+100\frac{16}{d^4}} \tag{6.6}$$

This equation relates the first critical value of *d* (in which the two-dimensional topological insulator starts becoming three-dimensional). By plugging the estimated $k_F$ above, the solution of (6.6) gives a *d*=3.86*nm*, something that indicates that our tentative value of *d* (of 10*nm*) is in a region where interesting effects might be expected and that, generally speaking, strengthens the necessity of a more careful treatment of this system.

## 7. Conclusions

An exact solution providing analytical expressions for the magnetic thermodynamic functions of an interface or film in a perpendicular magnetic film (with rigid walls) has been presented, in a picture of noninteracting electrons. Interactions were later taken into account by following the same method for the Landau ş -levels in a picture of noninteracting Composite Fermions. The method used, different from standard density of states methods and grandcanonical and semiclassical approaches, is exact but also physically transparent at every step, providing hence the possibility of applying it to more involved systems such as 3D topological insulators (where the thickness-related modes are strongly-coupled to the planar motion). Even for conventional systems it has been found that the finite thickness is not as innocent as widely believed or implied. Its presence does not merely provide just another variable and just another label to the wavefunctions and energy spectra (and this is basically due to the fact that the Pauli principle can be momentarily circumvented at every step, these steps forming a sequence that leads to interesting (and rather unpredictable) behaviors). The finite thickness has been found here to induce certain ñinternal transitionsò (at partial Landau Level filling) of magnetization that are not captured by earlier approaches and that violate the standard de Haas-van Alphen periodicities. The correctness of all these results has been tested against an independent exact analytical solution of the full 3D problem, which apparently also leads to certain behaviors that have not been reported earlier. For topologically nontrivial systems, evidence that such effects may also be operative in the dimensionality crossover between 3D and 2D topological insulator wells has also been given. This suggests that the versatile method presented here needs to be carefully applied to such systems (a task that can be carried out numerically if the analytical patterns are too involved), something that is currently under investigation.

## APPENDIX A: Full analysis of the interface results

In the main text we have seen that the procedure of finding the ground state for this problem is rather complicated. It would be nice, if we had a generalized method that would provide general behavioral patterns for the expressions that come out for ground state energies for all *B*ôs and *d*ôs, and this is exactly what we are going to present here. Let us suppose that the magnetic field *B* lies in the following generalized window:

$$B \in \left[ \frac{1}{2\rho} n_A \Phi_o, \frac{1}{2(\rho-1)} n_A \Phi_o \right], \quad \rho = 1, 2, 3 \ldots, \tag{A0}$$

(with $\rho$ being the total number of combined states {n, $n_z$} involved (with degeneracy of each set implied (not counted separately)). For a random value of *d*, we may then have the following pattern for the occupied states, (calling with *l* the number of Landau Levels that are occupied)

$$\begin{array}{cccc} \{0,1\} & \{0,2\} & . \; . & \{0,k_1\} \\ \{1,1\} & \{1,2\} & . \; . & \{1,k_2\} \\ . & . & . \; . & . \\ . & . & . \; . & . \\ \{l-1,1\} & \{l-1,2\} & . \; . & \{l-1,k_l\} \end{array} \tag{A1}$$

with the constraint: $\sum_{i=1}^{l} k_i = \rho$ and $l = 1,2 \ldots \rho$, $k_l = 1,2 \ldots \rho$ being the possible values for $l$ and $k_l$. Ground state energy demands that the following inequalities hold: $k_1 > k_2 > k_3 > \ldots > k_l$. Equivalently, one may study the problem by tracking for each *d* how many *z*-states are occupied (let us call this number *k*):

$$\begin{array}{cccc} \{0,1\} & \{0,2\} & . \; . & \{0,k\} \\ \{1,1\} & \{1,2\} & . \; . & \{1,k\} \\ . & . & . \; . & . \\ . & . & . \; . & . \\ \{l_1-1,1\} & \{l_2-1,2\} & . \; . & \{l_k-1,k\} \end{array} \tag{A2}$$

with $\sum_{i=1}^{k} l_i = \rho$ and $l_1 > l_2 > l_3 > ... > l_k$ .

In the latter scheme, let us first examine the case $k = 1$ (only the lowest QW state is occupied). Then, the general pattern from (A2) is quite simple:

$\{0,1\},\{1,1\}...\{l_1 - 1,1\}$ . Because $l_1 = \rho$ , we have: $\{0,1\},\{1,1\}...\{\rho - 1,1\}$ . This pattern will not be valid for every *d*. Instead, it holds only for $d \leq \sqrt{3\pi\Phi_o / \left(4(\rho - 1)\text{B}\right)}$ . This critical value corresponds to the case when the state $\{0,2\}$ is occupied (when the energy of ($\rho$-1,1) is equal to the energy of (0.2)). The total energy is then:

$$\text{E} = \frac{2\Phi}{\Phi_o}\sum_{n=0}^{\rho-2}\hbar\omega\left(n + \tfrac{1}{2}\right) + \frac{2\Phi}{\Phi_o}(\rho - 1)\frac{\hbar^2\pi^2}{2md^2} + \left(N - (\rho - 1)\frac{2\Phi}{\Phi_o}\right)\frac{\hbar^2\pi^2}{2md^2} + \left(N - (\rho - 1)\frac{2\Phi}{\Phi_o}\right)\hbar\omega\left(\rho - \tfrac{1}{2}\right)$$

or, in units of 2D Fermi energy:

$$E = NE_f\left[-2\rho(\rho - 1)\left(\frac{B}{n_A\Phi_o}\right)^2 + 2\left(\frac{B}{n_A\Phi_o}\right)(\rho - \frac{1}{2}) + \frac{\pi}{2n_Ad^2}\right] \tag{A3}$$

In the limit $\rho \to \infty$ , then $B = \frac{1}{2\rho} n_A\Phi_o$ (which is very small (goes to zero)) and (A3) becomes:

$$E = NE_f\left[-2\rho^2\left(\frac{B}{n_A\Phi_o}\right)^2 + 2\left(\frac{B}{n_A\Phi_o}\right)\rho + \frac{\pi}{2n_Ad^2}\right] = NE_f\left[-\tfrac{2}{4} + 1 + \frac{\pi}{2n_Ad^2}\right] = NE_f\left[\tfrac{1}{2} + \frac{\pi}{2n_Ad^2}\right]$$

The factor ½ in the above expression indicates that a Fermi circle has been created in $n_z$=1, in full correspondence with (3.7). So we see that the above scheme reproduces (for *k*=1, only one QW-level involved) the *B*=0 result in the limit of $\rho \to \infty$ . But what about the chemical potential of the system? This is defined to be

$$\mu = \varepsilon\left(\rho - 1,1\right) = \hbar\omega\left(\rho - \tfrac{1}{2}\right) + \frac{\hbar^2\pi^2}{2md^2} \xrightarrow{\rho\to\infty} \hbar\omega\rho + \frac{\hbar^2\pi^2}{2md^2} = \frac{\hbar^2 2\pi n_A}{2m} + \frac{\hbar^2\pi^2}{2md^2}$$ , as expected.

Now let us make things slightly more complicated. If $d \geq \sqrt{3\pi\Phi_o / \left(4(\rho - 1)\text{B}\right)}$ then $k = 2$ and the general pattern (A2) now is:

$$\begin{matrix}\{0,1\},\{1,1\}...........\{l_1 - 1,1\} \\ \{0,2\},\{1,2\}...\{l_2 - 1,2\}\end{matrix} \quad \text{or} \quad \begin{matrix}\{0,1\},\{1,1\}...........\{l_1 - 1,1\} \\ \{0,2\},\{1,2\}...\{\rho - l_1 - 1,2\}\end{matrix} \quad \text{with } \rho \geq 2 \text{ and } l_1 \geq \rho - l_1$$

If we want to restrict ourselves to occupation of only 2 QW levels, we must seek the appropriate values of $l_1$. Let us see some examples of the possible occupational scenarios for a certain value of $\rho$:

$\rho$ =5

$$\begin{matrix}\{0,1\},\{1,1\},\{2,1\},\{3,1\},\{0,2\} & \sqrt{3/4} \leq d \leq \sqrt{3/3} \\ \{0,1\},\{1,1\},\{2,1\},\{0,2\},\{3,1\} & \sqrt{3/3} \leq d \leq \sqrt{3/2} \\ \{0,1\},\{1,1\},\{0,2\},\{2,1\},\{1,2\} & \sqrt{3/2} \leq d \leq \sqrt{3/1} \\ \{0,1\},\{0,2\},\{1,1\},\{1,2\},\{2,1\} & \sqrt{3/1} \leq d \leq \sqrt{4}\end{matrix} \qquad \text{(d in units of } \sqrt{\pi\Phi_o/4B}\text{)} \tag{A4}$$

**$J$ =6**

$$\begin{array}{ll} \{0,1\},\{1,1\},\{2,1\},\{3,1\},\{4,1\},\{0,2\} & \sqrt{3/5} \le d \le \sqrt{3/4} \\ \{0,1\},\{1,1\},\{2,1\},\{3,1\},\{0,2\},\{4,1\} & \sqrt{3/4} \le d \le \sqrt{3/3} \\ \{0,1\},\{1,1\},\{2,1\},\{0,2\},\{3,1\},\{1,2\} & \sqrt{3/3} \le d \le \sqrt{3/2} \\ \{0,1\},\{1,1\},\{0,2\},\{2,1\},\{1,2\},\{3,1\} & \sqrt{3/2} \le d \le \sqrt{8/3} \end{array} \qquad \text{(d in units of } \sqrt{\pi\Phi_o/4B}\text{)} \qquad \text{(A5)}$$

**$J$ =7**

$$\begin{array}{ll} \{0,1\},\{1,1\},\{2,1\},\{3,1\},\{4,1\},\{5,1\},\{0,2\} & \sqrt{3/6} \le d \le \sqrt{3/5} \\ \{0,1\},\{1,1\},\{2,1\},\{3,1\},\{4,1\},\{0,2\},\{5,1\} & \sqrt{3/5} \le d \le \sqrt{3/4} \\ \{0,1\},\{1,1\},\{2,1\},\{3,1\},\{0,2\},\{4,1\},\{1,2\} & \sqrt{3/4} \le d \le \sqrt{3/3} \\ \{0,1\},\{1,1\},\{2,1\},\{0,2\},\{3,1\},\{1,2\},\{4,1\} & \sqrt{3/3} \le d \le \sqrt{3/2} \\ \{0,1\},\{1,1\},\{0,2\},\{2,1\},\{1,2\},\{3,1\},\{2,2\} & \sqrt{3/2} \le d \le \sqrt{5/2} \end{array} \qquad (d \text{ in units of } \sqrt{\pi\Phi_o/4B}) \qquad \text{(A6)}$$

The states are written in such a way that the single electron energy is gradually increased when moving from left to the right. Of course, there are more $d$ windows which are not written in each $J$, because we would then have more than two QW states occupied. The next step is to try to find in general (for every $J$) the values of integer $l_1$ which restricts the system in the two lowest QW levels. As one can verify from the above simple examples there must be one maximum value of $l_1$ and one minimum value (before state $\{0,3\}$ is occupied). The maximum value is $l_{1\max} = \rho - 1$. For the minimum value of $l_1$, we must see that for some $J$ the chemical potential of the system is either the state $\{l_{1\min} - 1,1\}$ or $\{\rho - l_{1\min} - 1,2\}$. For example, if $\rho = 6$ then from (A5) we have $l_{1\min} = 4$ or if $\rho = 7$ we have from (A6) that $l_{1\min} = 4$ again.

Let us suppose that, just before $\{0,3\}$ is occupied, the chemical potential is $\{l_{1\min} - 1,1\}$. So, when $\varepsilon\{l_{1\min} - 1,1\} = \varepsilon\{0,3\}$ or $d = \sqrt{8/(l_{1\min} - 1)}$ the state $\{0,3\}$ is occupied. This thickness must definitely be smaller than the one defined by the equation $\varepsilon\{l_{1\min} - 1,1\} = \varepsilon\{\rho - l_{1\min},2\}$ or $d = \sqrt{3/(2l_{1\min} - \rho - 1)}$. So, it must hold that:

$$l_{1\min} \le \frac{8\rho + 5}{13}$$

At the same time, it must definitely be greater than the one defined by the equation $\varepsilon\{l_{1\min} - 1,1\} = \varepsilon\{\rho - l_{1\min} - 1,2\}$ or $d = \sqrt{3/(2l_{1\min} - \rho)}$ :

$$l_{1\min} > \frac{8\rho - 3}{13}$$

Now, if there is an integer number inside this window of values of $l_{1\min}$, for some values of $\rho$ it defines $l_{1\min}$ and sets $\{l_{1\min} - 1,1\}$ as the chemical potential of the system. Equivalently,

$$\frac{13n}{8} + 1 \le \rho < \frac{13n}{8} + 2 \text{ with } l_{1\min} = n + 1 \text{ and } n = 1,2,3\ldots\infty \qquad \text{(A7)}$$

$$\begin{array}{lll} n=1 & \rho=2 & l_{1\min}=2 \\ n=2 & \rho=3 & l_{1\min}=3 \\ n=3 & \rho=6 & l_{1\min}=4 \\ n=4 & \rho=8 & l_{1\min}=5 \\ n=5 & \rho=10 & l_{1\min}=6 \end{array} \tag{A8}$$

then the above is automatically satisfied. In conclusion, when $\sqrt{8\pi\Phi_o / 4\left(l_{1\min}-1\right)B} \geq d \geq \sqrt{3\pi\Phi_o / \left(4(\rho-1)B\right)}$ then two (and only two) QW levels are occupied, where, in the final $d$-window the chemical potential is the state $\{l_{1\min}-1,1\}$. Note that as $\rho \to \infty$, $l_{1\min} = 8\rho/13$ (not being necessarily an integer), $B = \left(1/2\rho\right) n_A \Phi_o$ and then $\sqrt{13\pi/2n_A} \geq d \geq \sqrt{3\pi/2n_A}$, in full accordance with (3.24). The number of $d$-windows is $NW = 2\left(\rho - l_{1\min}\right)$ where when $\rho \to \infty$ then $NW = 10\rho/13$. We then conclude that the magnetic field splits the usual $d$-window (3.24) in $2\left(\rho - l_{1\min}\right) - 1$ separate windows which do not exist if $B$=0. But if the state $\{\rho - l_{1\min} - 1, 2\}$ is the chemical potential of the system just before $\{0,3\}$ is occupied, then we find that

$$\frac{8\rho-3}{13} \geq l_{1\min} > \frac{8\rho-8}{13} \tag{A9}$$

Let us now examine the case: $l$=1. Then, the pattern will obviously be: $\{0,1\}$ $\{0,2\}$ . . $\{0,\rho\}$, that is, only the lowest Landau level is occupied. Here is obvious that $k_1 = \rho$ when $l$=1. It is easy to determine the critical value of the thickness necessary for this pattern to exist: we compare the chemical potentials of states $\{0,\rho\}$ and $\{1,1\}$:

$$d \geq \sqrt{\frac{\pi(\rho^2-1)\Phi_o}{4\mathrm{B}}} \tag{A10}$$

with total energy:

$$E = NE_f \left[ \left( \frac{B}{n_A \Phi_o} \right) + \frac{\pi\rho^2}{2n_A d^2} - \left( \frac{B}{n_A \Phi_o} \right) \left( \frac{\pi}{n_A d^2} \right) \frac{\rho(\rho-1)(4\rho+1)}{6} \right] \tag{A11}$$

A final interesting observation concerns the limit when the thickness $d$ goes to infinity, in a manner that the volume density remains always a constant. Let us examine eq. (A11) which is valid for large thicknesses. We take $d \to \infty$ and at the same time $\rho \to \infty$ so that $n_V = n_A / d$ remains constant. Eq. (A11) then becomes:

$$E/N = \frac{E_f(3D)}{\left(3\pi^2\right)^{2/3}} \left[ \left( \frac{2\pi B}{n_V^{2/3} \Phi_o} \right) + \frac{\pi^2}{n_V^{2/3}} \left( \frac{\rho}{d} \right)^2 - \frac{4}{3} \left( \frac{B}{\Phi_o} \right) \left( \frac{\pi^2}{n_V^{5/3}} \right) \left( \frac{\rho}{d} \right)^3 \right], \tag{A12}$$

where $E_f\left(3D\right) = \hbar^2 \left(3\pi^2 n_V\right)^{2/3} / 2m$ is the 3D Fermi energy. From (A0) for $n_A = nd$ we have (when $\rho \to \infty$):

$$B = \frac{1}{2\rho} n_V d \Phi_o \Rightarrow \frac{\rho}{d} = \frac{n_V \Phi_o}{2B},$$

which means that the previous window is now shrunk into a single value. Substituting this in (A12) we have:

$$E/N = \frac{E_f(3D)}{\left(3\pi^2\right)^{2/3}} \left[ \left( \frac{2\pi B}{n_V^{2/3} \Phi_o} \right) + \frac{\pi^2}{12} \left( \frac{n_V^{2/3} \Phi_o}{B} \right)^2 \right] \tag{A13}$$

At the same time, from (A10) it must hold that:

$$d \geq \sqrt{\frac{\pi\rho^2\Phi_o}{4B}} \Rightarrow B \geq \frac{\pi\Phi_o}{4}\left(\frac{\rho}{d}\right)^2 \Rightarrow B \geq \frac{\pi\Phi_o}{4}\frac{n_V^2\Phi_o{}^2}{4B^2} \Rightarrow B \geq \left(\frac{\pi}{16}\right)^{1/3} n_V^{2/3}\Phi_o\,, \qquad \text{(A14)}$$

which is just (5.5), namely, the first critical value of $B$ before the second LL is about to be filled by electrons. The above results are in a full agreement with the corresponding 3D results that were presented in Section 5, for the first window of values of $B$ (when only one LL is occupied). In Section 5 the electronic problem is studied when the thickness is infinitely large, but the only difference is the boundary conditions that are taken to be periodic. The conclusion is once again that when the available space is very large boundary conditions do not have an influence on thermodynamic properties (at least for this topologically conventional system).